\documentclass[useAMS,usenatbib]{mn2e}
\usepackage{graphicx}
\usepackage{epstopdf}

\def\Hii{\hbox{H~{\sc ii}}}

\title[Chemical differentiation in HMSF regions]{Chemical differentiation in regions of high mass star formation II. Molecular multiline and dust continuum studies of selected
objects}

\author[I. Zinchenko, P. Caselli \& L. Pirogov]{I. Zinchenko$^{1,2,3}$\thanks{E-mail:
zin@appl.sci-nnov.ru (IAP)},
P. Caselli$^{4}$
and L. Pirogov$^{1}$\\
$^1$Institute of Applied Physics of the Russian Academy of Sciences, Ulyanova 46,
603950 Nizhny Novgorod, Russia \\
$^2$Nizhny Novgorod University , Gagarin av. 23,
603950 Nizhny Novgorod, Russia \\
$^3$Helsinki University Observatory, T\"ahtitorninm\"aki, P.O. Box 14, FIN-00014
University of Helsinki, Finland\\
$^4$School of Physics and Astronomy, University of Leeds, Leeds LS2 9JT, UK
}

\begin{document}

\date{ }

\pagerange{\pageref{firstpage}--\pageref{lastpage}} \pubyear{2009}

\maketitle

\label{firstpage}

\begin{abstract}
{The aim of this study is to investigate systematic chemical differentiation of molecules in regions of high mass star formation.}{We observed five prominent sites of high mass star formation in HCN, HNC, HCO$^+$, their isotopes, C$^{18}$O, C$^{34}$S and some other molecular lines, for some sources both at 3 and 1.3~mm and in continuum at 1.3~mm. Taking into account earlier obtained data for N$_2$H$^+$ we derive molecular abundances and physical parameters of the sources (mass, density, ionization fraction, etc.). The kinetic temperature is estimated from CH$_3$C$_2$H observations. Then we analyze correlations between molecular abundances and physical parameters and discuss chemical models applicable to these species.}{The typical physical parameters for the sources in our sample are the following: kinetic temperature in the range $\sim 30-50$~K (it is systematically higher than that obtained from ammonia observations and is rather close to dust temperature), masses from tens to hundreds solar masses, gas densities $\sim 10^5$~cm$^{-3}$, ionization fraction $\sim 10^{-7}$. In most cases the ionization fraction slightly (a few times) increases towards the embedded YSOs. The observed clumps are close to gravitational equilibrium. There are systematic differences in distributions of various molecules. The abundances of CO, CS and HCN are more or less constant. There is no sign of CO and/or CS depletion as in cold cores. At the same time the abundances of HCO$^+$, HNC and especially N$_2$H$^+$ strongly vary in these objects. They anti-correlate with the ionization fraction and as a result decrease towards the embedded YSOs. For N$_2$H$^+$ this can be explained by dissociative recombination to be the dominant destroying process. }{N$_2$H$^+$, HCO$^+$, and HNC are valuable indicators of massive protostars.}
\end{abstract}


\section{Introduction}

It is now well
established that the central parts of dense low mass cloud cores
suffer strong depletion of molecules onto dust grains.  Best
studied is CO, which has been shown to be depleted in, e.g., L1544
\citep{Caselli99},
IC 5146 \citep{Kramer99}, L1498
\citep*{Willacy98}, and L1689B \citep{Jessop01}.  Species related to CO, such as HCO$^+$,
are also expected to disappear at gas densities above
$\sim 10^5$ cm$^{-3}$ \citep{Caselli02}.  Moreover,
\citet{Tafalla02} and \citet{Bergin01} have shown that CS also depletes out in the central
parts of dense cores, suggesting that CS (so far considered a high
density tracer) does not actually probe the central core regions.
On the other hand, N$_2$H$^+$ is an excellent tracer of dust continuum
emission \citep{Caselli02}, implying that this species
does not deplete out (due to the volatility of the parent species N$_2$).

In more massive cores, depletion is probably active in dense regions
away from star forming sites where dust temperatures may be low enough
($T < 20$ K) for CO and CS abundances to drop and cause chemical differentiation \citep{Fontani06}.

Several years ago we mapped several tens of dense cores towards
water masers in CS(2--1) with the SEST-15m and Onsala 20m radio telescopes
\citep*{Zin95,Zin98}.
In 2000 many of them were mapped
in N$_2$H$^+$ \citep{Pirogov03}.
The goal was to identify dense clumps as local maxima in N$_2$H$^+$ maps
and to further investigate their properties. However, large differences between the N$_2$H$^+$ and CS distributions have been found.

In this situation it is important to understand which species better trace the total gas distribution: is it N$_2$H$^+$ as in low mass cores? What is the reason for this differentiation in warm clouds, where freeze-out is hardly effective? To answer these questions we
observed dust continuum emission and several additional molecular lines
towards selected sources which show significant differences between the
CS and N$_2$H$^+$ maps. The results for the southern sources (where we observed only N$_2$H$^+$ $J=1-0$, CS  $J=2-1$ and $J=5-4$ and dust continuum) have been
published separately \citep[][hereafter Paper~I]{Pirogov07}. In that paper we have shown that the differences in the CS and N$_2$H$^+$ maps cannot be explained by molecular excitation and/or line opacity effects but are caused by chemical differentiation of these species. We found that N$_2$H$^+$ abundance in many cases drops significantly towards embedded luminous YSOs. However, the reasons for this behavior were not clear. Two possible explanations were mentioned: an accelerated collapse model suggested by \citet{Lintott05} and dissociative recombination of N$_2$H$^+$.

Here we present and discuss the results for the northern sample where we observed also HCN, HNC, HCO$^+$, their isotopes, C$^{18}$O and some other molecular lines, for some sources both at 3 and 1.3~mm. These data help to understand better the chemical differentiation in these objects. In addition they give important information on their physical properties.

\section{Observations}

\subsection{Sources}
The sources for this investigation were selected from the sample of massive
cores studied by us earlier in various lines \citep*{Zin98,Zin00,Pirogov03}. The main criterium for this
selection was a presence of significant differences in the CS and N$_2$H$^+$
maps. The list of the sources is given in
Table~\ref{table:sources}. For S187, W3 and S140, the source coordinates correspond
to water masers. For S255, the IRAS position is used. In the DR-21 area we used the coordinates of an ammonia core as given by \citet*{Jijina99} which practically coincide (within a few arcsec) with the position of DR 21 (OH).
In the last column the distances to the sources
are indicated. The data sets obtained for these sources are different.

\begin{table}
%\centering
\caption{Source list.}
%\small
\begin{tabular}{llll}
%\noalign{\hrule}\noalign{\smallskip}
\hline%\hline
Source
          & $\alpha$(2000)    & $\delta$(2000)    & $D$    \\
          & (${\rm ^h\  ^m\  ^s }$) &($\degr$ $\arcmin$ $\arcsec$) & (kpc)\\
\noalign{\smallskip}\hline\noalign{\smallskip}
S~187         &01 23 15.0   &61 48 47    &  1.0 $^a$  \\
W3            &02 25 28.2   &62 06 58    &  2.1 $^b$  \\
%AFGL~6366 S   &06 05 40.1   &21 31 16    &  2.0 $^c$  \\
S~255         &06 12 53.3   &17 59 22    &  2.5 $^b$  \\
DR-21 NH$_3$  &20 39 00.4   &42 22 53    &  3.0 $^c$  \\
S~140         &22 19 18.2   &63 18 49    &  0.9 $^b$  \\
%\noalign{\smallskip}\noalign{\hrule}\noalign{\smallskip}}
\noalign{\smallskip}\hline\noalign{\smallskip}
\end{tabular}

$^a$\citet{fich},
$^b$\citet*{blitz},
%$^c$\citet{carpenter},
$^c$\citet*{harvey1}

\label{table:sources}
\end{table}

\subsection{Instruments and frequencies}

The sources were observed with the 20-m
Onsala, 12-m NRAO (which belongs now to the Arizona Radio Observatory) and 30-m IRAM radio telescopes in the 3 mm and
1.3 mm wavebands. Several molecular transitions were observed in each
waveband.
%The list of the lines and the basic observational parameters are presented in Table~\ref{table:obs}.
At IRAM 30m also the dust continuum emission was
mapped at 1.2 mm. The details of observations at each instrument are
given below. A part of these observations has been already published
(some of the CS and C$^{34}$S $J=2-1$ data by \citet{Zin98}
and the N$_2$H$^+$ $J=1-0$ data by \citet{Pirogov03}).
We express
the results of the line observations in units of main beam brightness temperature ($T_{\rm
mb}$) assuming the main beam efficiencies ($\eta_{\rm mb}$) as provided by the telescope documentation.

\subsubsection{Onsala observations}
The observations were performed with SIS receiver in a single-sideband (SSB)
mode using either dual beam switching with a
beam throw of 11{\farcm}5 or frequency switching.
As a backend until the year 2000 we used 2 filter spectrometers (usually
in parallel):
a 256 channel filterbank with 250~kHz resolution and a 512 channel
filterbank with 1~MHz resolution.
Since 2000 we used mainly the autocorrelator spectrometer tuned to 50~kHz
resolution.
Pointing was checked periodically by observations of nearby SiO
masers; the pointing accuracy was typically $\la 5$\arcsec. The half-power beam width (HPBW) was from about 35\arcsec\ at the highest frequencies to about 40\arcsec\ at the lowest frequencies.

The standard chopper-wheel technique was used for the calibration. The system temperature varied in a wide range depending on the weather and observing frequency, from $\sim 200$~K at lower frequencies in good weather to $\sim 1000$~K and more at higher frequencies and under cloudy conditions.

\subsubsection{NRAO 12m observations}
At the NRAO 12-m telescope only two sources from those listed in
Table~\ref{table:sources}
were observed: S187 and S255 (in the C$^{18}$O, CS, C$^{34}$S, SiO and methanol lines at 1.3 mm). As a result, the data sets for these sources
are the most complete ones.

The observations were perfomed in 2000 with the SIS receiver and two backends
in parallel: the filter bank with the 0.5~MHz resolution and MAC
autocorrelator with the 100~kHz resolution. We used frequency switching and position switching
observing modes. The pointing and focus were checked periodically
by observations of planets. The HPBW was 26--27\arcsec. The system temperature varied from $\sim 300$~K in good weather conditions to $\sim 1500$~K with increasing humidity.

\subsubsection{IRAM 30m observations}

At the IRAM 30m telescope we obtained maps of the continuum emission
from our sources at 1.2 mm and their maps in several components of the
CH$_3$C$_2$H $J=13-12$ transition. The continuum observations were
performed with the MAMBO bolometer array (the details of this instrument
are available at the IRAM Web site) and reduced with the MOPSIC package.
The spectral line observations
were done with the HERA (Heterodyne Receiver Array) and VESPA
autocorrelator backend. The pointing and focus
were checked periodically on nearby strong continuum sources. The antenna HPBW is about 12\arcsec\ at these frequencies. The typical system temperatures for HERA observations were in the range $\sim 200-400$~K.

\section{Observational results}

\begin{figure*}
\begin{minipage}[b]{0.49\textwidth}
\includegraphics[angle=-90,width=\textwidth]{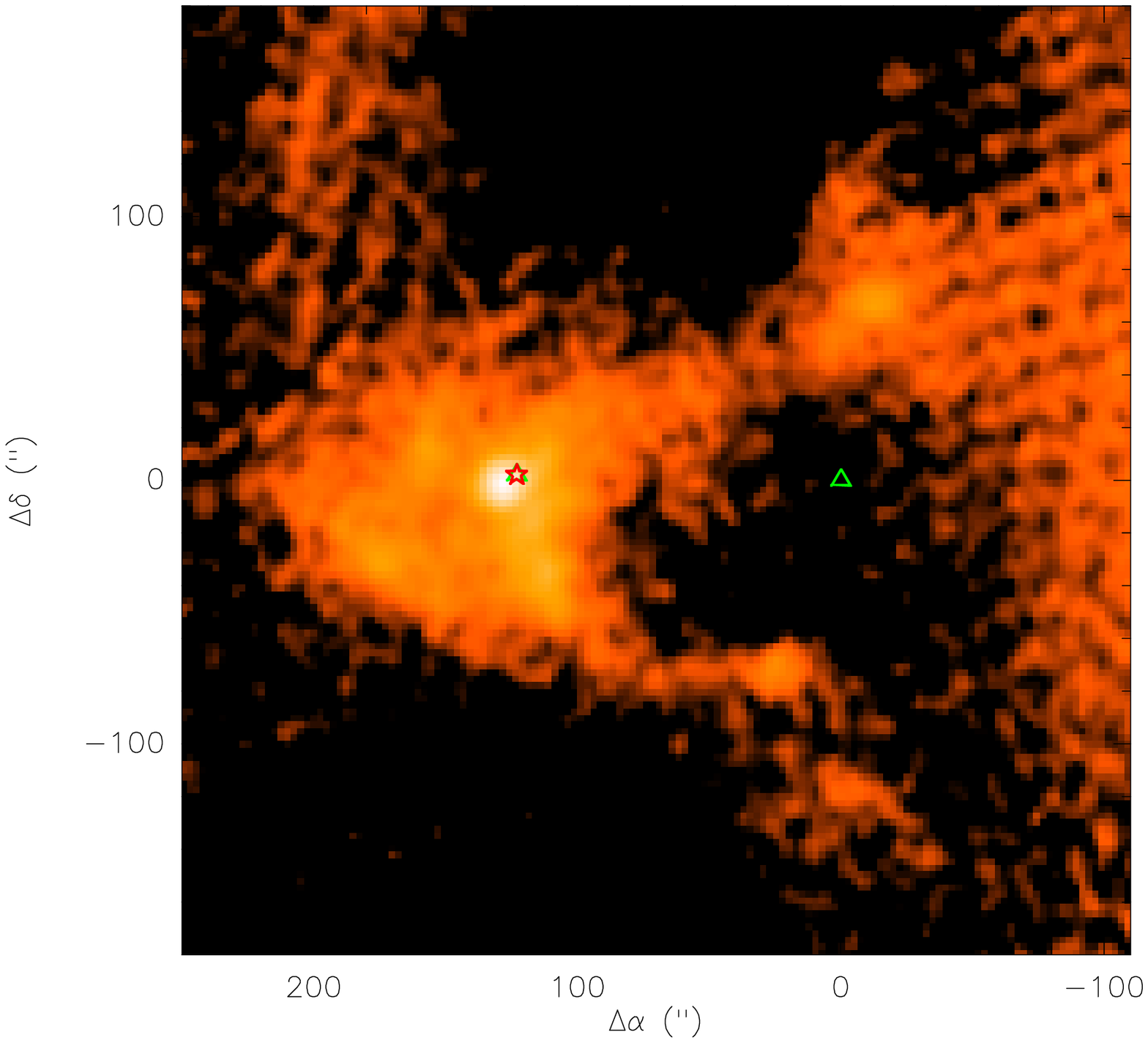}
\end{minipage}
\hfill
\begin{minipage}[b]{0.49\textwidth}
\includegraphics[angle=-90,width=\textwidth]{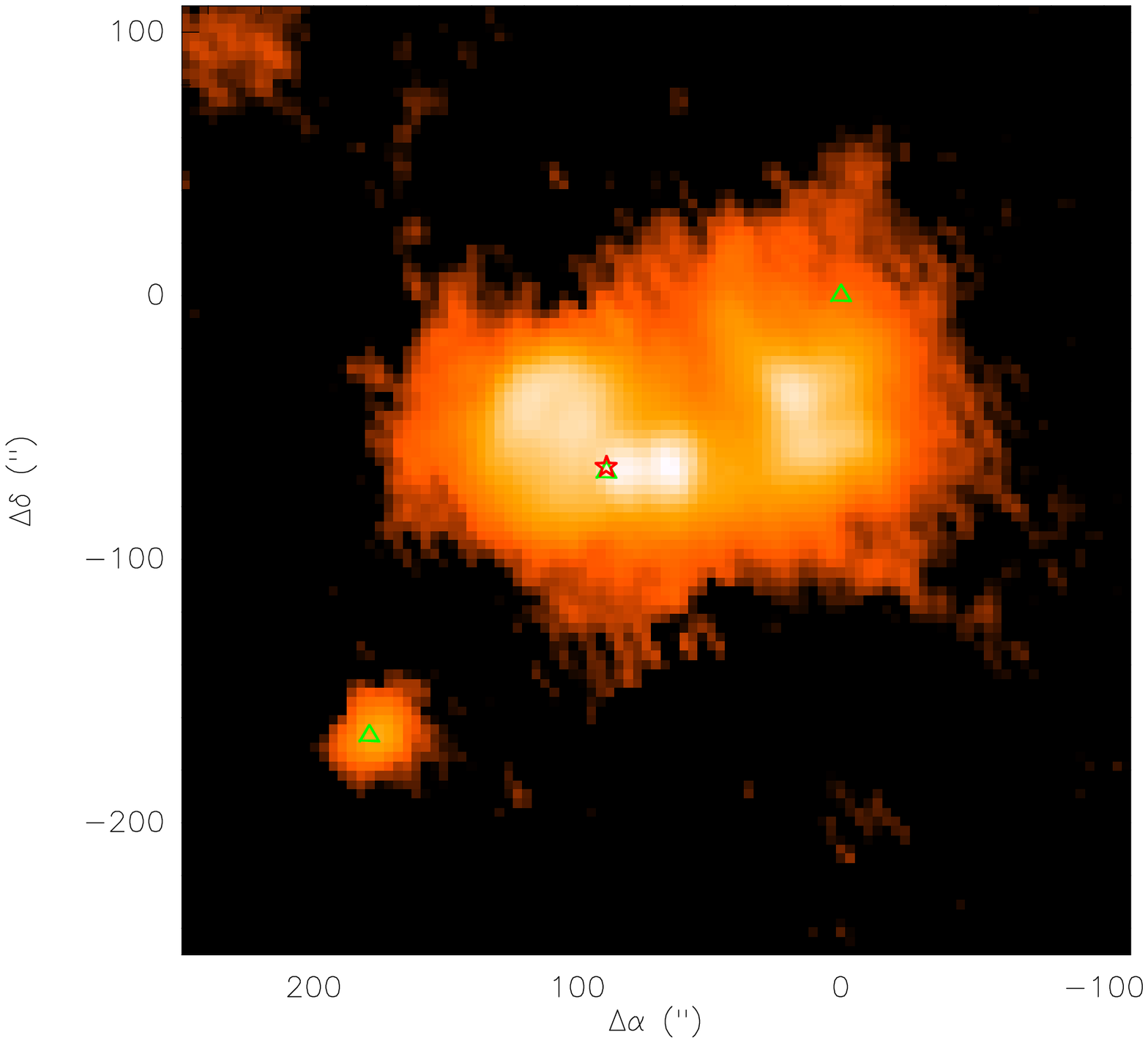}
\end{minipage}\\[2ex]
\begin{minipage}[b]{0.49\textwidth}
\includegraphics[angle=-90,width=\textwidth]{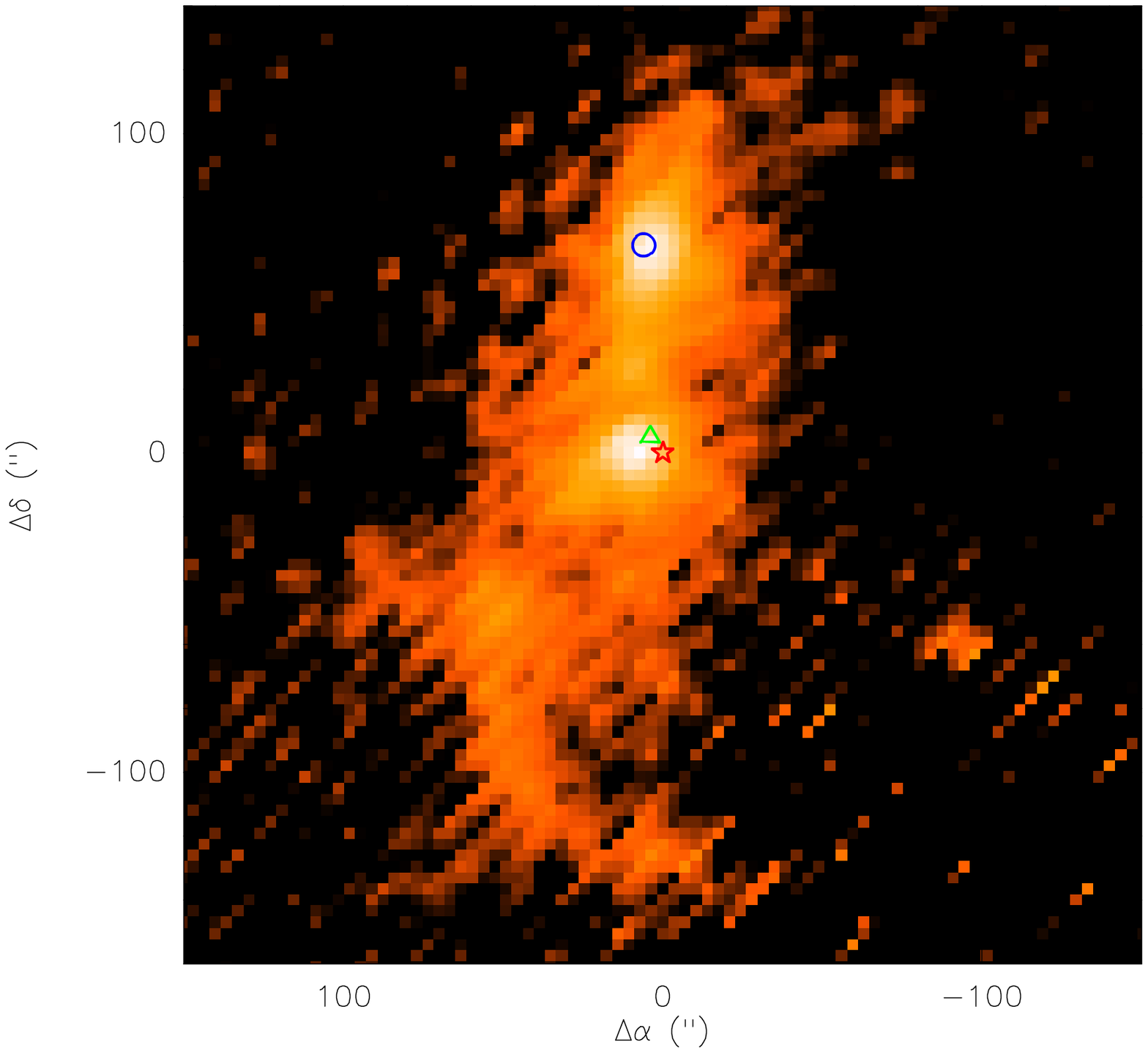}
\end{minipage}
\hfill
\begin{minipage}[b]{0.49\textwidth}
\includegraphics[angle=-90,width=\textwidth]{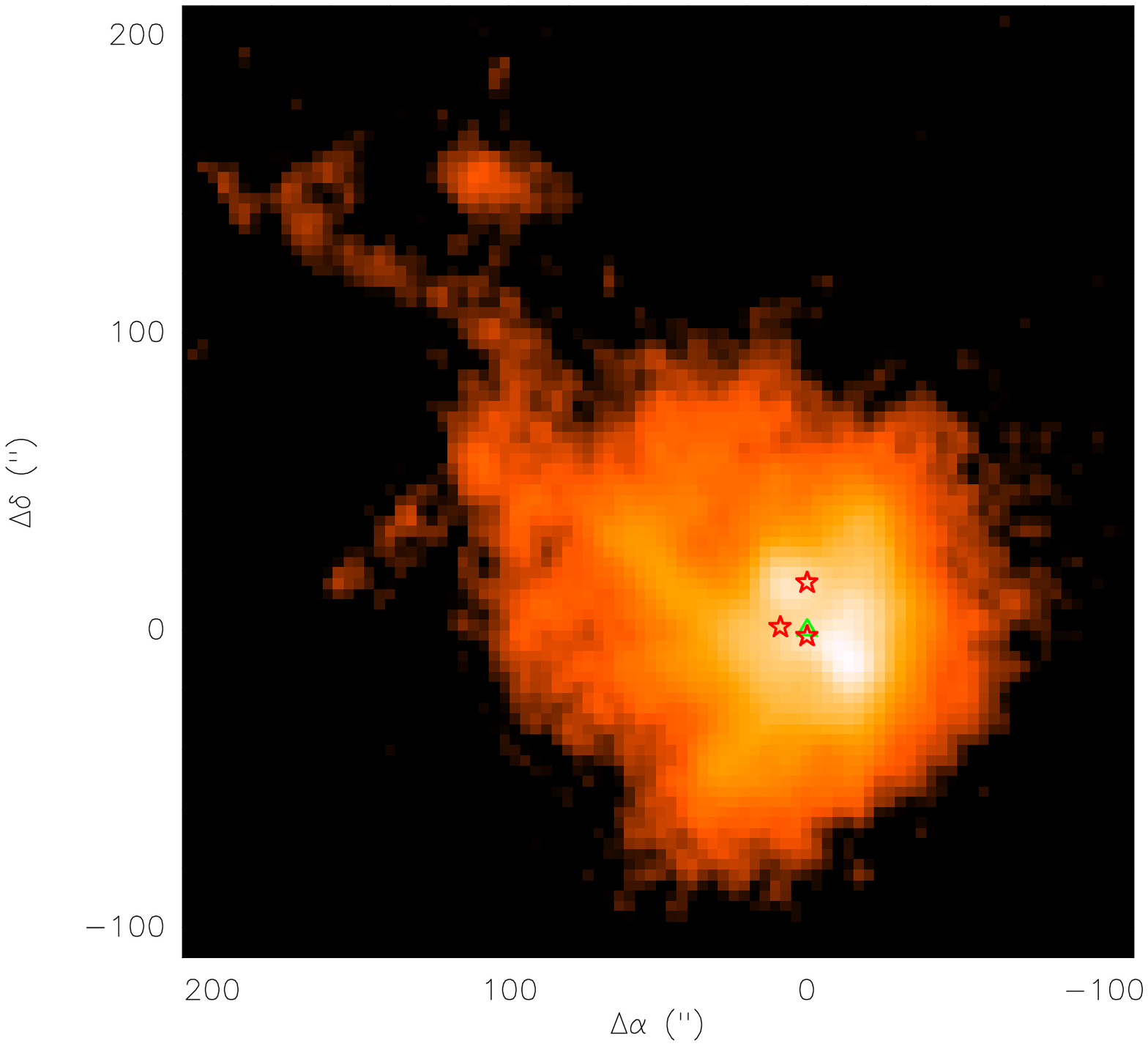}
\end{minipage}
\caption{Maps of the sample sources in continuum at 1.2 mm obtained at the IRAM 30-m telescope. The sources are (from left to right and from top to bottom): S187, W3, S255 and S140. The intensity scale is logarithmic. The triangles mark water masers, the stars indicate point IR sources (IRAS point sources for S187, S255 and W3; IRS1, IRS2 and IRS3 for S140). The circle corresponds to the submillimetre source SMA1 in the S255~N clump area \citep*{Cyganowski07}.} \label{fig:cont-maps}
\end{figure*}

We present the observational results in the form of maps
and tables where the line parameters at selected positions
are given. These positions correspond to different emission peaks in the
sources which are identified in the maps of molecular emission and in most cases can be seen in the continuum maps presented in Fig.~\ref{fig:cont-maps}. These maps are plotted using logarithmic scale for intensity in order to emphasize weak features. The continuum brightness, fluxes and angular sizes of the emission clumps towards selected positions
are summarized in Table~\ref{table:cont-res}. The sources are rather
extended in continuum. In order to provide a better comparison with
the molecular data, most of which were obtained with $\sim 30''$ (HPBW)
beams, we give the fluxes integrated over 1\arcmin\ circles
centered at the selected positions. These positions are labelled as ``CS'' and ``N$_2$H$^+$'' emission peaks according to our previous CS and N$_2$H$^+$ surveys \citep{Zin98,Pirogov03} and other available data. The sizes are derived from these fluxes and brightness as $\theta = \sqrt{4F/\pi B}$. For Gaussian brightness distribution this corresponds to size at the $1/e$ level which is about 20\% larger than the size at the half maximum level ($\theta_{0.5}$). In S140 it is hard to see a distinct continuum clump at the N$_2$H$^+$ emission peak, one can see rather an extended filament here. Nevertheless, we provide brightness and flux for this position, too.
The specific features of every source are briefly described
below. The spectral data reduction was performed with the GILDAS and XSpec
(Onsala data) software packages.

\begin{table}
\caption[]{Results of the continuum observations: brightness ($B$), fluxes ($F$) and angular sizes ($\theta$) of the emission clumps towards selected positions.}
\begin{tabular}{lrrllll}
\hline%\hline
\noalign{\smallskip}
Source &$\Delta\alpha$ &$\Delta\delta$ &$B$ &$F$ &$\theta$ &Peak\\
&($''$) &($''$) &(Jy/beam) &(Jy) &($''$) \\
\noalign{\smallskip}\hline\noalign{\smallskip}
S~187         &+160 &0  &0.15 &2.1 &45 &CS\\
              &0    &+80&0.09 &0.46 &28 &N$_2$H$^+$ \\
W3            &20 &$-$40&2.03 &19 &38 &CS \\
              &+160&$-$160&0.70 &2.23 &22 &N$_2$H$^+$ \\
%AFGL~6366 S   &0  &+20  &1.15 &6.2  & \\
%              &+100&+60 &0.33 &2.9  & \\
S~255         &0   &0   &1.35 &6.4 &27 &CS\\
              &0   &+60 &1.24 &6.4 &28 &N$_2$H$^+$ \\
S~140         &0   &0   &1.58 &15.9 &39 &CS\\
              &+40 &+20 &0.34 &5.3 &49 &N$_2$H$^+$ \\
%\noalign{\smallskip}\noalign{\hrule}\noalign{\smallskip}}
\noalign{\smallskip}\hline\noalign{\smallskip}
\end{tabular}
\label{table:cont-res}
\end{table}

\subsection{Molecular maps and line parameters}

\subsubsection{S187}
\begin{figure*}
\begin{minipage}[b]{0.33\textwidth}
\centering
\resizebox{\hsize}{!}{\rotatebox{-90}{\includegraphics{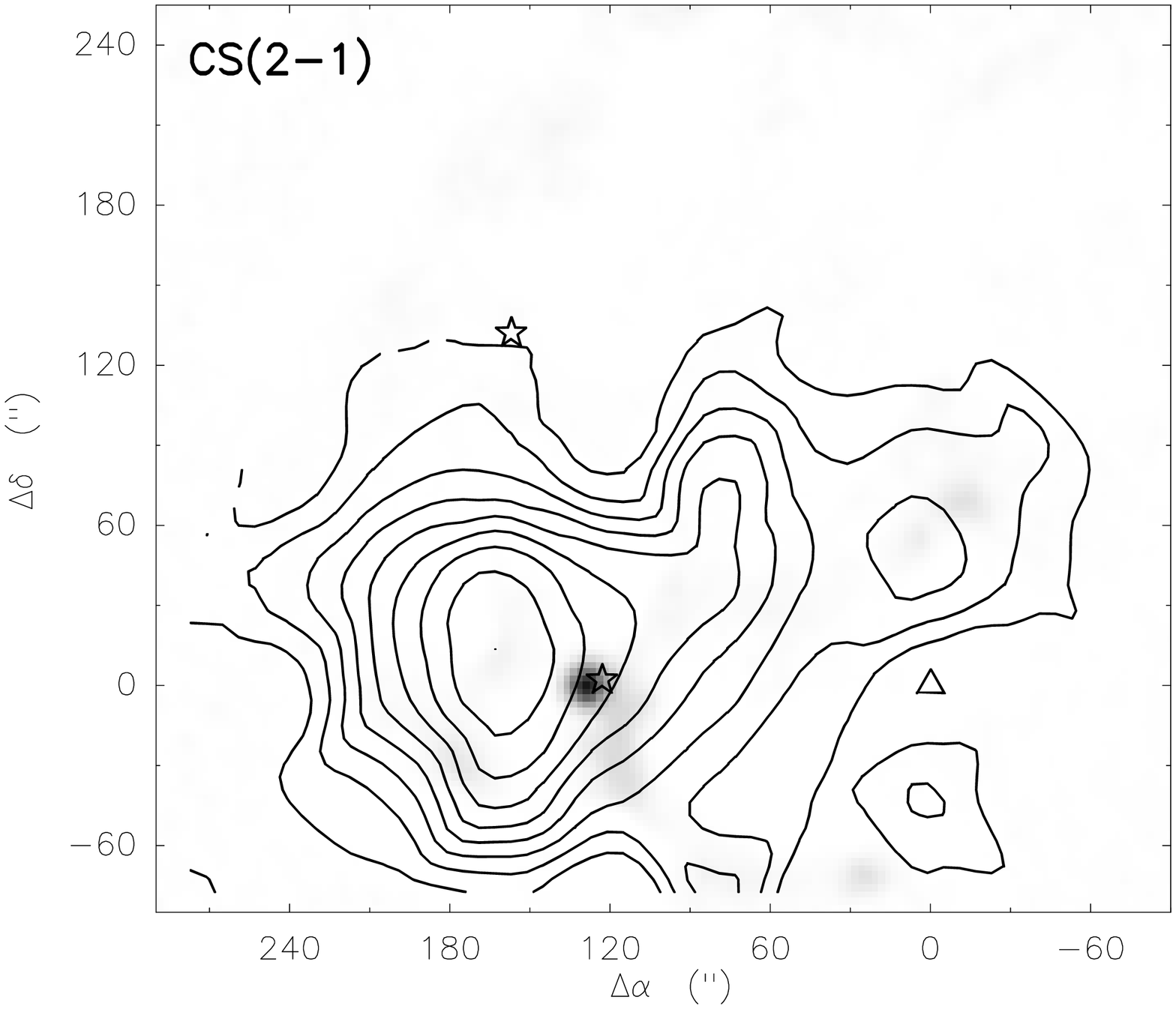}}}
\end{minipage}
\hfill
\begin{minipage}[b]{0.33\textwidth}
\centering
\resizebox{\hsize}{!}{\rotatebox{-90}{\includegraphics{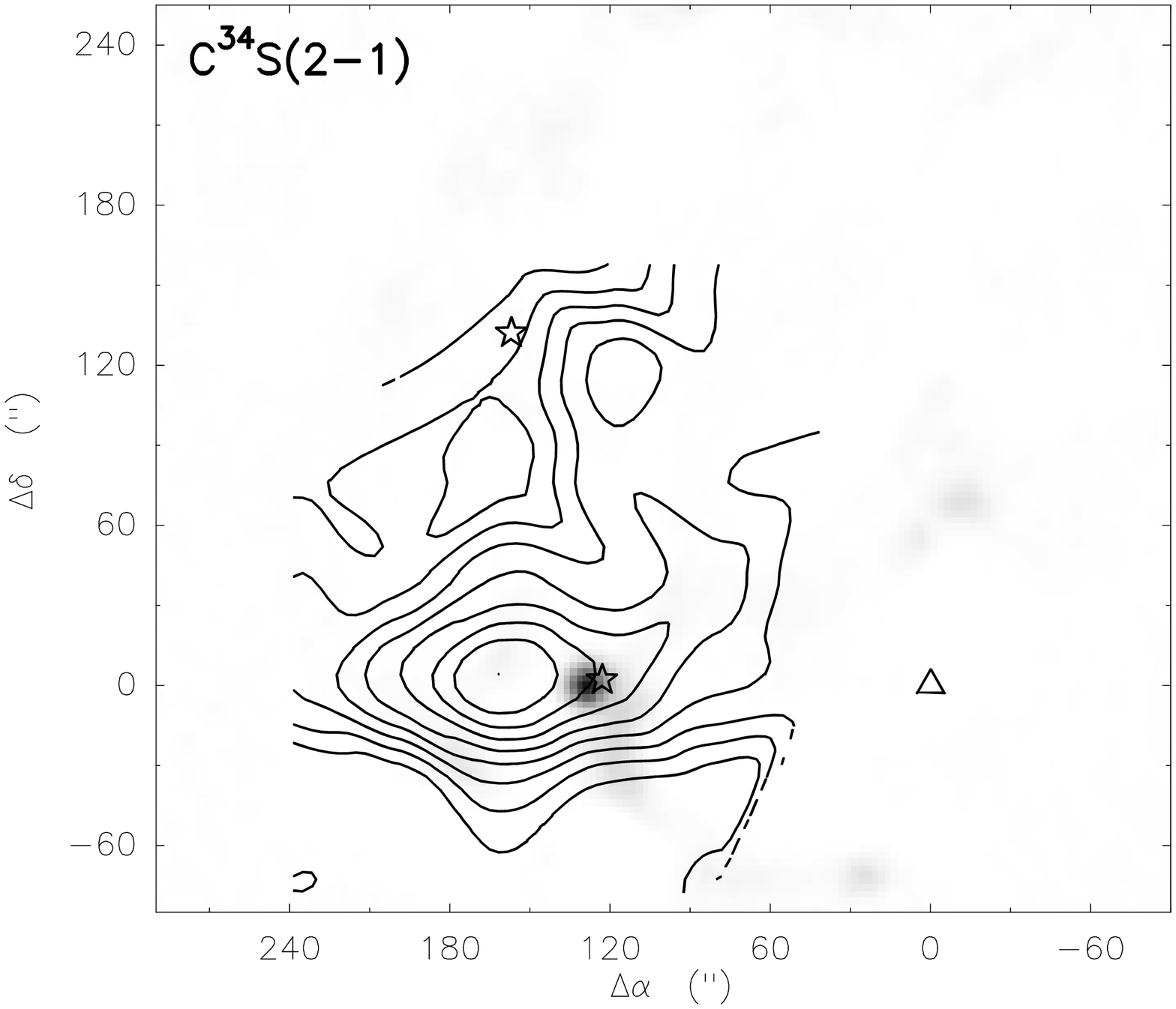}}}
\end{minipage}
\hfill
\begin{minipage}[b]{0.33\textwidth}
\centering
\resizebox{\hsize}{!}{\rotatebox{-90}{\includegraphics{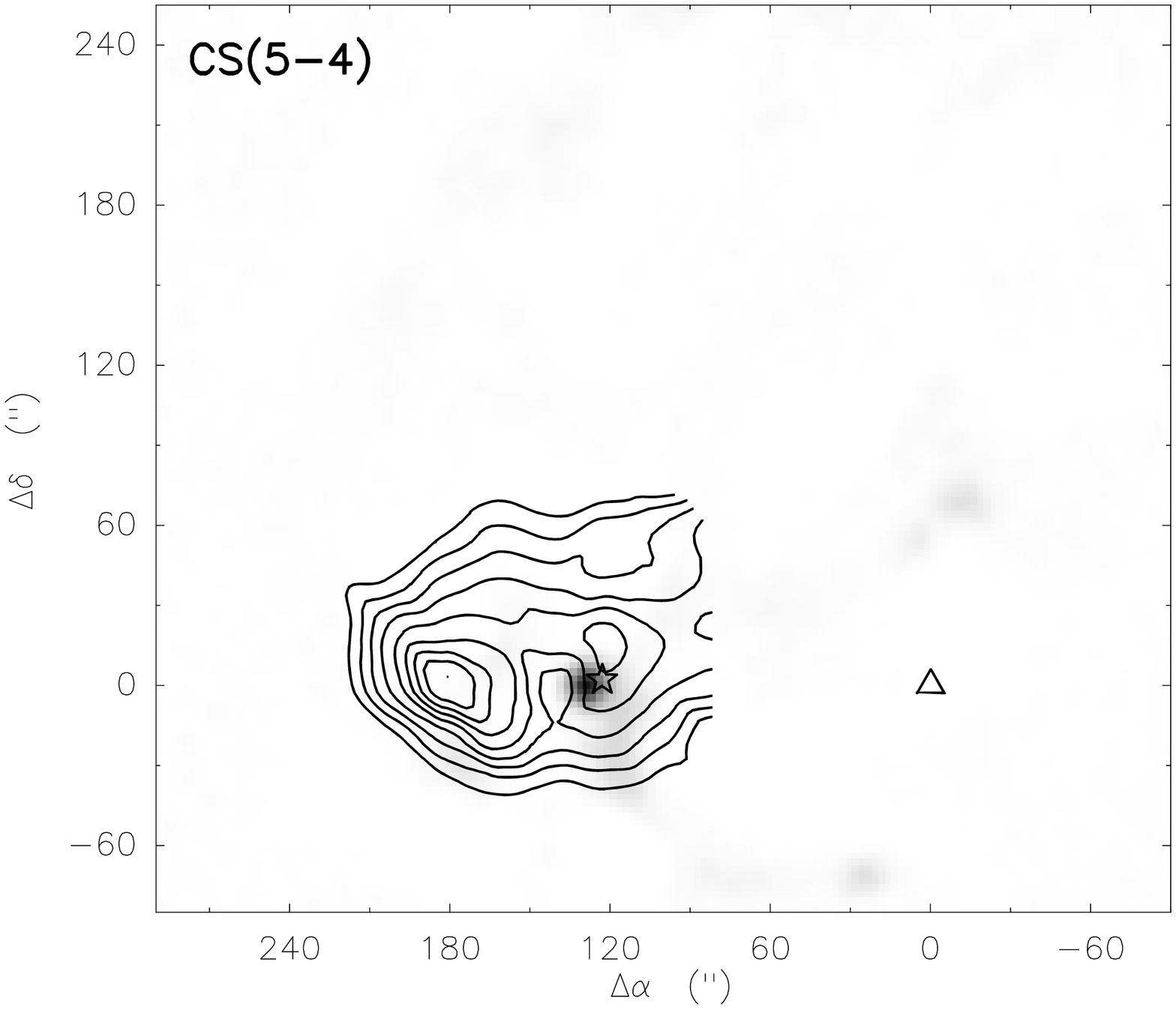}}}
\end{minipage}
\\[1ex]
\begin{minipage}[b]{0.33\textwidth}
\centering
\resizebox{\hsize}{!}{\rotatebox{-90}{\includegraphics{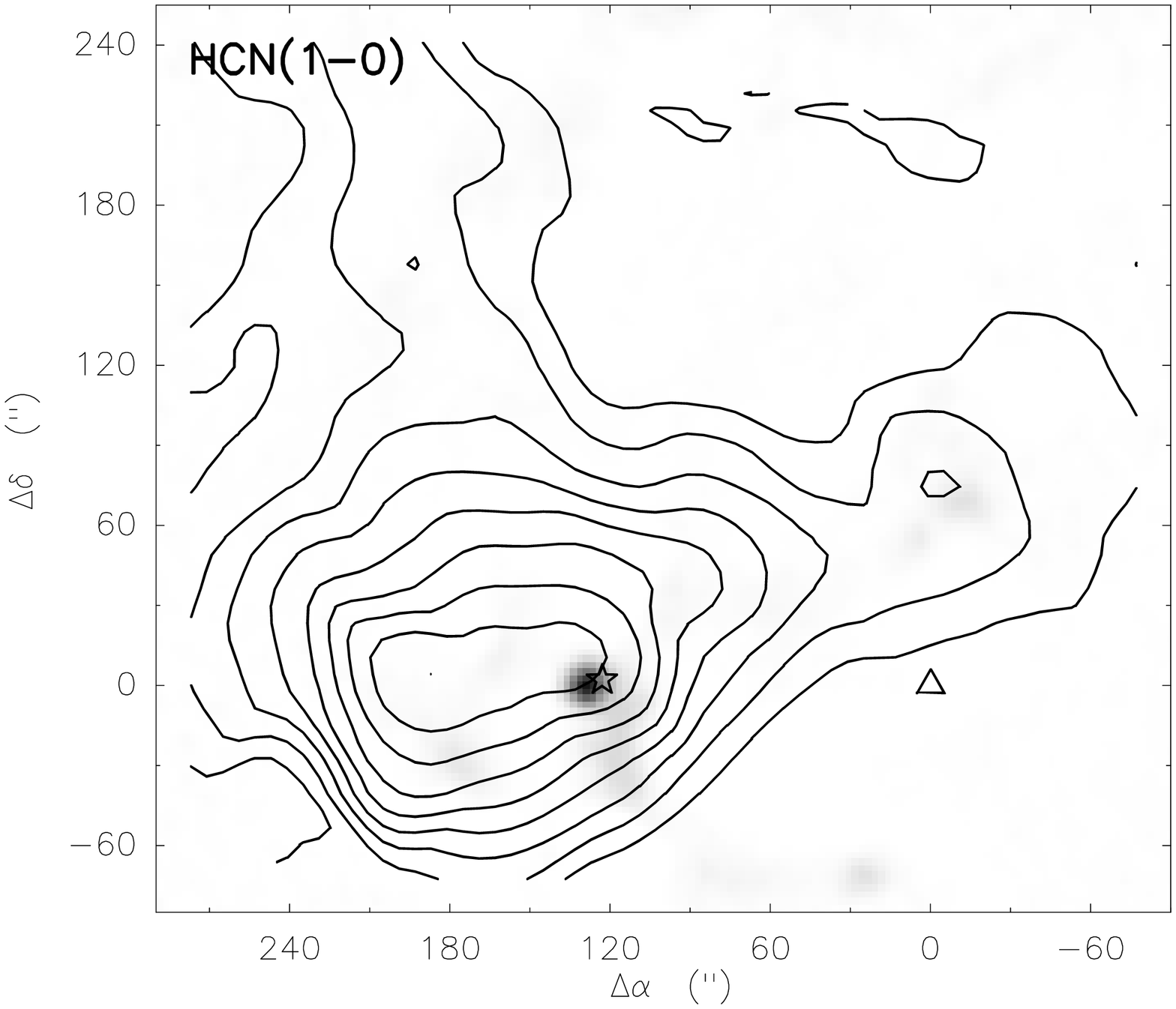}}}
\end{minipage}
\hfill
\begin{minipage}[b]{0.33\textwidth}
\centering
\resizebox{\hsize}{!}{\rotatebox{-90}{\includegraphics{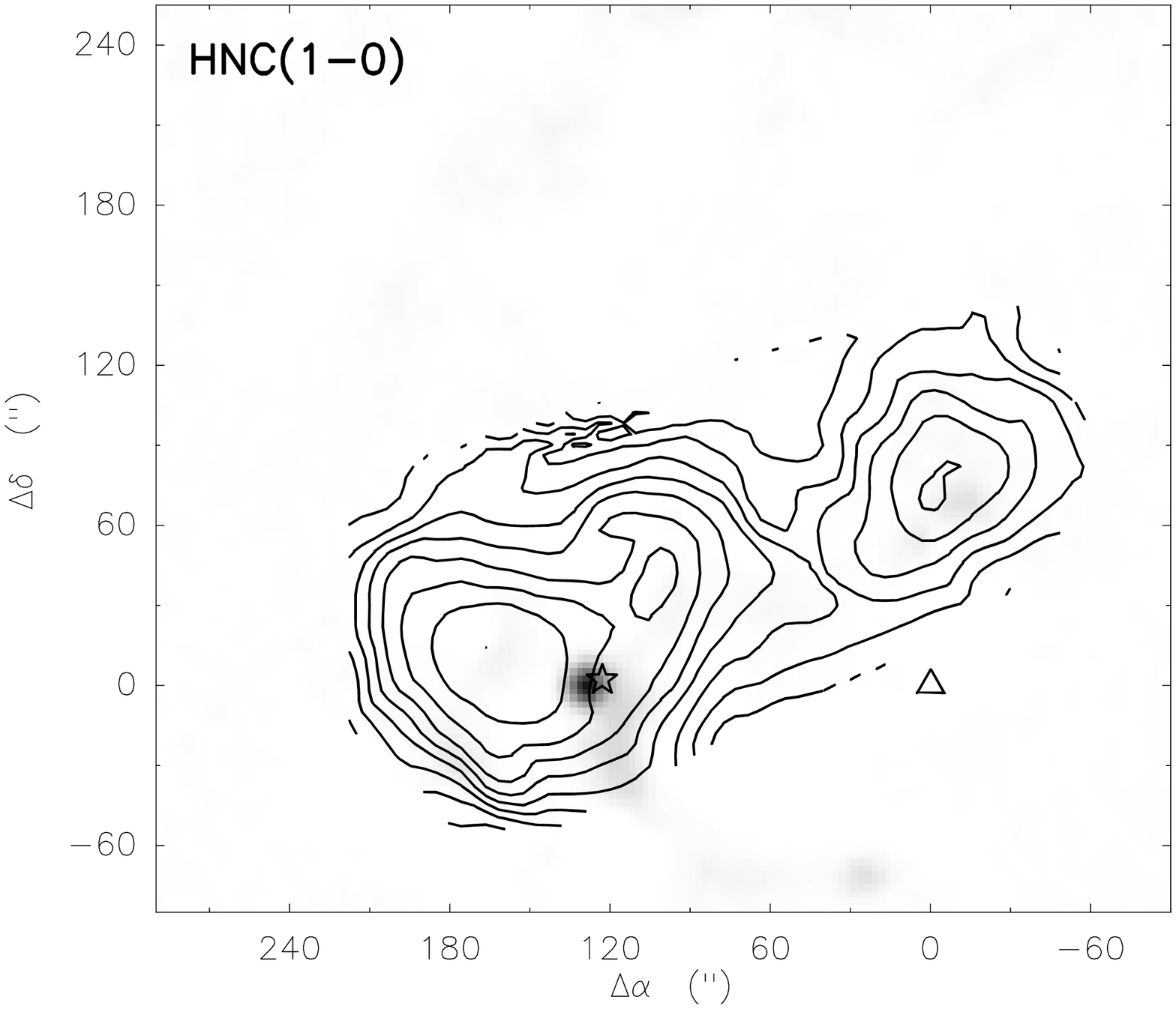}}}
\end{minipage}
\hfill
\begin{minipage}[b]{0.33\textwidth}
\centering
\resizebox{\hsize}{!}{\rotatebox{-90}{\includegraphics{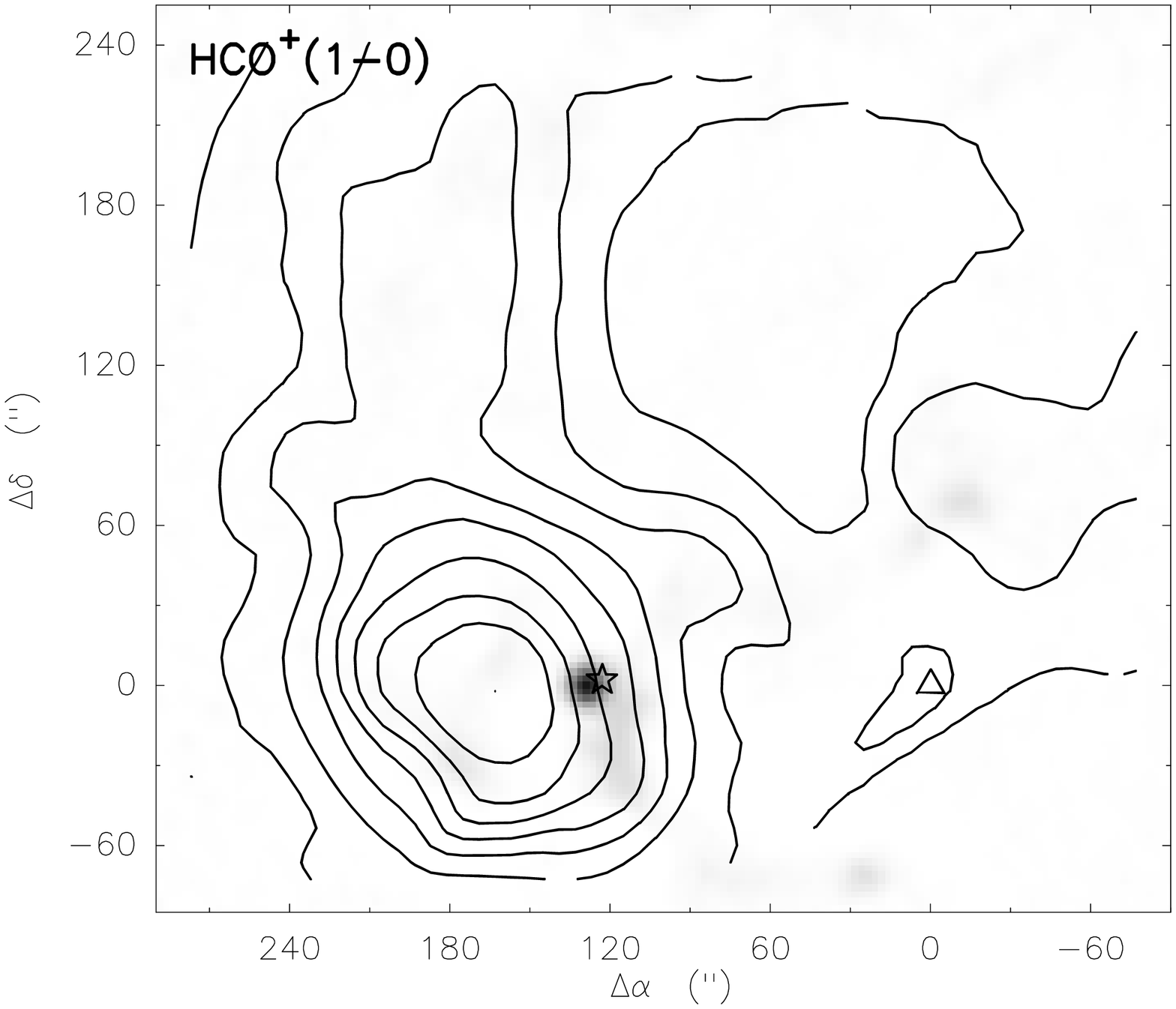}}}
\end{minipage}
\\[1ex]
\begin{minipage}[b]{0.33\textwidth}
\centering
\resizebox{\hsize}{!}{\rotatebox{-90}{\includegraphics{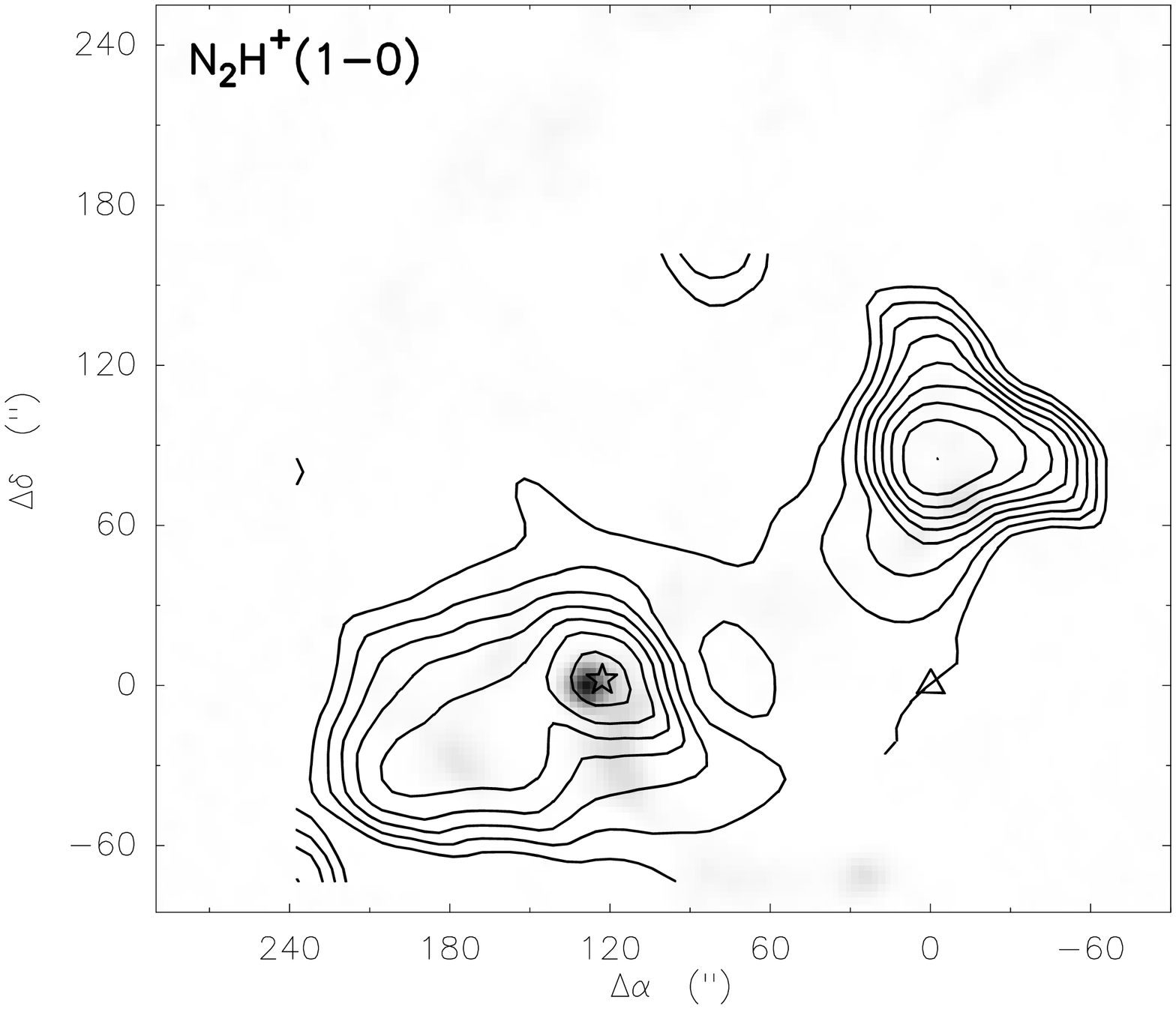}}}
\end{minipage}
\hfill
\begin{minipage}[b]{0.33\textwidth}
\centering
\resizebox{\hsize}{!}{\rotatebox{-90}{\includegraphics{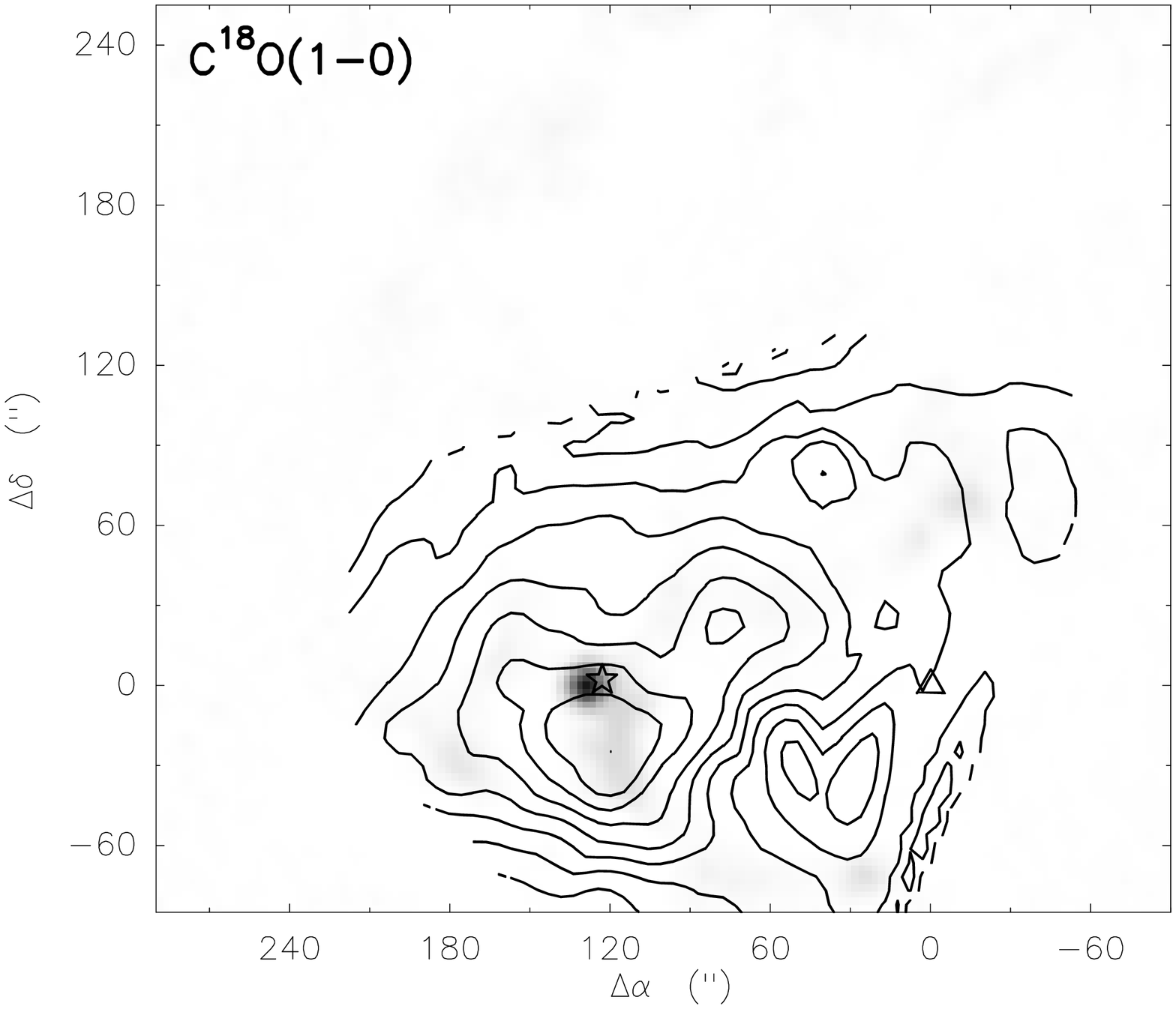}}}
\end{minipage}
\hfill
\begin{minipage}[b]{0.33\textwidth}
\centering
\resizebox{\hsize}{!}{\rotatebox{-90}{\includegraphics{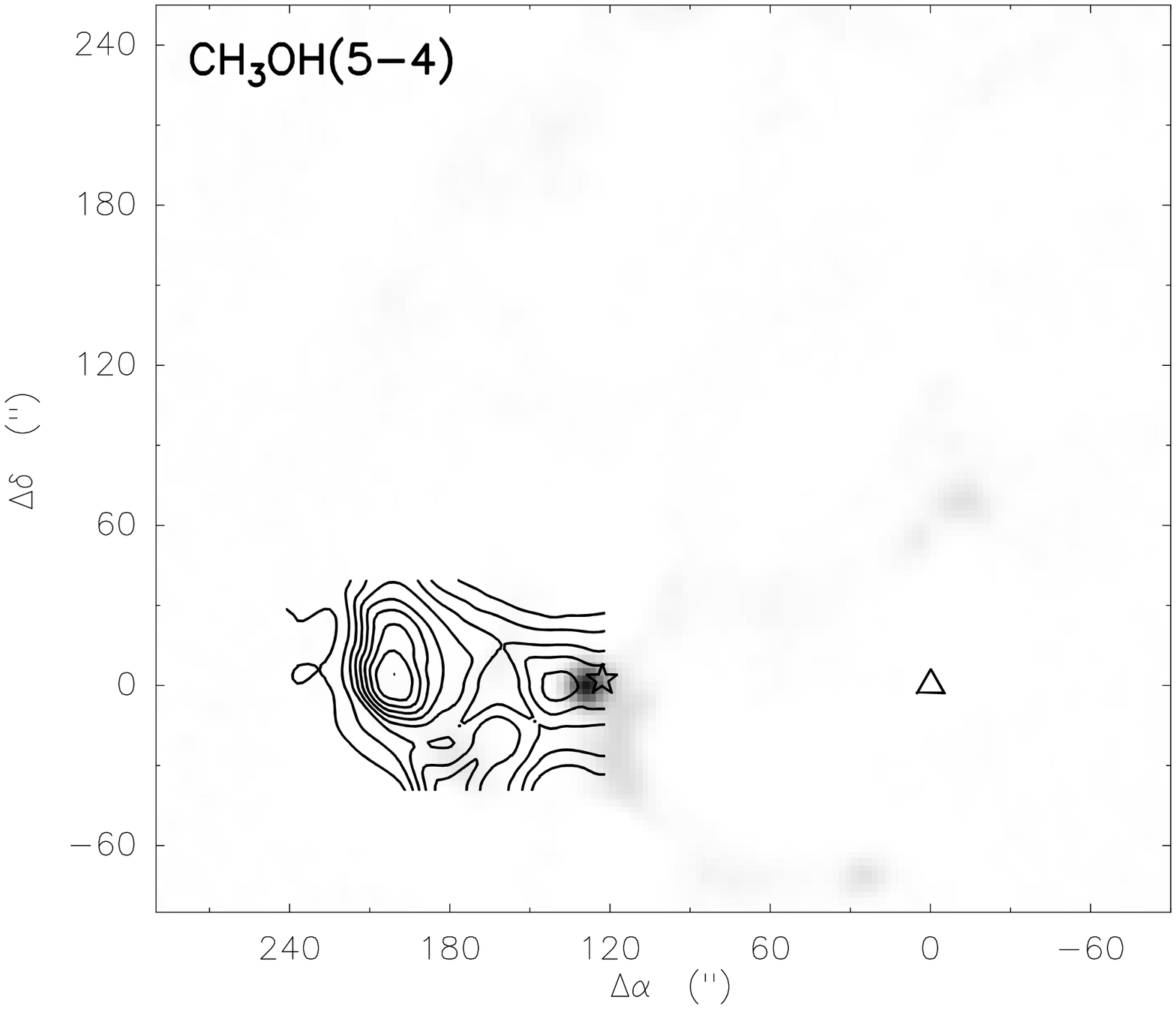}}}
\end{minipage}
\caption{Maps of S187 in various molecular lines (contours) overlaid on
the map of 1.2 mm dust continuum emission (grayscale). The contour levels
span from 20\% to 100\% of the peak intensity in steps of 10\%.}
\label{fig:maps-s187}
\end{figure*}

Maps of S187 in various lines overlaid on the greyscale map of the
1.2 mm continuum emission are presented in Fig.~\ref{fig:maps-s187}.
The CS(2--1) and N$_2$H$^+$(1--0) maps have been published earlier
\citep{Zin98,Pirogov03}.
There is a striking difference between the various maps. Two separated clumps are clearly seen in N$_2$H$^+$(1--0), which peaks North-West of the IRAS~01202+6133 source. The two peaks are still visible in the HNC(1--0) map, although they are not as well separated as in the N$_2$H$^+$(1--0) map. All the other maps have a cometary or more irregular morphology, with peaks all shifted from the continuum peak.
The gaussian line parameters at the CS and N$_2$H$^+$ emission peaks
are summarized in Table~\ref{table:lines-s187}. In C$^{18}$O $J=1-0$
and $J=2-1$ emission towards the CS peak there is an additional narrow
($\sim 0.7$~km/s) component at about $-$15.5~km/s. However, it is not
pronounced in the lines of other high density tracers and we do not
include it in Table~\ref{table:lines-s187}.

The molecular data indicate the presence of at least 3 clumps in the
area. There are several IRAS point sources and molecular masers
here. The strongest IRAS point source, IRAS~01202+6133, is located
at about 2\arcmin\ to the east from our central position and
coincides with the main 1.2~mm continuum peak. Here an OH maser and
UC \Hii\ region are present \citep*{Argon00}.
The secondary N$_2$H$^+$ peak coincides with this IRAS position.
Also a weaker CS clump is located here as is clearly seen from the
CS $J=5-4$ data. The main CS, HCO$^+$ and HCN emission peaks are
shifted by about 1\arcmin\ further to the east. No IR sources or
masers are known in this area. It is worth noting that the methanol
emission peak is shifted still further to the east. At the same time
C$^{18}$O emission peaks near IRAS~01202+6133.

The strongest N$_2$H$^+$ peak coincides with a relatively weak 1.2~mm
continuum clump. It is shifted by about 0{\farcm}5
from the strong near IR source NIRS~60 \citep*{Salas98}.

\begin{table*}
\caption{Molecular line parameters at the CS and N$_2$H$^+$ emission
peaks in S187.}
\label{table:lines-s187}
\begin{tabular}{llllllll}
\hline%\hline
&\multicolumn{3}{l}{(+160$''$,0)}  &&\multicolumn{3}{l}{(0,+80$''$)}\\
\cline{2-4}
\cline{6-8}
Line &$T_{\rm mb}$ &$V_{\rm LSR}$ &$\Delta V$ & &$T_{\rm mb}$ &$V_{\rm LSR}$ &$\Delta V$ \\
&(K)          &(km/s)    &(km/s)     & &(K)          &(km/s)    &(km/s)     \\
\hline
C$^{18}$O(1--0) &5.47(18) &$-$13.92(03) &1.71(09) &&3.24(18) &$-$13.78(04) &1.67(10) \\
C$^{18}$O(2--1) &5.85(19) &$-$14.05(02) &1.69(05) \\
CS(2--1) &4.29(10) &$-$14.26(03) &2.30(07) &&1.20(20) &$-$13.63(27) &3.24(70)\\
C$^{34}$S(2--1) &1.00(04) &$-$14.05(04) &2.25(11) \\
CS(5--4) &3.58(06) &$-$14.09(01) &1.86(03) \\
C$^{34}$S(5--4) &$<$0.4 \\
HCN(1--0) &5.14(07) &$-$14.31(02) &2.33(04) &&1.92(08) &$-$13.81(04) &1.63(07)\\
H$^{13}$CN(1--0) &0.50(05) &$-$14.17(10) &1.83(21)\\
HCO$^+$(1--0) &5.54(08) &$-$14.49(02) &2.22(04) &&2.41(08) &$-$14.03(03) &1.85(08)\\
H$^{13}$CO$^+$(1--0) &0.58(07) &$-$14.15(09) &1.56(22)
&&1.28(10) &$-$13.51(03) &0.83(07)\\
HNC(1--0) &5.44(12) &$-$14.29(02) &2.27(06) &&5.70(15) &$-$13.58(02) &1.39(04)\\
HN$^{13}$C(1--0) &0.33(04) &$-$13.88(09) &1.66(20) &&0.63(05) &$-$13.35(03) &0.76(07)\\
N$_2$H$^+$(1--0) &0.61(08) &$-$14.09(07) &1.45(20) &&1.59(12) &$-$13.33(03) &0.90(06)\\
\hline
\end{tabular}
\end{table*}

\subsubsection{W3}
\begin{figure*}
\begin{minipage}[b]{0.24\textwidth}
\centering
\resizebox{\hsize}{!}{\rotatebox{-90}{\includegraphics{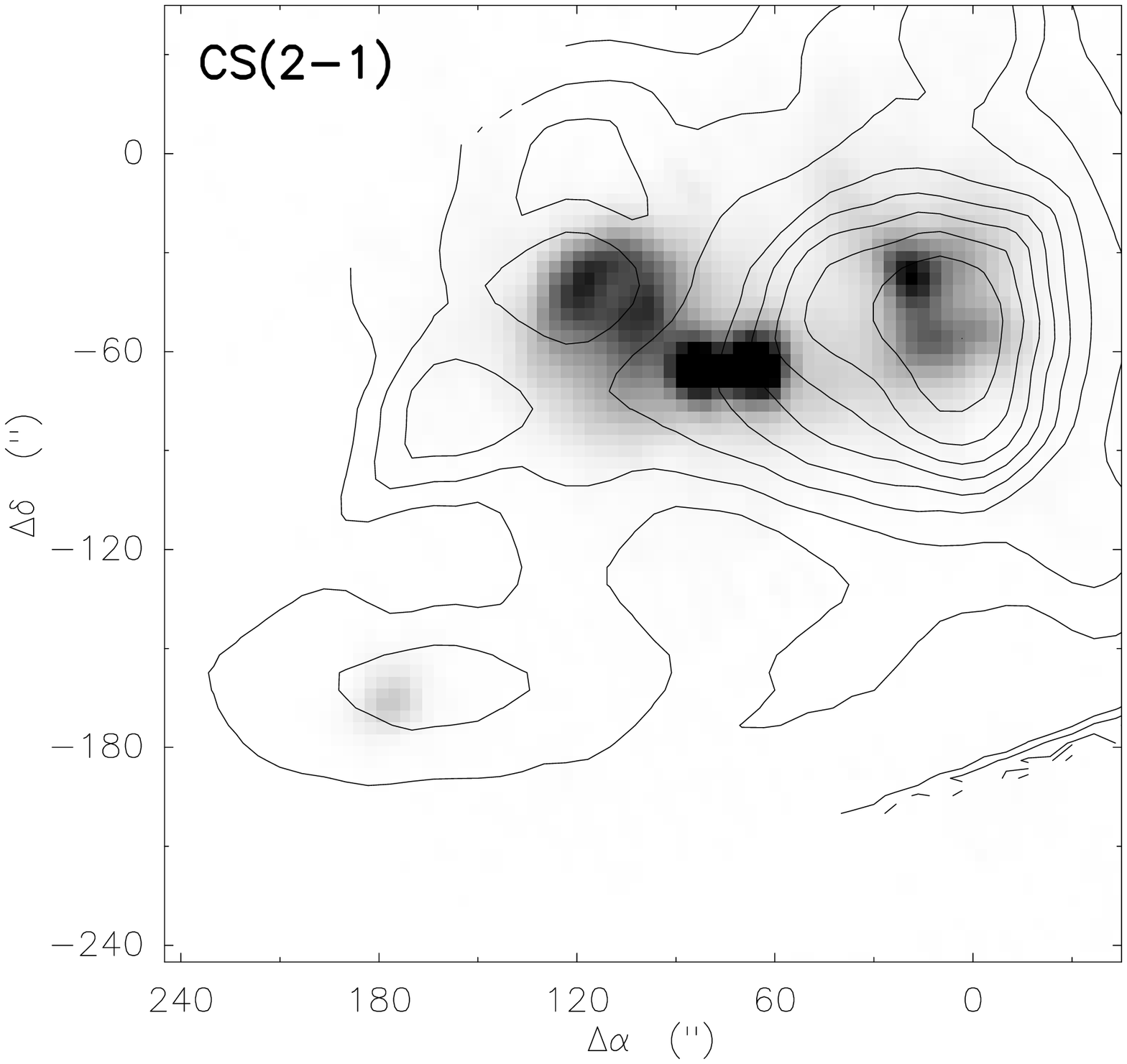}}}
\end{minipage}
\hfill
\begin{minipage}[b]{0.24\textwidth}
\centering
\resizebox{\hsize}{!}{\rotatebox{-90}{\includegraphics{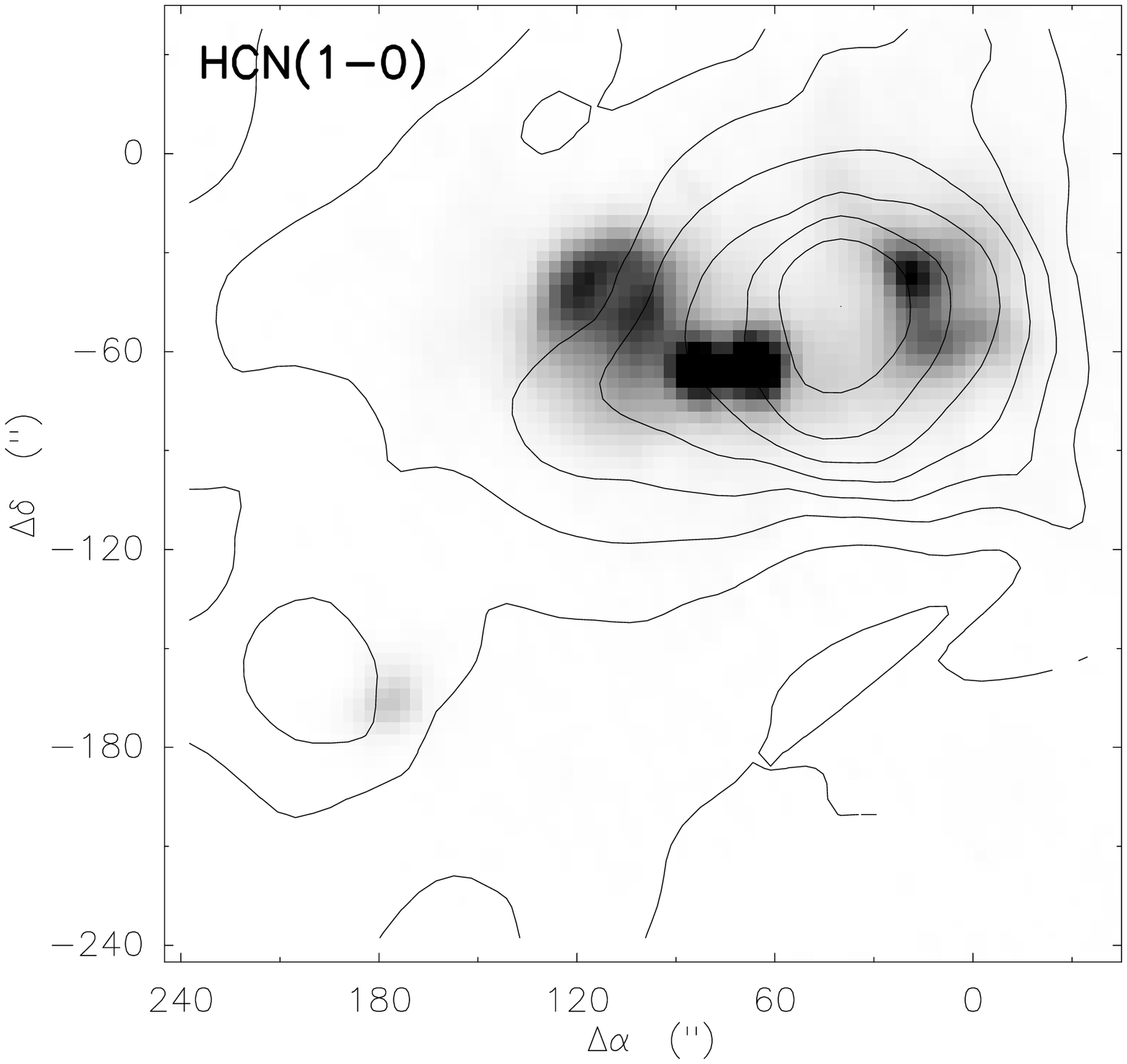}}}
\end{minipage}
\hfill
\begin{minipage}[b]{0.24\textwidth}
\centering
\resizebox{\hsize}{!}{\rotatebox{-90}{\includegraphics{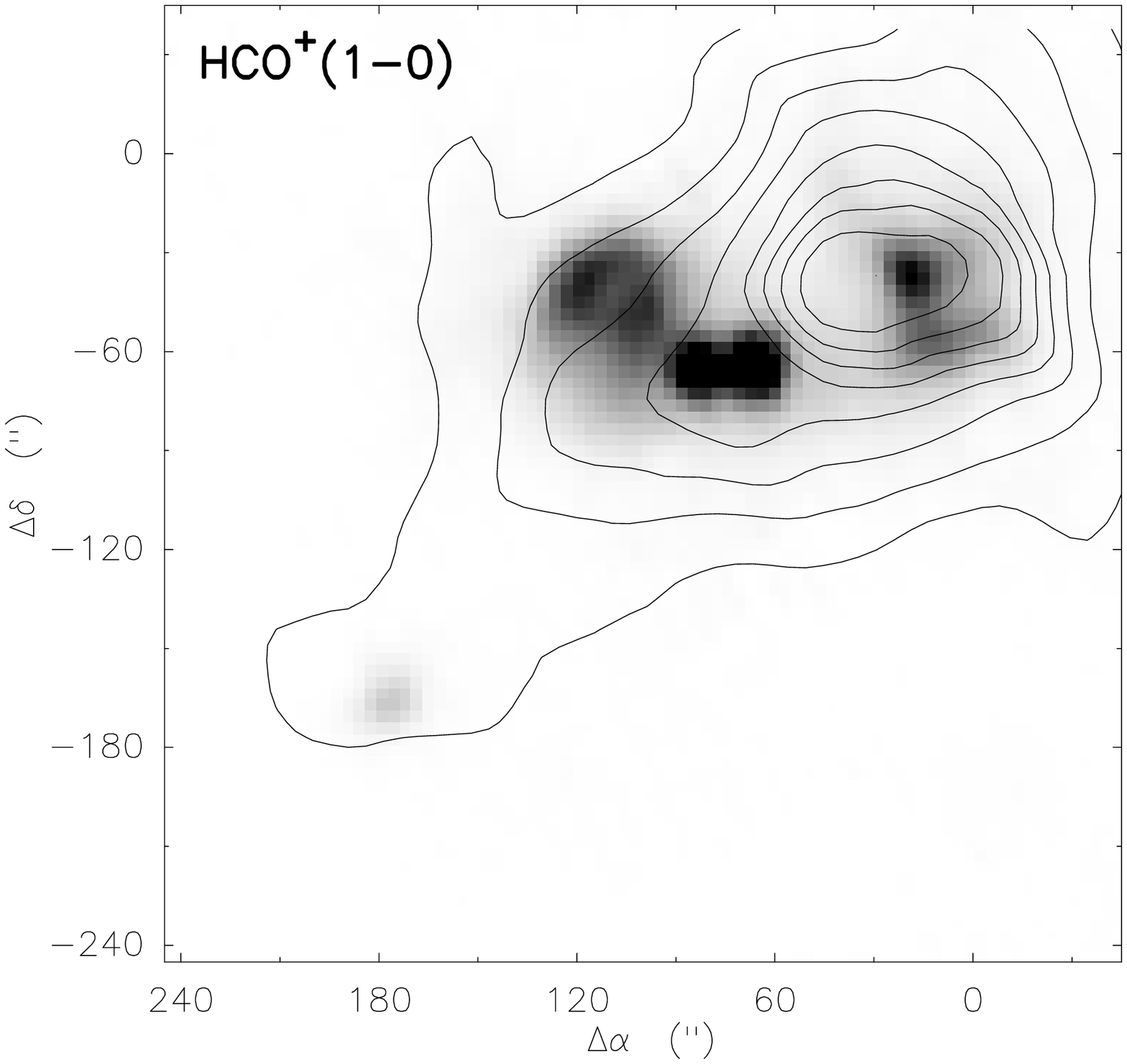}}}
\end{minipage}
\hfill
\begin{minipage}[b]{0.24\textwidth}
\centering
\resizebox{\hsize}{!}{\rotatebox{-90}{\includegraphics{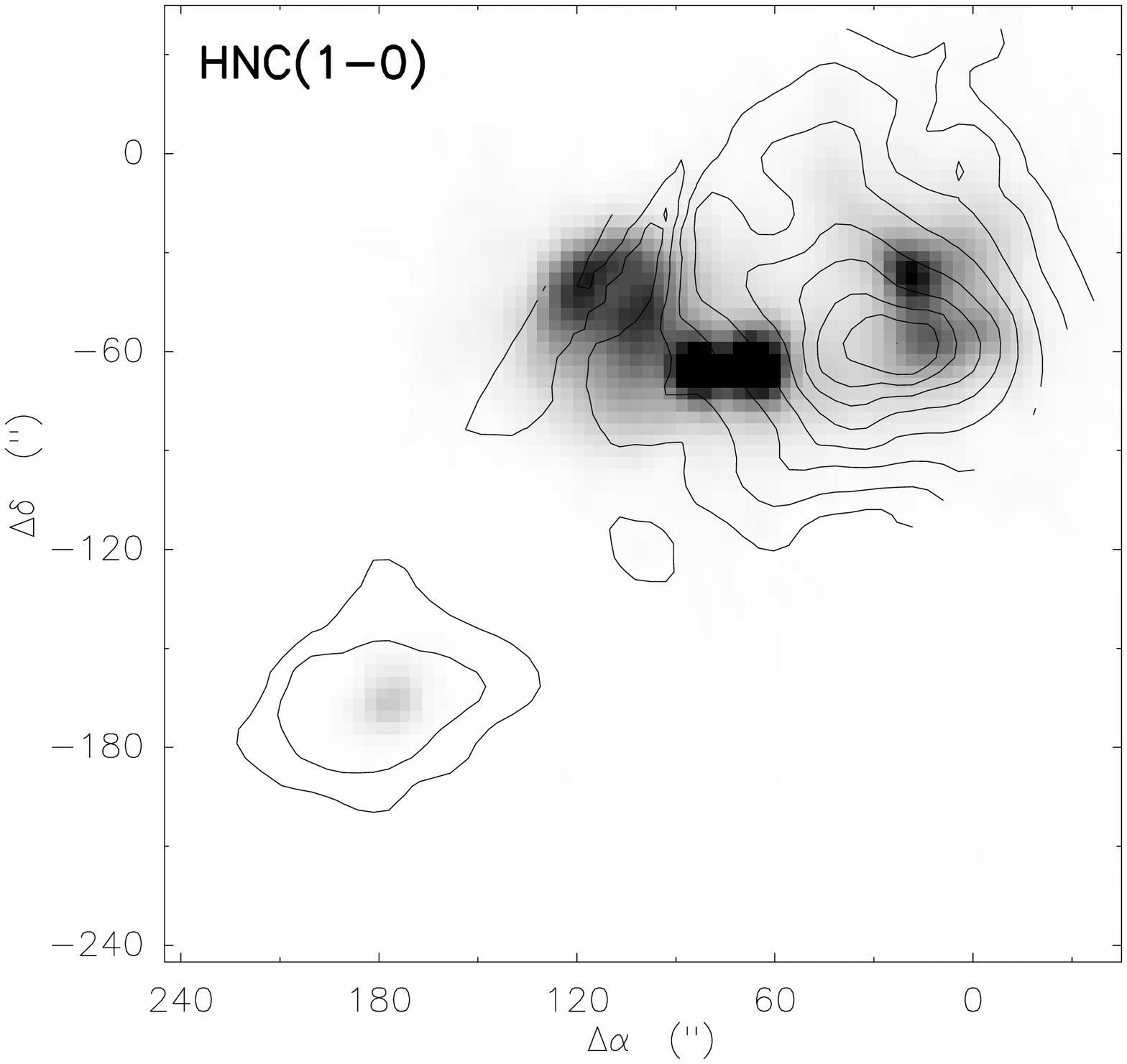}}}
\end{minipage}
\\[1ex]
\begin{minipage}[b]{0.24\textwidth}
\centering
\resizebox{\hsize}{!}{\rotatebox{-90}{\includegraphics{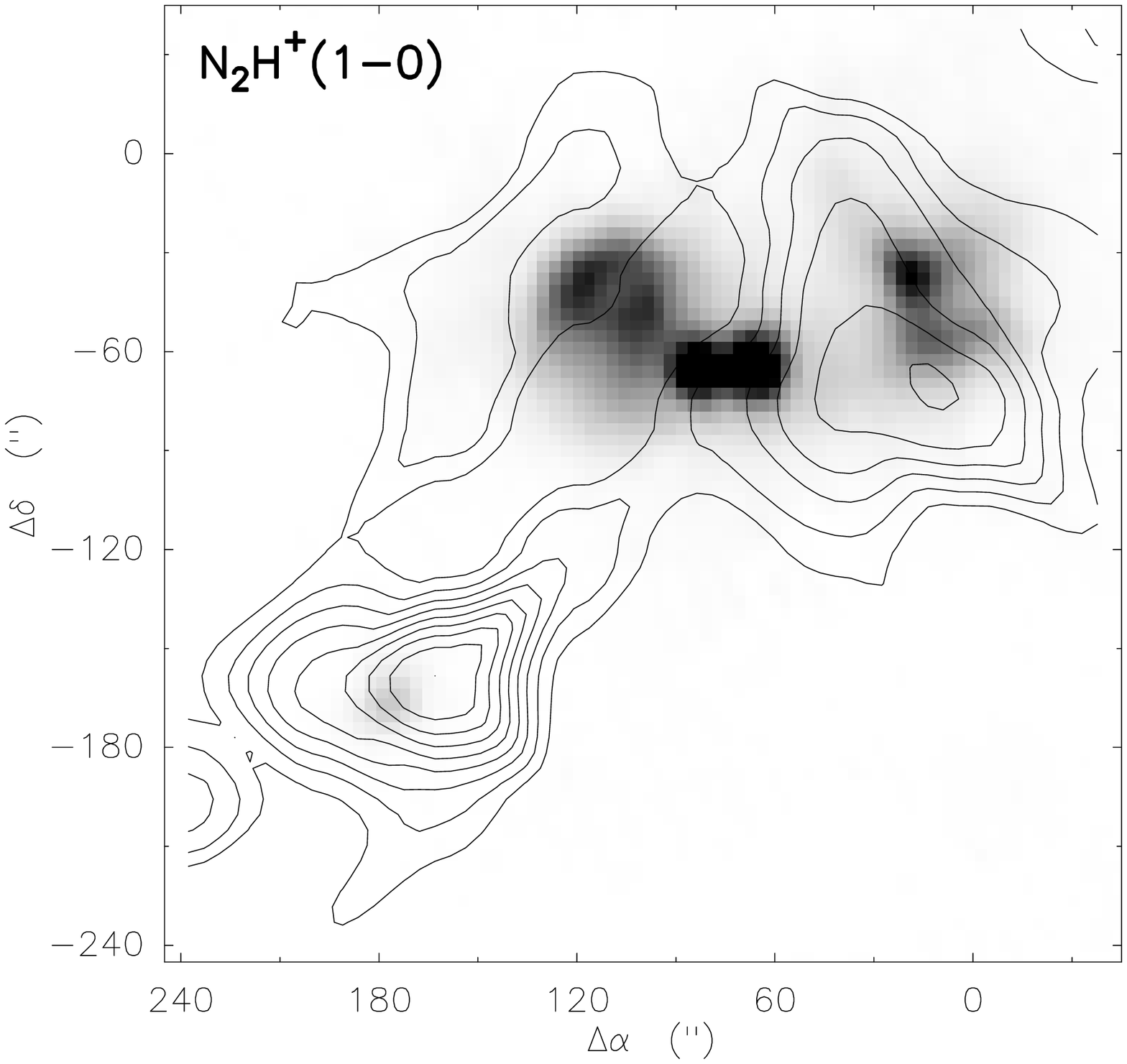}}}
\end{minipage}
\hfill
\begin{minipage}[b]{0.24\textwidth}
\centering
\resizebox{\hsize}{!}{\rotatebox{-90}{\includegraphics{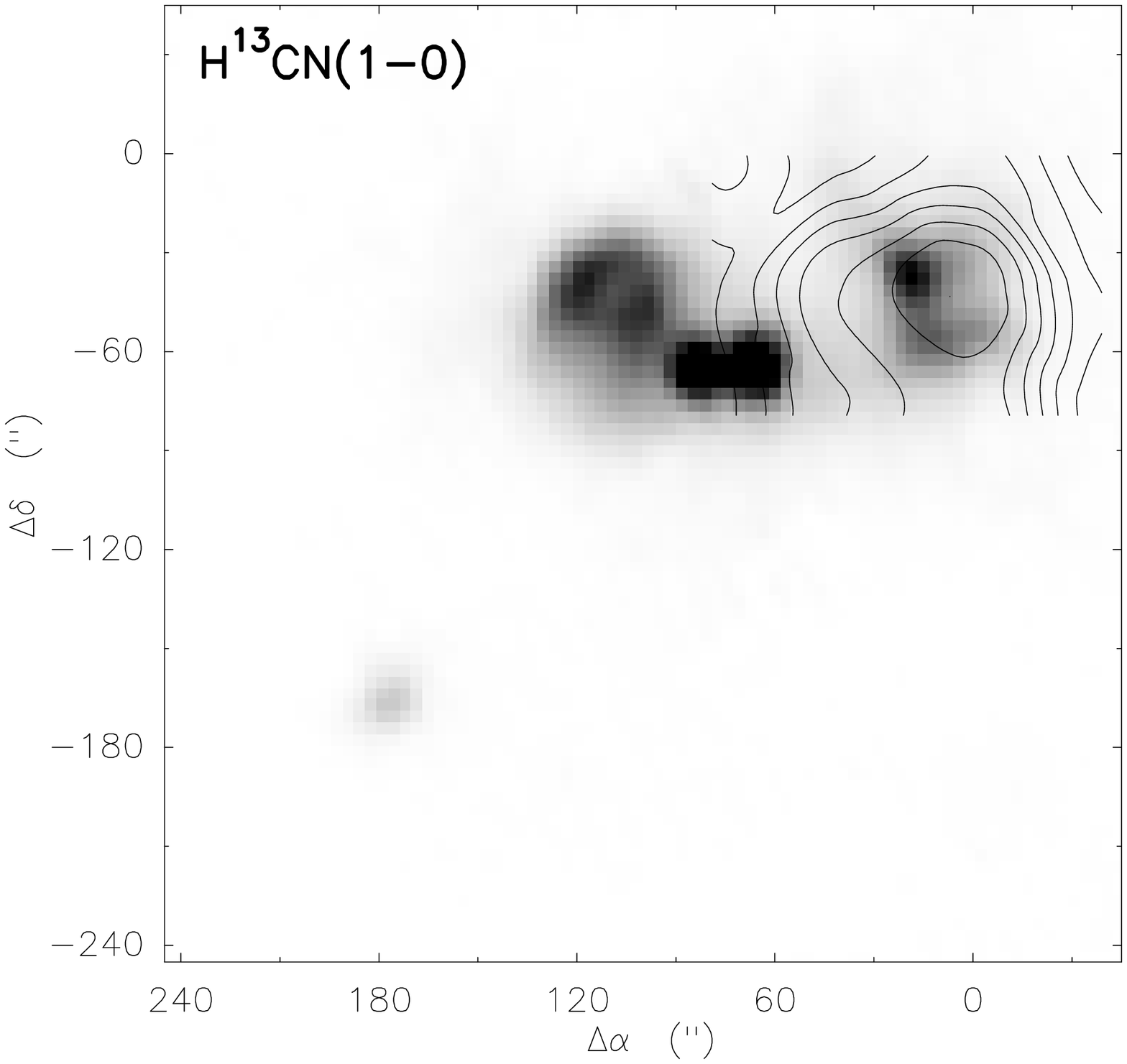}}}
\end{minipage}
\hfill
\begin{minipage}[b]{0.24\textwidth}
\centering
\resizebox{\hsize}{!}{\rotatebox{-90}{\includegraphics{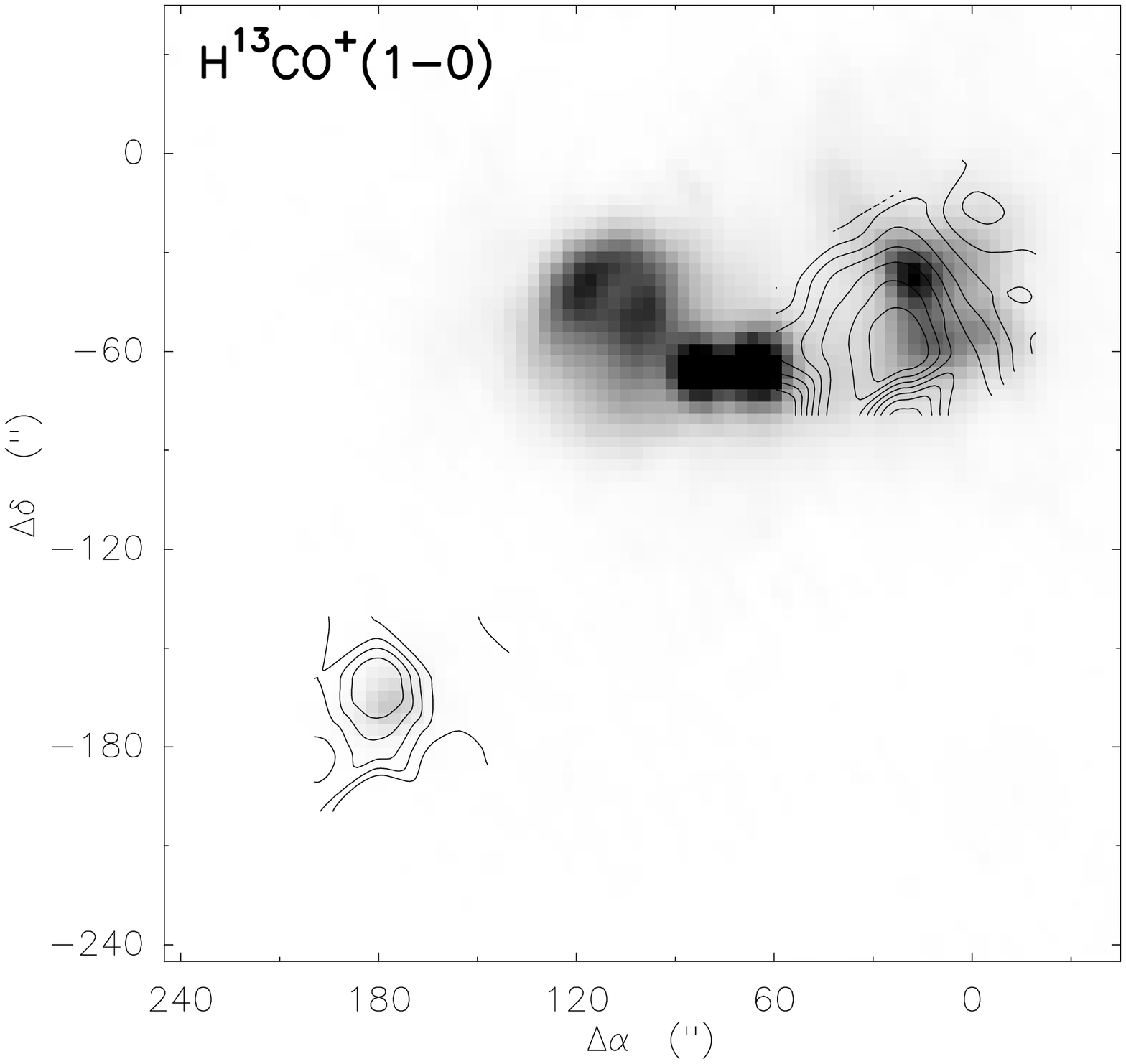}}}
\end{minipage}
\hfill
\begin{minipage}[b]{0.24\textwidth}
\centering
%\hspace*{\textwidth}
\resizebox{\hsize}{!}{\rotatebox{-90}{\includegraphics{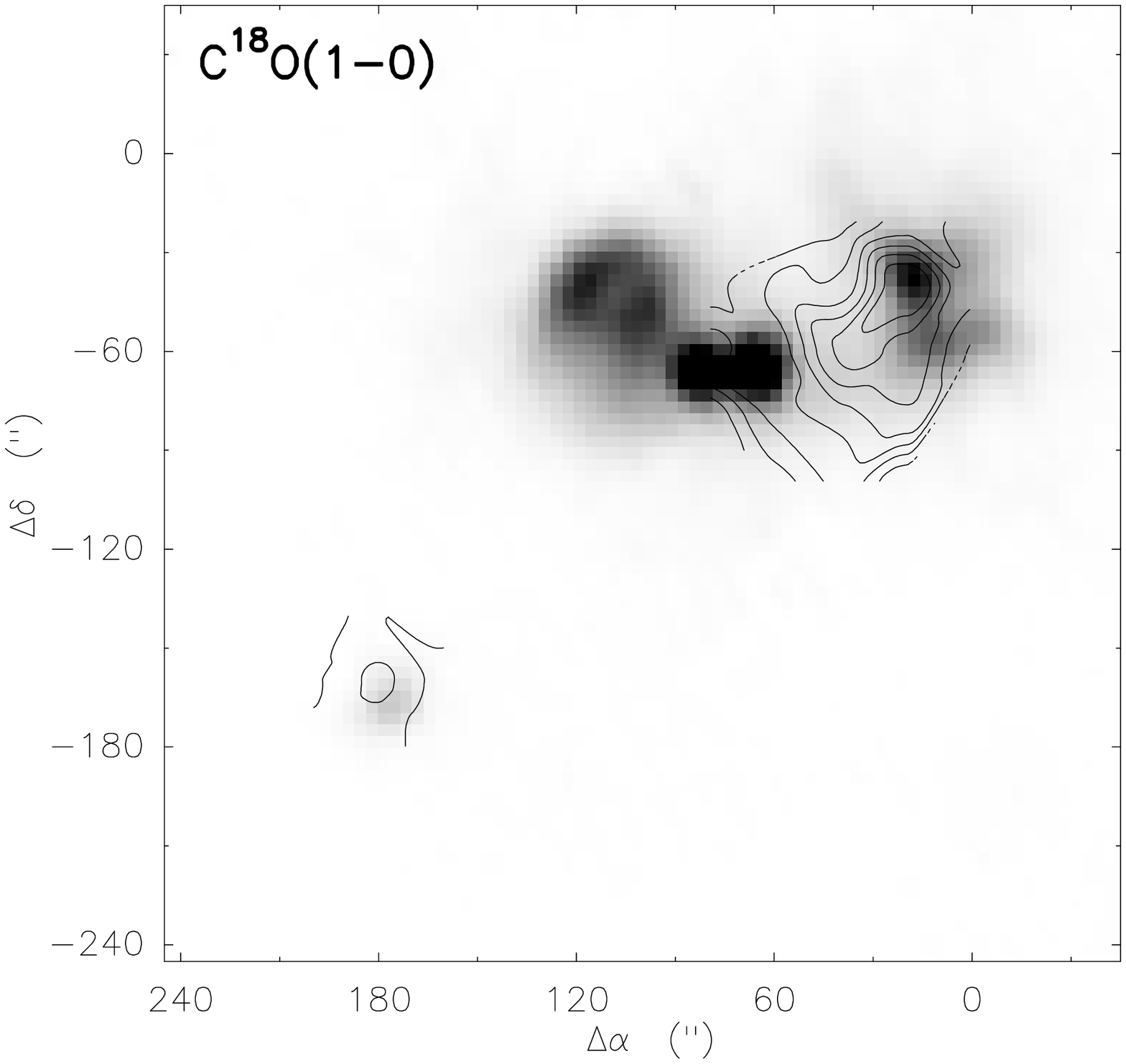}}}
\end{minipage}
\caption{Maps of W3 in various molecular lines (contours) overlaid on
the map of 1.2 mm continuum emission (grayscale). The contour levels
span from 20\% to 100\% of the peak intensity in steps of 10\%.}
\label{fig:maps-w3}
\end{figure*}

The molecular line maps of W3 are shown in Fig.~\ref{fig:maps-w3} overlaid with the 1.2 mm continuum emission map. The CS(2--1) and N$_2$H$^+$(1--0) maps have been
published earlier \citep{Zin98,Pirogov03}. The brightest continuum peak is attributed to
free-free emission from a compact \Hii\ region \citep[e.g.][]{Tieftrunk97}.
There are
two main molecular emission peaks in this area associated with two
dust clumps. As in the S187 case, the N$_2$H$^+$ map looks very
different from most other maps. The N$_2$H$^+$ peak coincides with
a relatively weak south-east (SE) clump at the approximately
(160$''$,$-$160$''$) position. The gaussian line parameters at the
CS and N$_2$H$^+$ emission peaks are summarized in
Table~\ref{table:lines-w3}. Some spectra clearly show non-gaussian
features: broad wings at the CS peak position and
red-shifted self-absorption in HCO$^+$ at the N$_2$H$^+$ peak.

\begin{table*}
\caption{Molecular line parameters at the CS and N$_2$H$^+$ emission
peaks in W3.}
\label{table:lines-w3}
\begin{tabular}{llllllll}
\hline%\hline
&\multicolumn{3}{l}{(0,$-$40$''$)}
&
&\multicolumn{3}{l}{(+160$''$,$-$160$''$)}\\
\cline{2-4}
\cline{6-8}
Line &$T_{\rm mb}$ &$V_{\rm LSR}$ &$\Delta V$ & &$T_{\rm mb}$ &$V_{\rm LSR}$ &$\Delta V$ \\
&(K)          &(km/s)    &(km/s)     & &(K)          &(km/s)    &(km/s)     \\
\hline
C$^{18}$O(1--0) &2.05(20) &$-$42.70(27) &5.32(71)
&&2.36(37) &$-$38.58(11) &1.49(27)\\
CS(2--1) &8.82(10) &$-$43.09(03) &4.76(06) &&3.19(03) &$-$38.66(19) &4.43(53)\\
%C$^{34}$S(2--1) &1.53(03) &$-$42.88(04) &4.70(09)\\
HCN(1--0) &7.96(11) &$-$43.08(02) &3.41(05) &&1.65(13) &$-$38.55(40) &8.08(38)\\
H$^{13}$CN(1--0) &1.51(07) &$-$42.75(12) &4.62(20) &&\\
HCO$^+$(1--0) &16.26(07) &$-$43.40(01) &3.86(02) &&4.05(06) &$-$38.97(04) &4.73(09)\\
H$^{13}$CO$^+$(1--0) &0.88(17) &$-$43.11(21) &2.44(51)
&&0.95(18) &$-$38.33(19) &2.02(45)\\
HNC(1--0) &5.36(11) &$-$42.92(04) &3.83(09) &&4.41(12) &$-$38.74(04) &3.12(10)\\
HN$^{13}$C(1--0) &0.37(03) &$-$42.63(09) &2.80(21) &&0.17(01) &$-$38.38(05) &2.95(23)\\
N$_2$H$^+$(1--0) &0.66(09) &$-$42.22(11) &2.04(30) &&1.37(10) &$-$38.76(10) &1.75(14)\\
\hline
\end{tabular}
\end{table*}

%\subsubsection{AFGL 6366 S}

\subsubsection{S255}
\begin{figure*}
\begin{minipage}[b]{0.19\textwidth}
\centering
\resizebox{\hsize}{!}{\rotatebox{-90}{\includegraphics{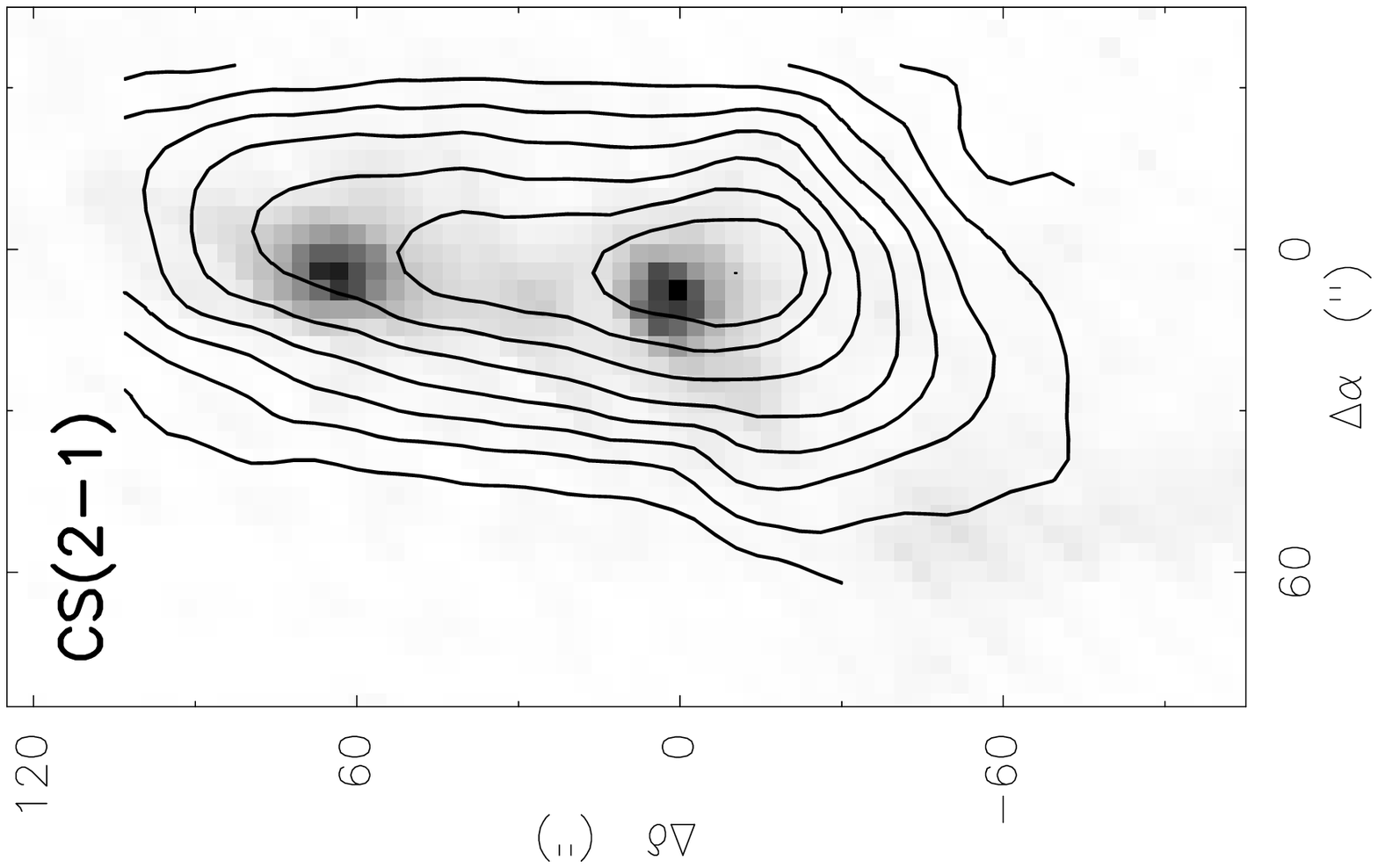}}}
\end{minipage}
\hfill
\begin{minipage}[b]{0.19\textwidth}
\centering
\resizebox{\hsize}{!}{\rotatebox{-90}{\includegraphics{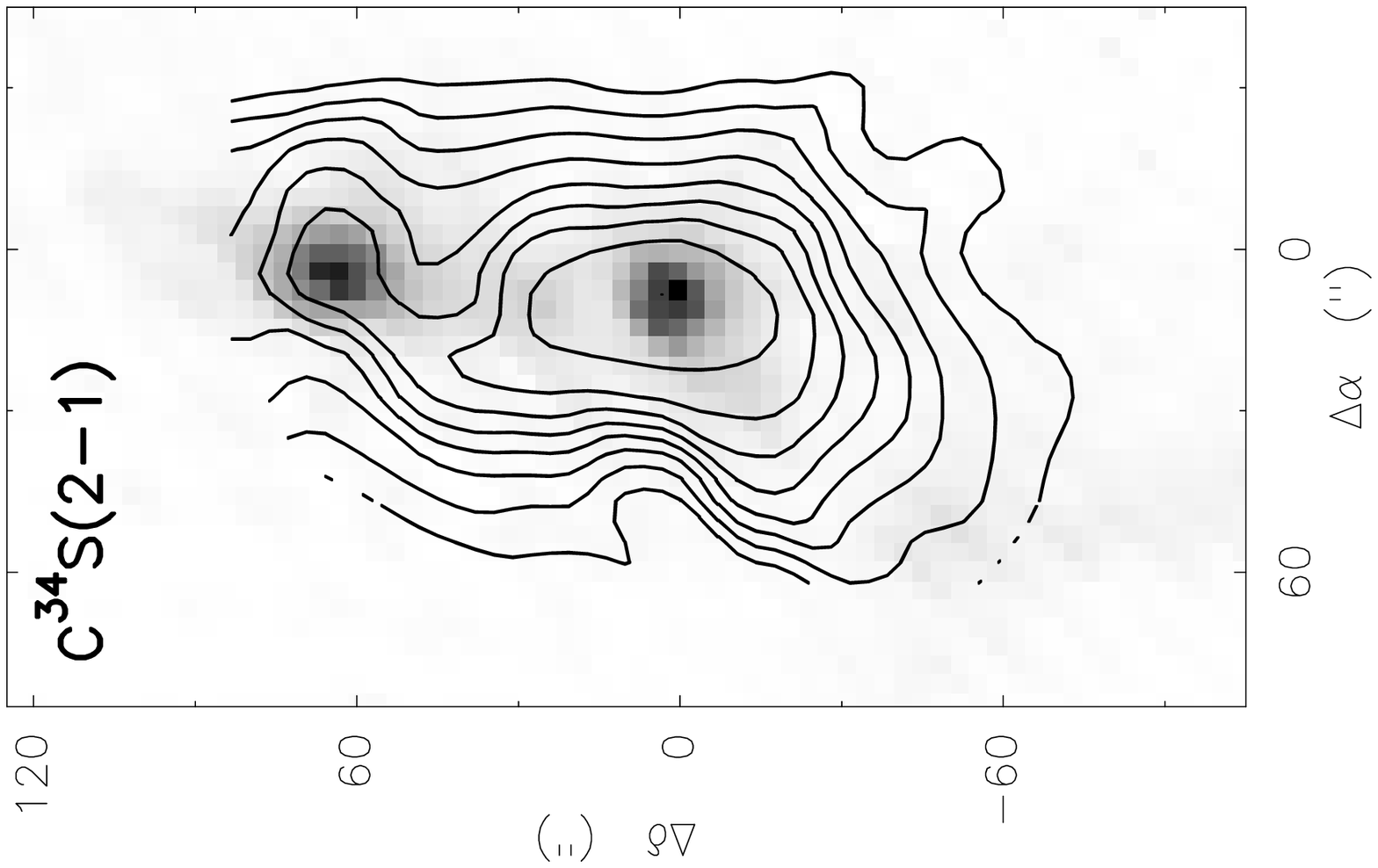}}}
\end{minipage}
\hfill
\begin{minipage}[b]{0.19\textwidth}
\centering
\resizebox{\hsize}{!}{\rotatebox{-90}{\includegraphics{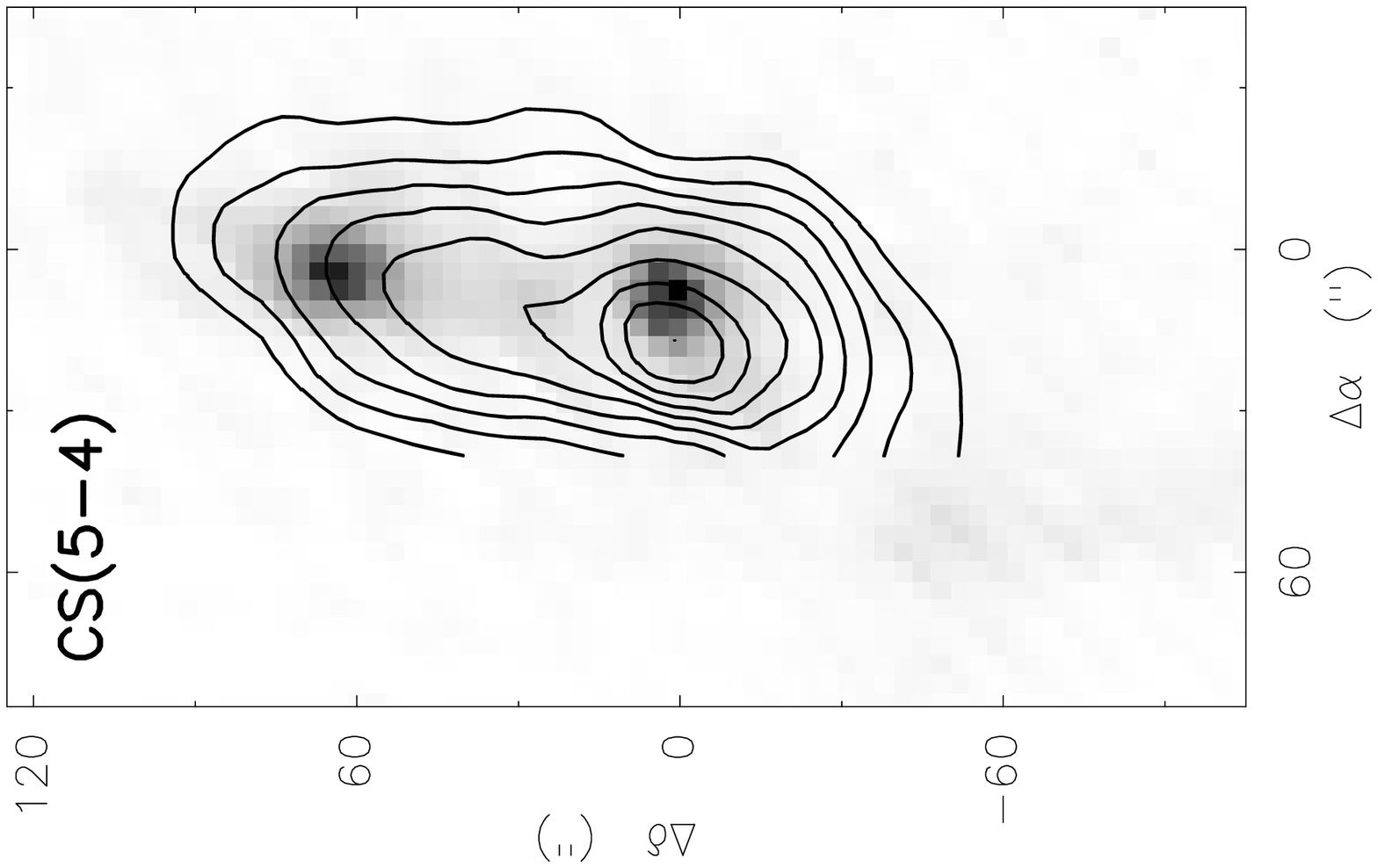}}}
\end{minipage}
\hfill
\begin{minipage}[b]{0.19\textwidth}
\centering
\resizebox{\hsize}{!}{\rotatebox{-90}{\includegraphics{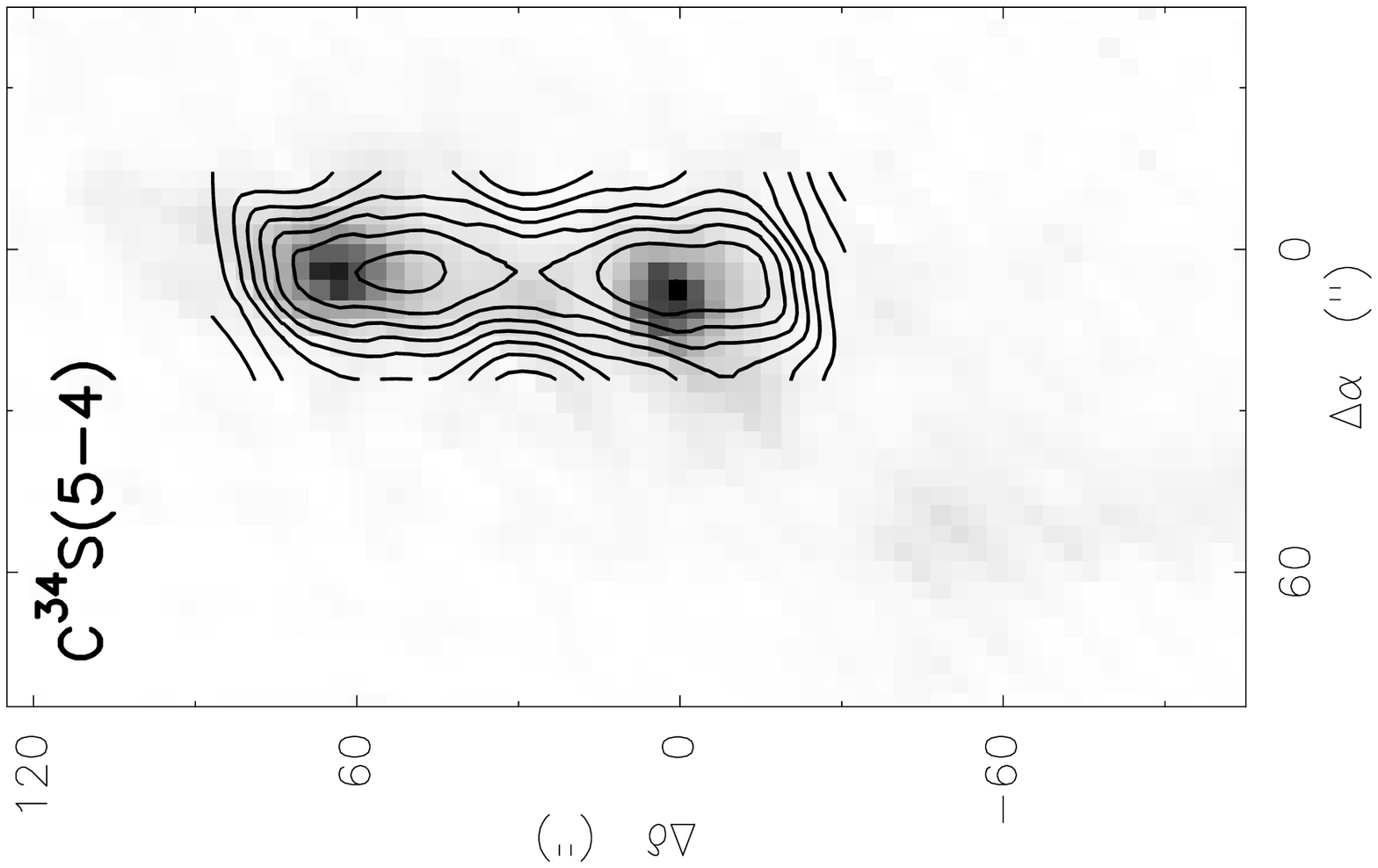}}}
\end{minipage}
\hfill
\begin{minipage}[b]{0.19\textwidth}
\centering
\resizebox{\hsize}{!}{\rotatebox{-90}{\includegraphics{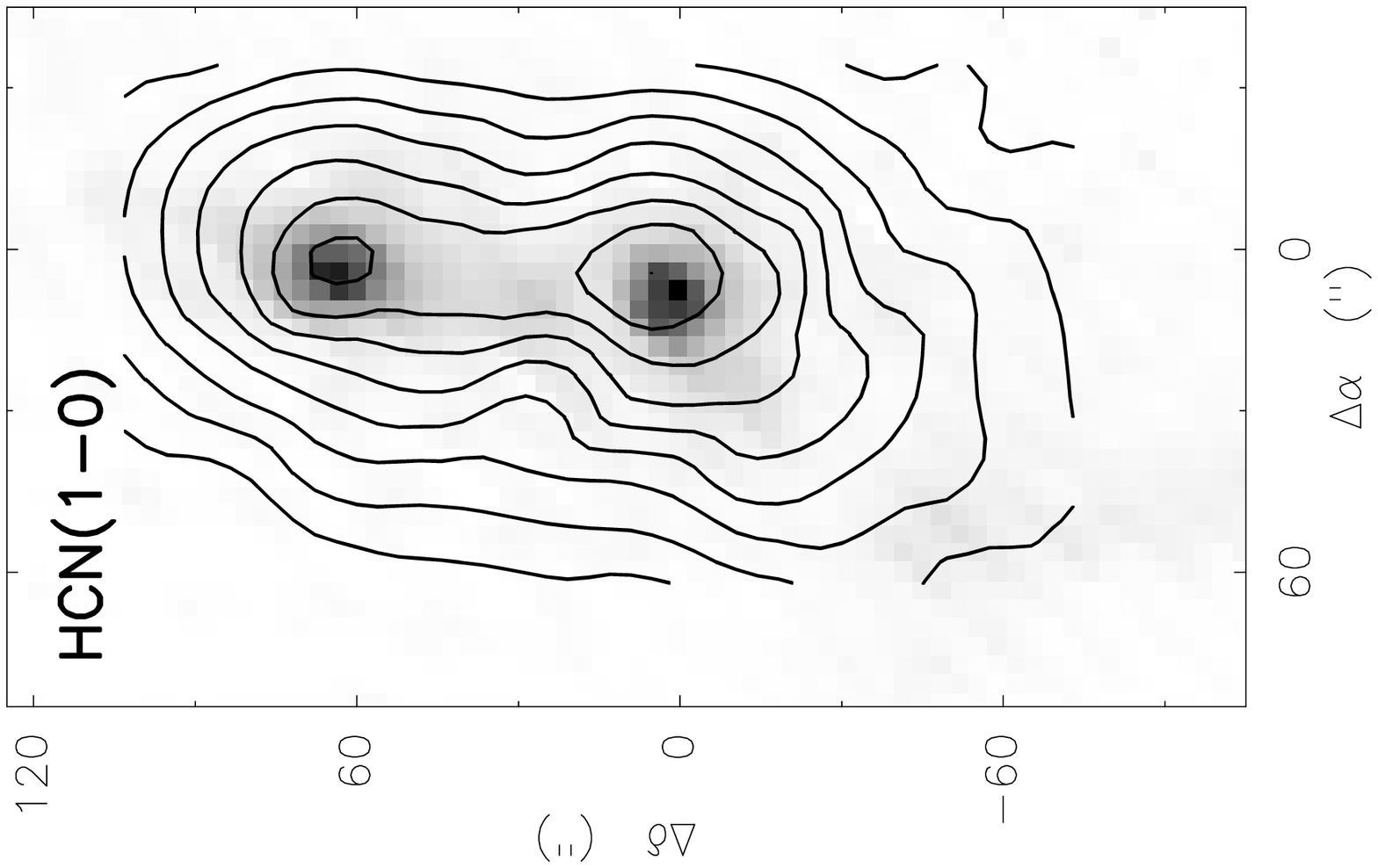}}}
\end{minipage}
\\[1ex]
\begin{minipage}[b]{0.19\textwidth}
\centering
\resizebox{\hsize}{!}{\rotatebox{-90}{\includegraphics{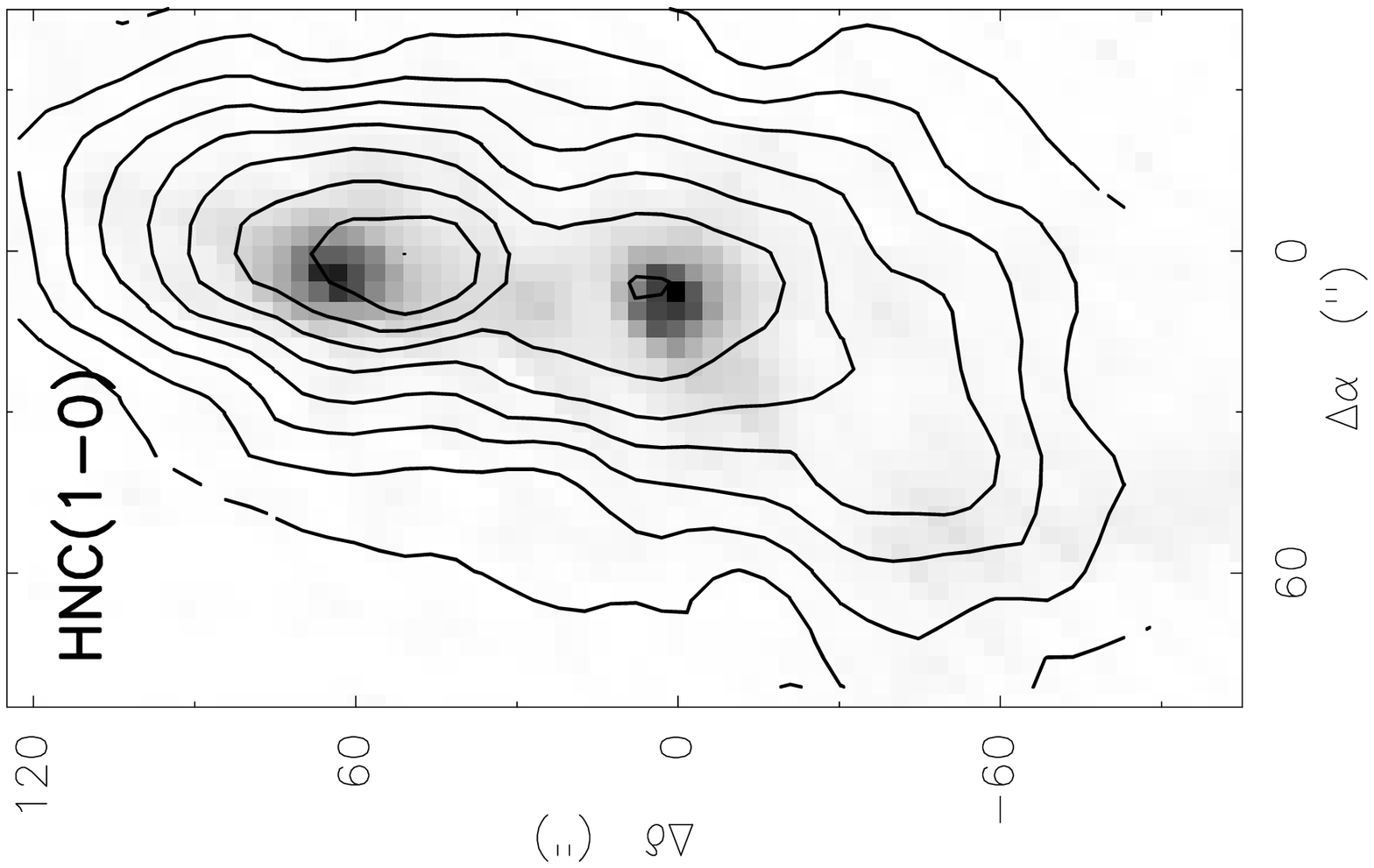}}}
\end{minipage}
\hfill
\begin{minipage}[b]{0.19\textwidth}
\centering
\resizebox{\hsize}{!}{\rotatebox{-90}{\includegraphics{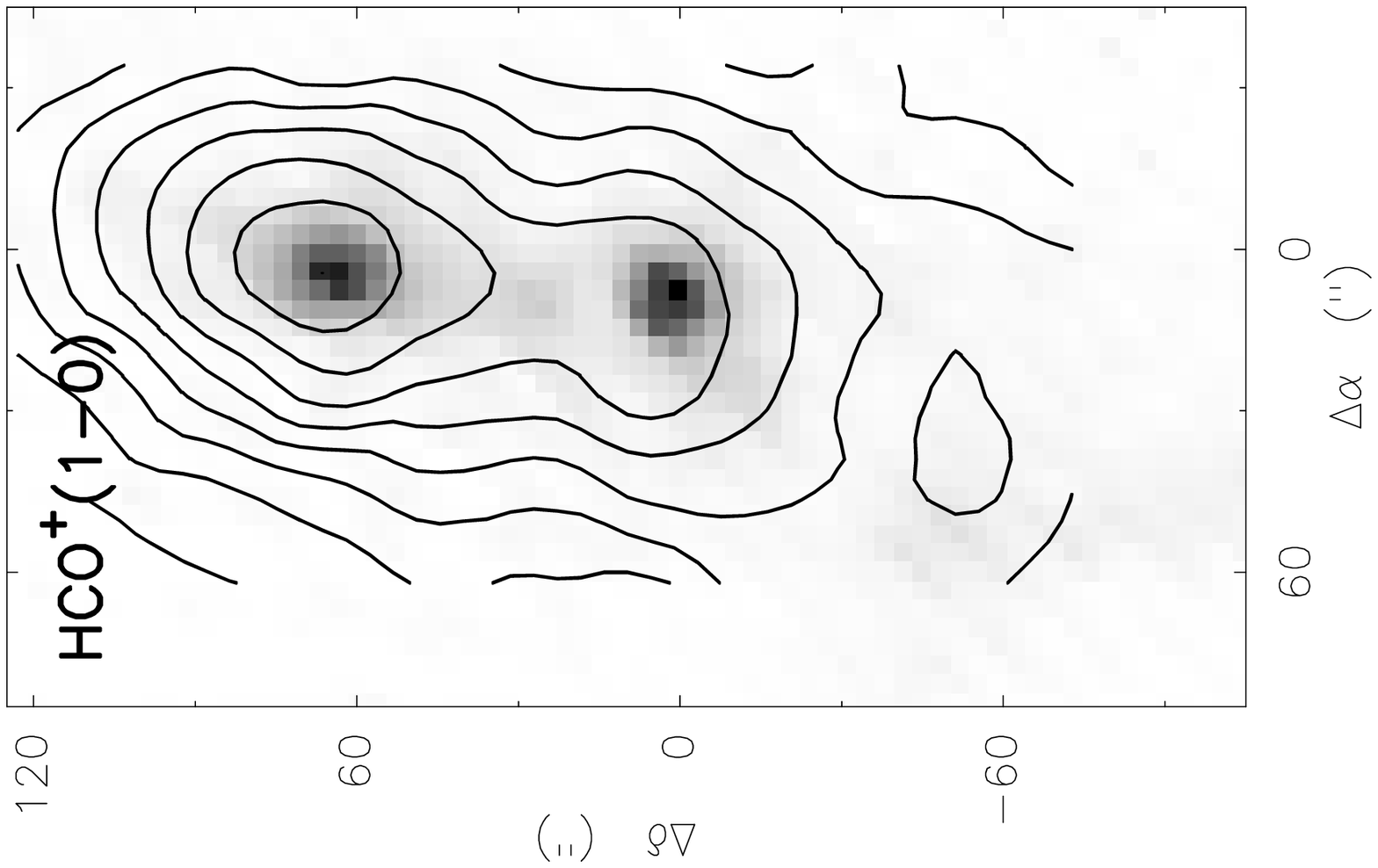}}}
\end{minipage}
\hfill
\begin{minipage}[b]{0.19\textwidth}
\centering
\resizebox{\hsize}{!}{\rotatebox{-90}{\includegraphics{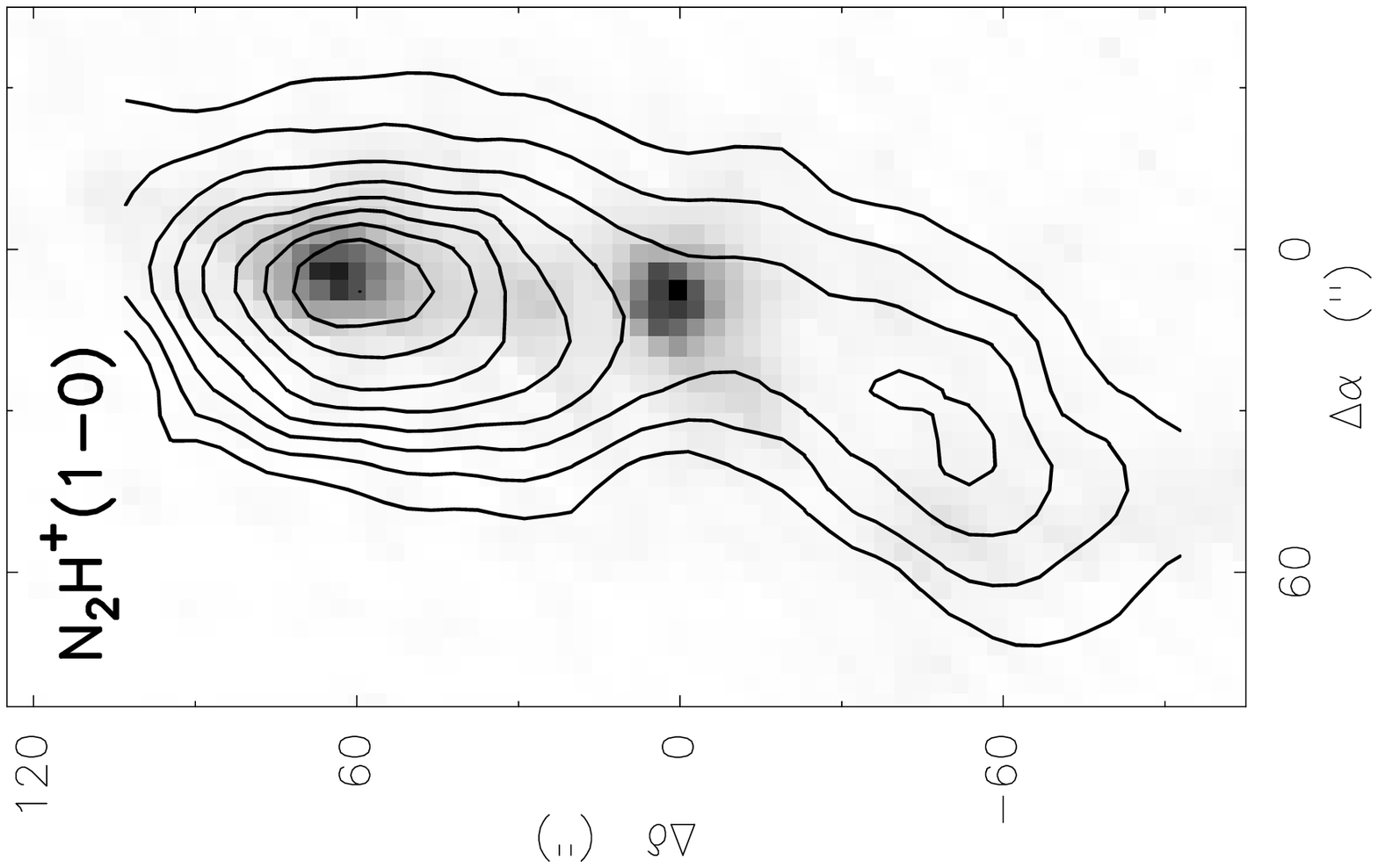}}}
\end{minipage}
\hfill
\begin{minipage}[b]{0.19\textwidth}
\centering
\resizebox{\hsize}{!}{\rotatebox{-90}{\includegraphics{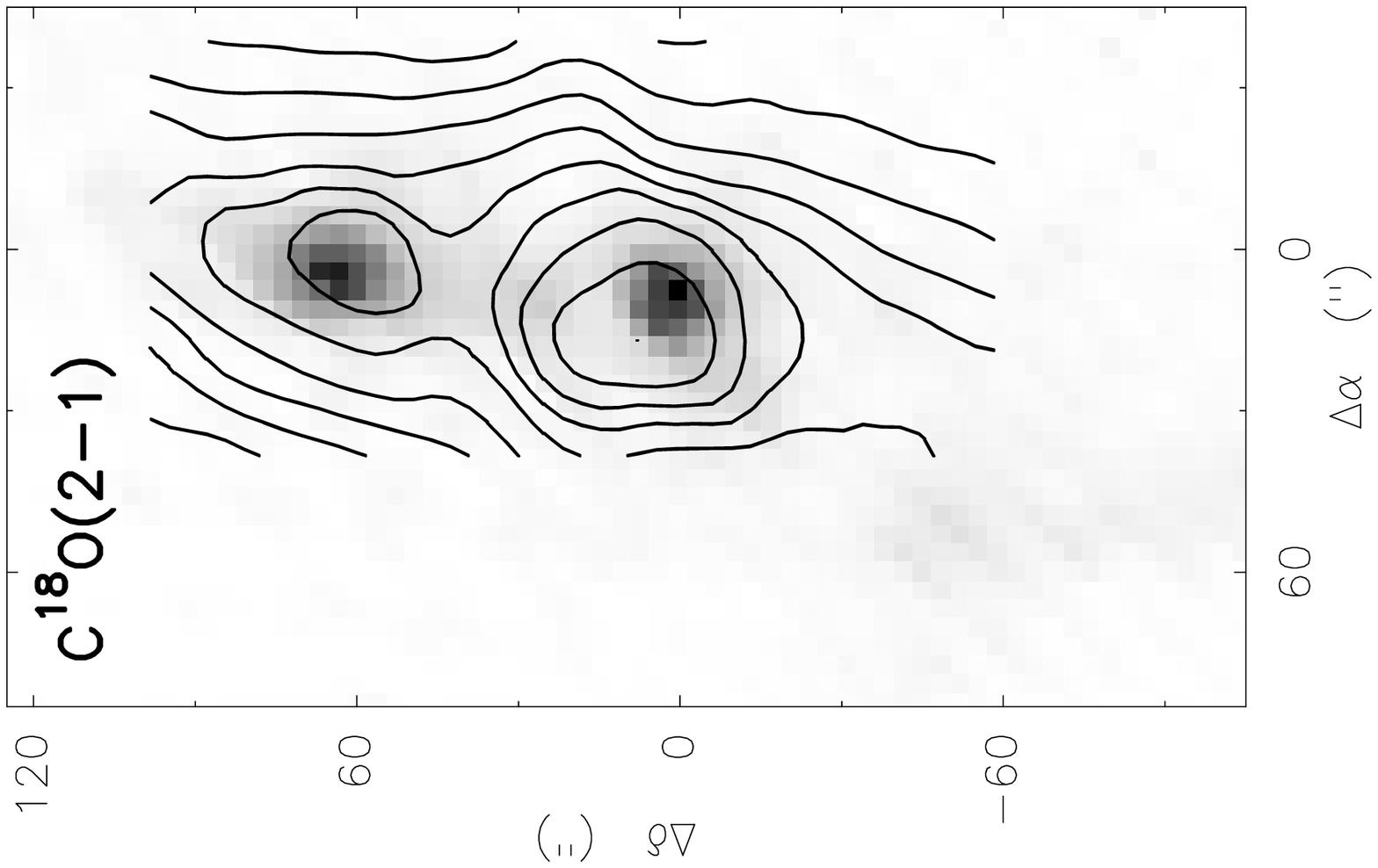}}}
\end{minipage}
\hfill
\begin{minipage}[b]{0.19\textwidth}
\centering
\resizebox{\hsize}{!}{\rotatebox{-90}{\includegraphics{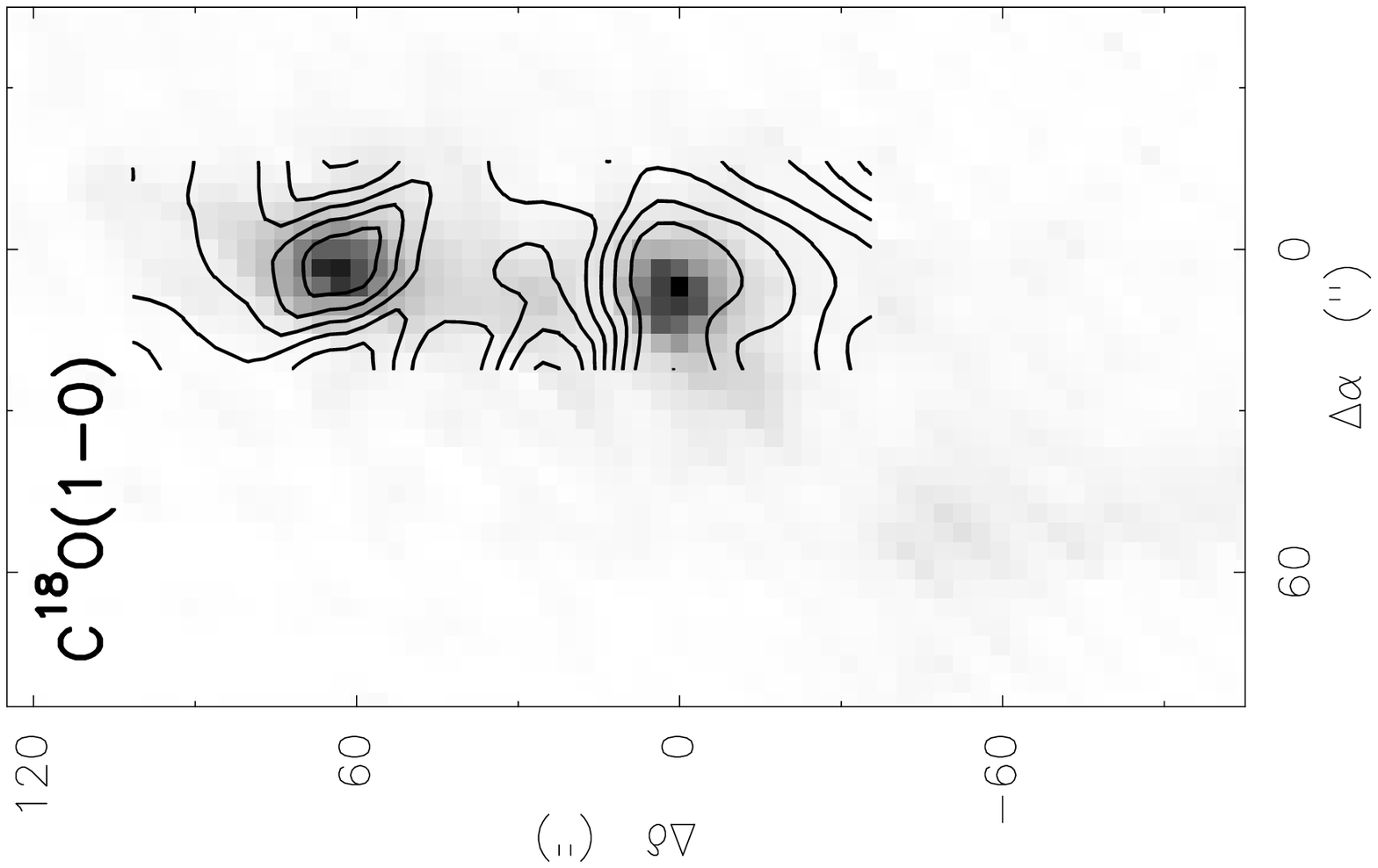}}}
\end{minipage}
\\[1ex]
\begin{minipage}[b]{0.19\textwidth}
\centering
\resizebox{\hsize}{!}{\rotatebox{-90}{\includegraphics{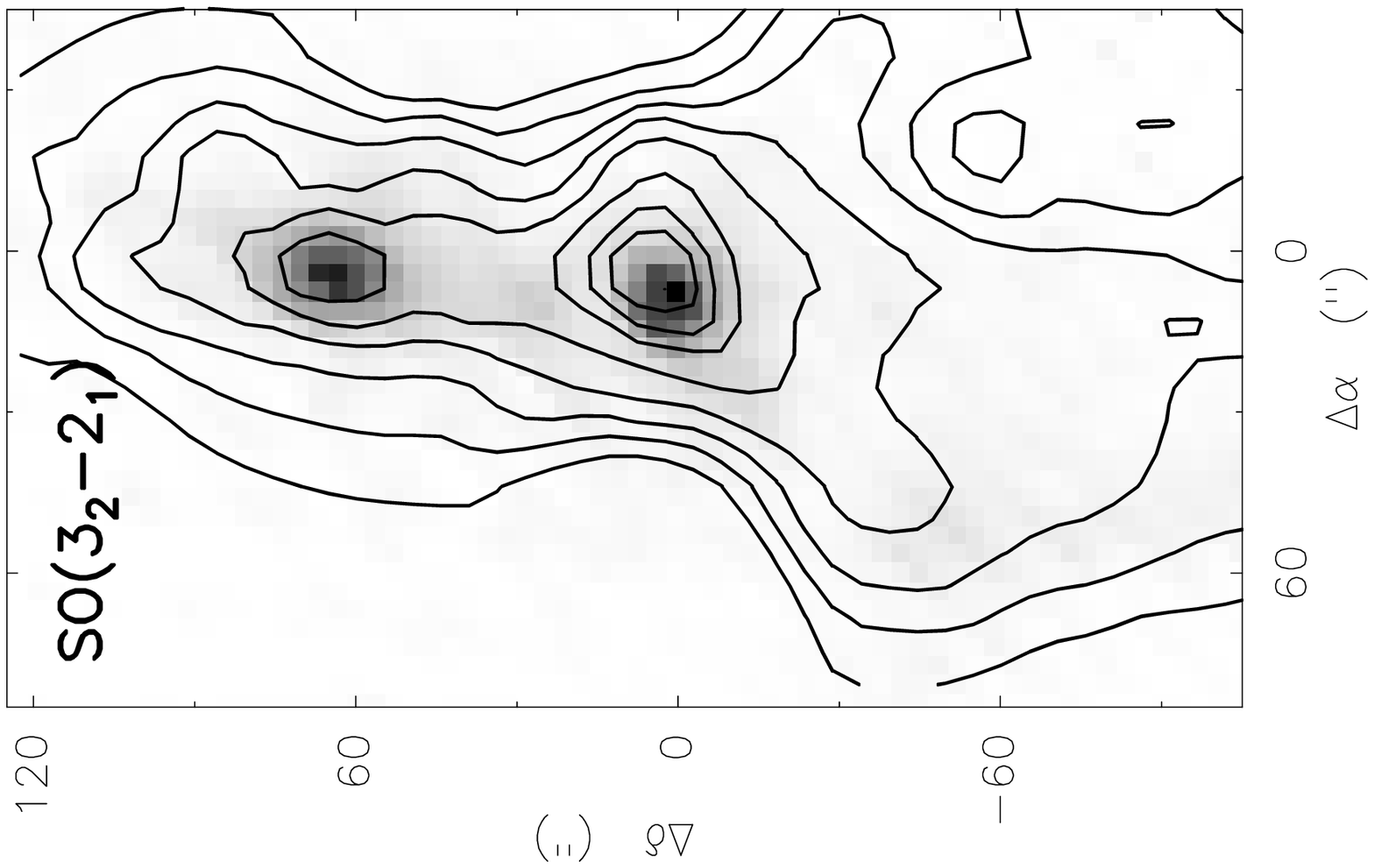}}}
\end{minipage}
\hfill
\begin{minipage}[b]{0.19\textwidth}
\centering
\resizebox{\hsize}{!}{\rotatebox{-90}{\includegraphics{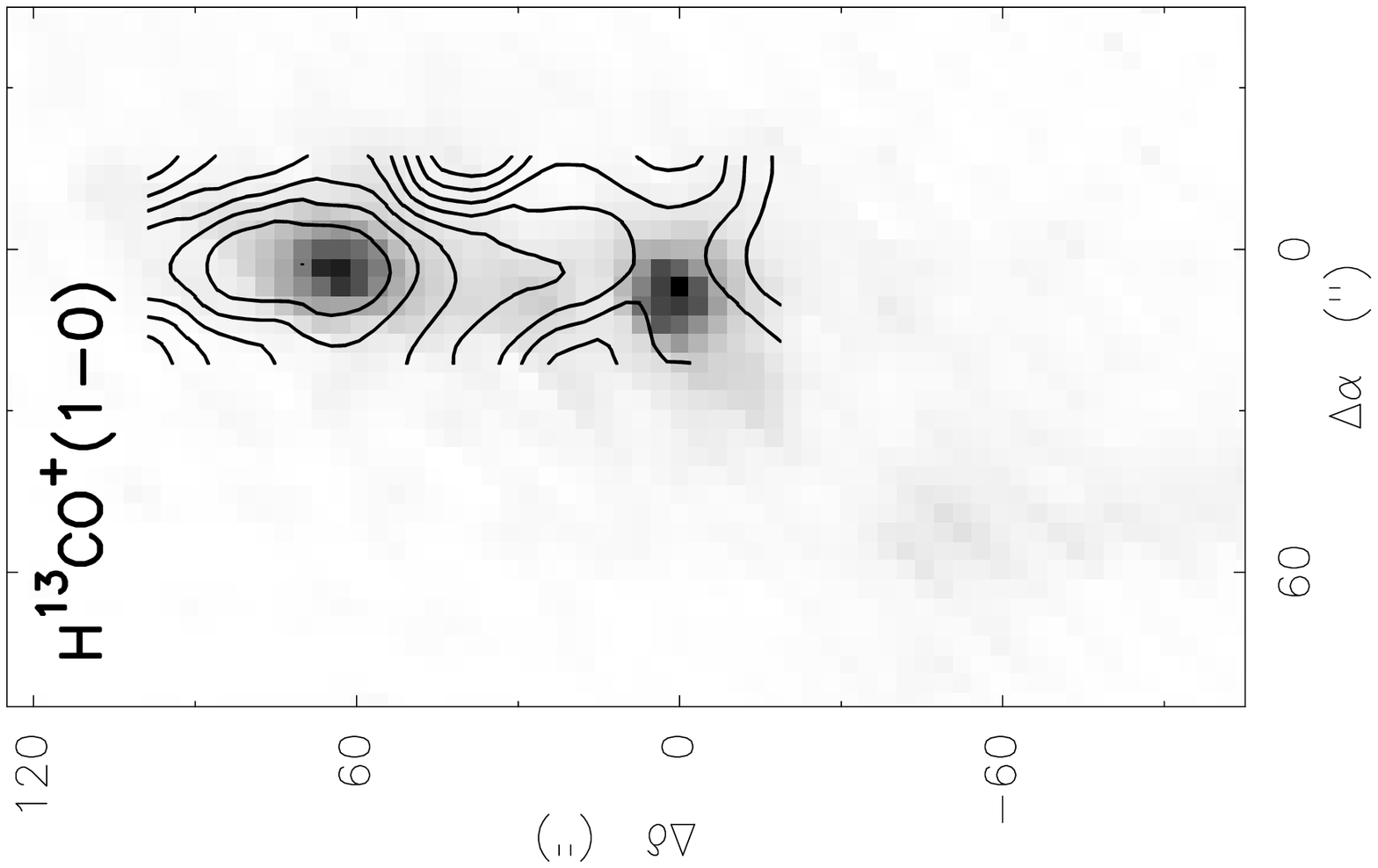}}}
\end{minipage}
\hfill
\begin{minipage}[b]{0.19\textwidth}
\centering
\resizebox{\hsize}{!}{\rotatebox{-90}{\includegraphics{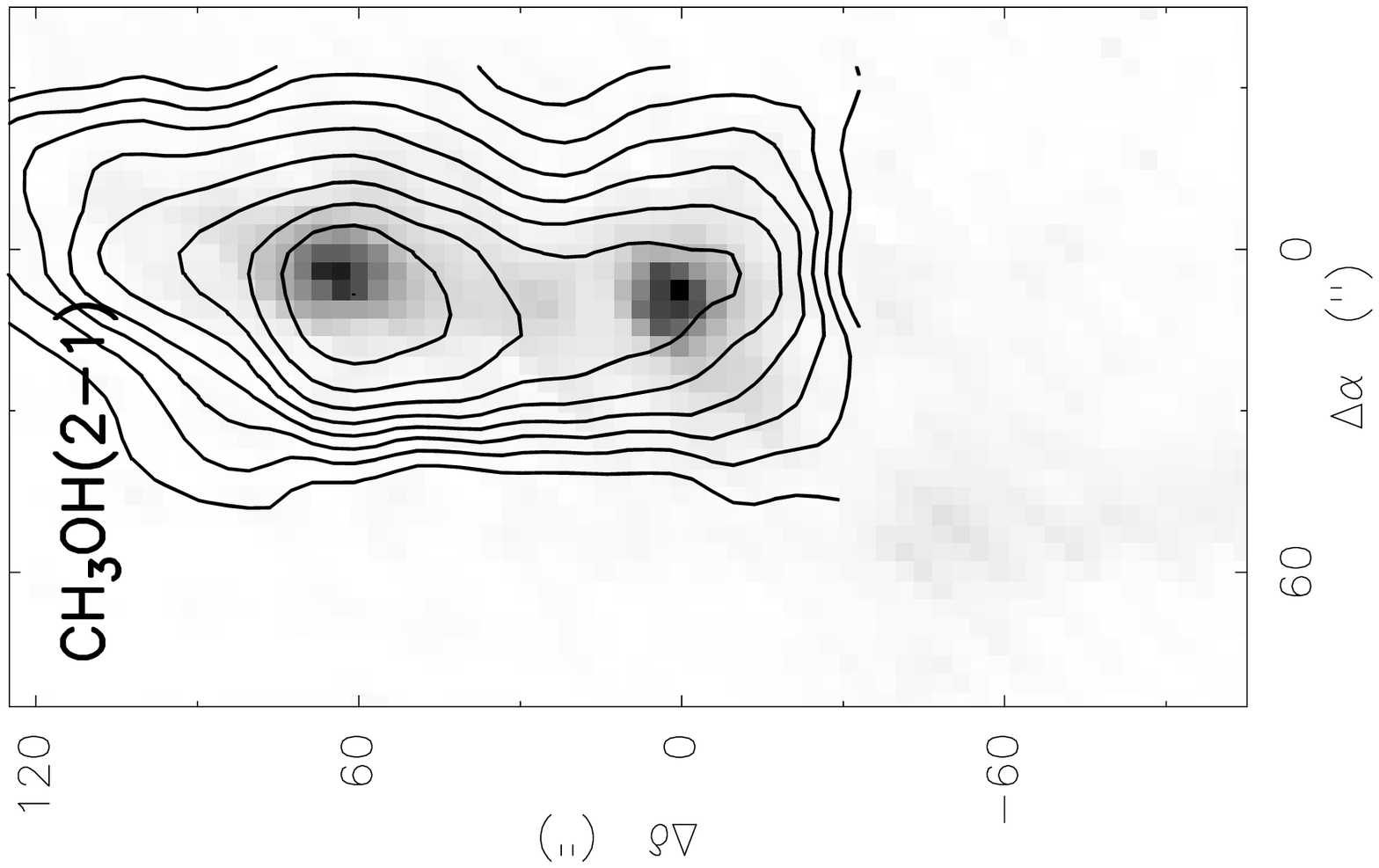}}}
\end{minipage}
\hfill
\begin{minipage}[b]{0.19\textwidth}
\centering
\resizebox{\hsize}{!}{\rotatebox{-90}{\includegraphics{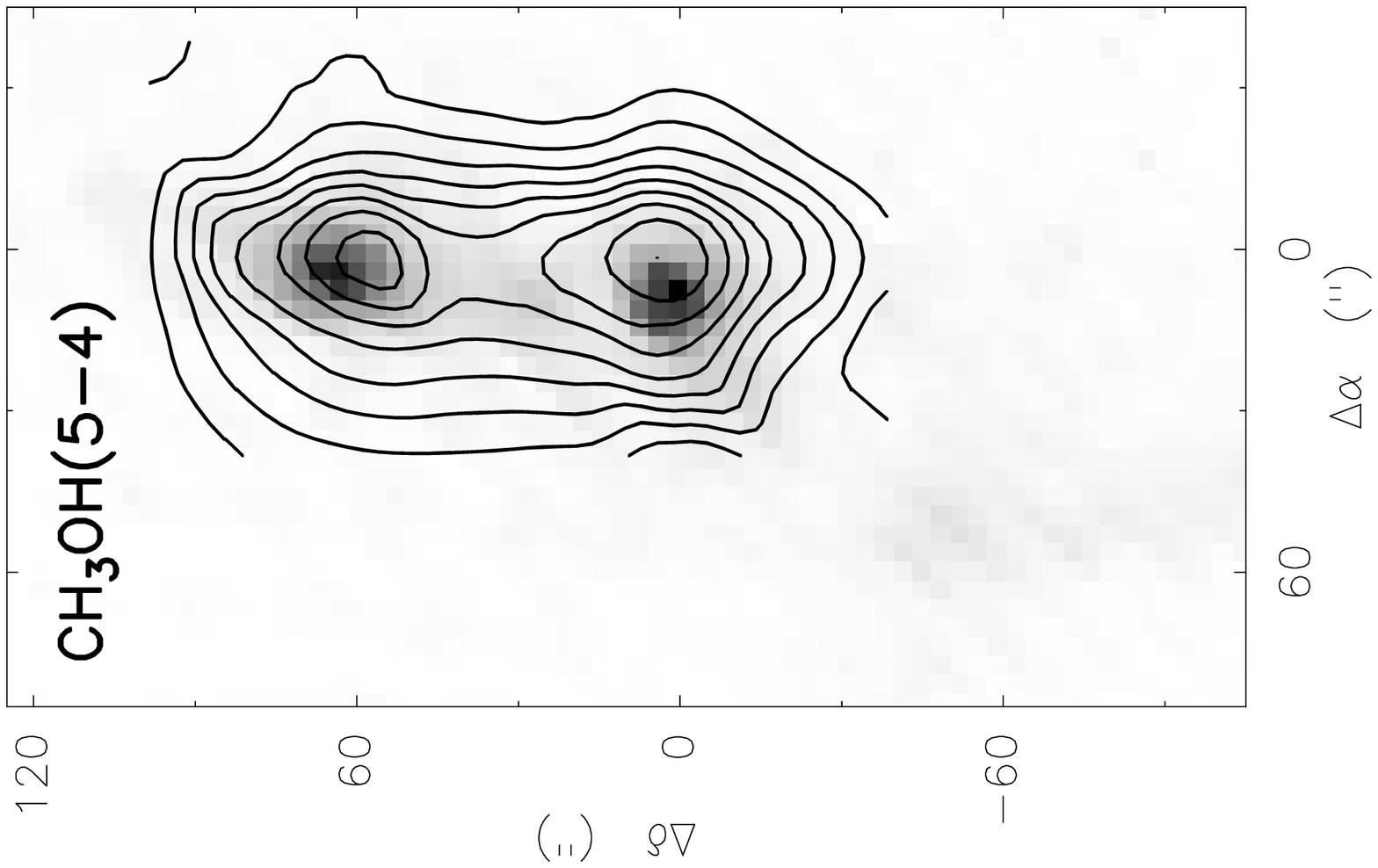}}}
\end{minipage}
\hfill
\begin{minipage}[b]{0.19\textwidth}
\centering
\resizebox{\hsize}{!}{\rotatebox{-90}{\includegraphics{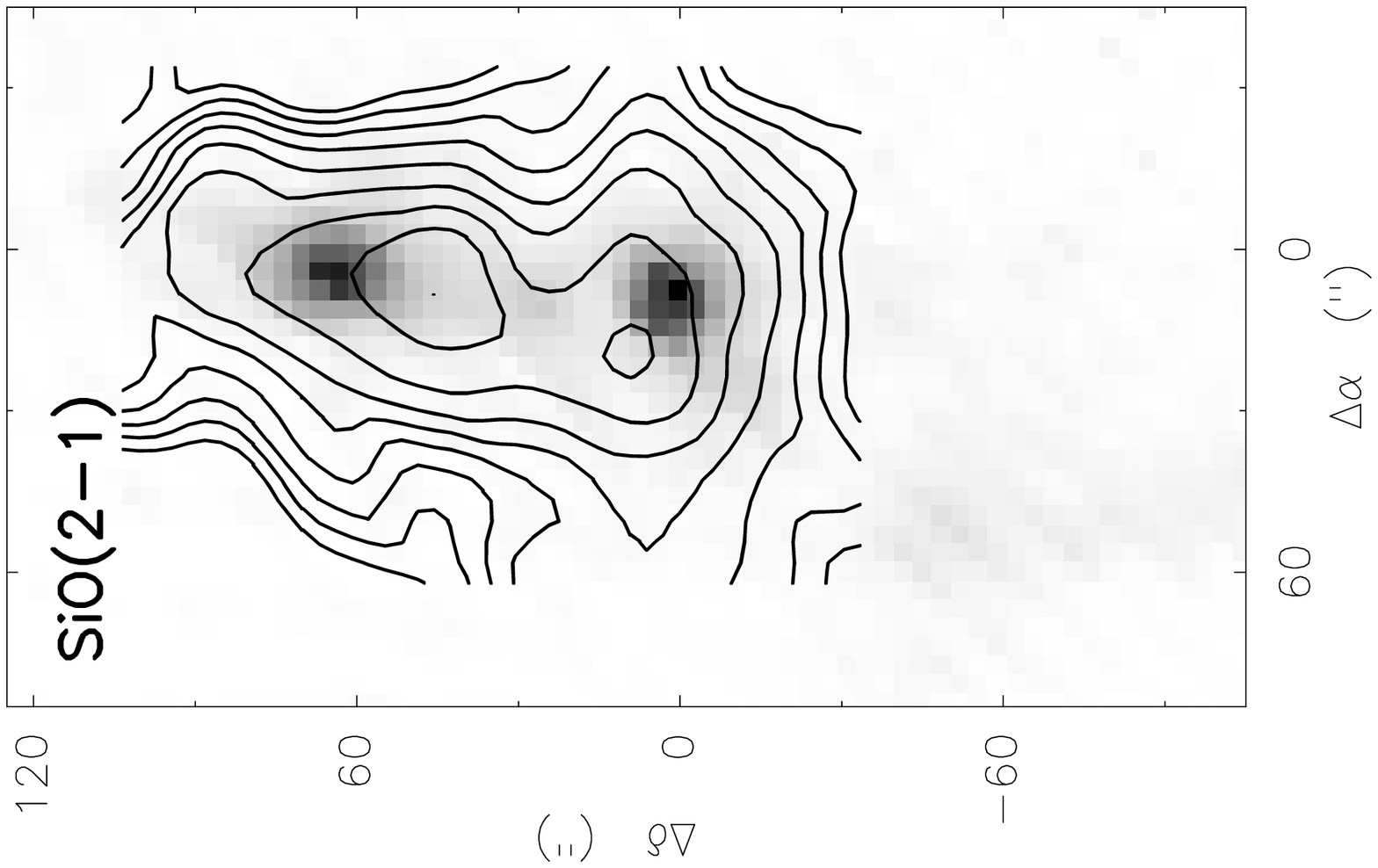}}}
\end{minipage}
\caption{Maps of S255 in various molecular lines (contours) overlaid on
the map of 1.2 mm dust continuum emission (grayscale). The contour levels
span from 20\% to 100\% of the peak intensity in steps of 10\%.}
\label{fig:maps-s255}
\end{figure*}

For S255 we have the most complete data set. In addition to the
common set of molecular transitions it includes SO($3_2-2_1$), SiO (2--1)
and (5--4), several methanol 2--1 and 5--4 series lines. Earlier we mapped it
also in ammomia (1,1) and (2,2) \citep*{Zin97}.
The molecular  maps overlaid on the greyscale map of the
1.2 mm continuum emission are presented in Fig.~\ref{fig:maps-s255}.
This source was mapped at 1.2 mm in continuum at the
IRAM 30m telescope by \citet{Mezger88}. Our observations
with the new array receiver provide a better sensitivity and a wider
map area, although the basic features of our map are consistent with
those previous results.

The maps show two main peaks of molecular and dust continuum
emission, around the (0,0) and (0,+60$''$) positions. The commonly accepted names for these clumps are S255IR and S255N. There is also a third
(southern) peak at about (+20$''$,$-$50$''$) noticeable in the
continuum map and in several molecular maps, at least in N$_2$H$^+$(1--0) and HCO$^+$(1--0). N$_2$H$^+$ and ammonia
are significantly stronger at the northern peak (S255N) while most other
species are either stronger at the central position (S255IR) or comparable at
both central and northern clumps. The dust emission is almost equal
for these clumps. The nature of these two components is different.
The central one is associated with a luminous cluster of IR sources,
whereas toward the northern one an ultracompact \Hii\ region
(G192.58-0.04) was detected. Recently several compact submillimetre continuum clumps were detected there with SMA \citep{Cyganowski07}. This object is extremely red in the
mid-IR band \citep{Crowther03}. \citet{Mezger88} derived almost exactly the same dust masses and
temperatures for both components from their 1.2 mm and $350\mu$m
observations. The gaussian line parameters at the central and
northern emission peaks are summarized in
Table~\ref{table:lines-s255}.

\begin{table*}
\caption{Molecular line parameters at the CS (central) and N$_2$H$^+$ (northern) emission
peaks in S255.}
\label{table:lines-s255}
\begin{tabular}{llllllll}
\hline%\hline
&\multicolumn{3}{l}{(0,0)}  &&\multicolumn{3}{l}{(0,+60$''$)}\\
\cline{2-4}
\cline{6-8}
Line &$T_{\rm mb}$ &$V_{\rm LSR}$ &$\Delta V$ & &$T_{\rm mb}$ &$V_{\rm LSR}$ &$\Delta V$ \\
&(K)          &(km/s)    &(km/s)     & &(K)          &(km/s)    &(km/s)     \\
\hline
C$^{18}$O(1--0) &1.99(12) &6.84(17) &4.31(43)
&&1.67(13) &8.74(14) &3.71(34)\\
C$^{18}$O(2--1) &6.45(08)  &7.18(01)  &2.88(03) &&5.89(08) &8.82(01) &2.95(03) \\
CS(2--1) &13.36(09)   &7.46(01)  &2.78(02) &&9.08(08) &8.40(01) &3.16(03)\\
C$^{34}$S(2--1) &1.93(11) &7.46(08) &2.78(18) &&1.33(10) &8.56(12) &3.20(29)\\
CS(5--4) &10.94(12) &7.49(01) &2.73(03) &&7.46(12) &8.55(02) &3.32(05) \\
C$^{34}$S(5--4) &1.13(06) &7.52(05) &2.33(12) &&0.75(08) &9.10(13) &3.57(26)\\
HCN(1--0) &13.79(11) &7.33(01) &3.25(03) &&11.52(11) &8.40(02) &3.60(03)\\
H$^{13}$CN(1--0) &1.41(07) &7.57(06) &2.58(12) &&1.37(07) &8.91(06) &2.12(11)\\
HCO$^+$(1--0) &7.34(10) &7.53(03) &3.90(06) &&11.27(11) &8.83(02) &3.21(04)\\
H$^{13}$CO$^+$(1--0) &0.74(10) &7.51(11) &1.69(26) &&1.39(08) &8.97(07) &2.79(18)\\
HNC(1--0) &11.87(18) &7.30(02) &2.73(05) &&11.30(17) &8.55(02) &3.32(06)\\
HN$^{13}$C(1--0) &0.18(02) &7.29(20) &3.30(47) &&0.28(02) &8.87(16) &4.78(37)\\
N$_2$H$^+$(1--0) &1.46(06) &7.44(04) &2.25(10) &&2.58(05) &8.97(02) &2.61(06)\\
SiO(2--1) &0.18(06) &7.96(52) &3.32(122) &&0.20(05) &8.06(61) &5.24(144)\\
\hline
\end{tabular}
\end{table*}

\subsubsection{DR-21 NH$_3$}

\begin{figure}
\resizebox{\hsize}{!}{\includegraphics{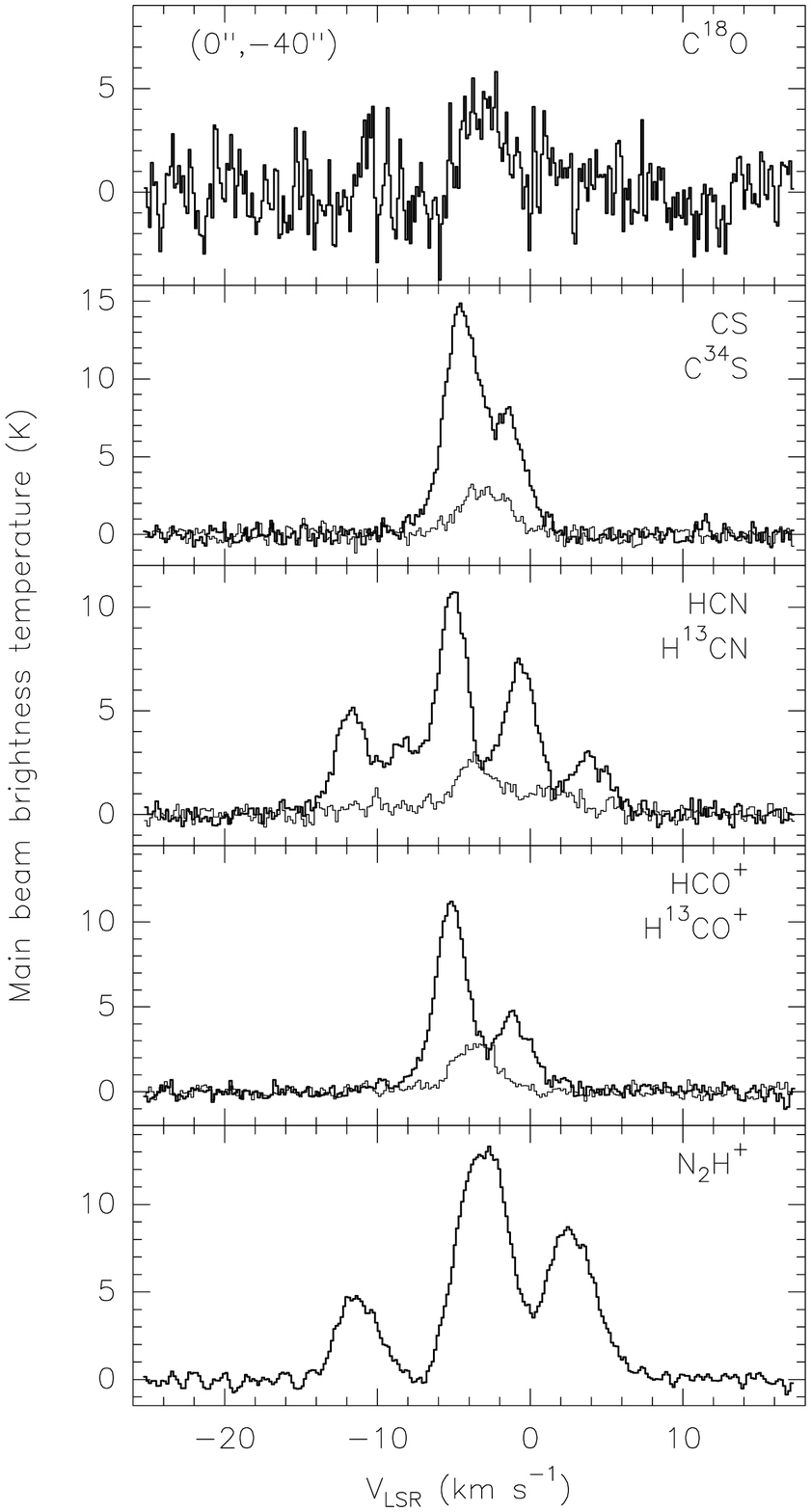}}
\caption{Molecular line spectra towards the N$_2$H$^+$ emission peak
in DR-21 NH$_3$.}
\label{fig:dr21-spectra}
\end{figure}

\begin{figure*}
\begin{minipage}[b]{0.24\textwidth}
\centering
\resizebox{\hsize}{!}{\rotatebox{-90}{\includegraphics{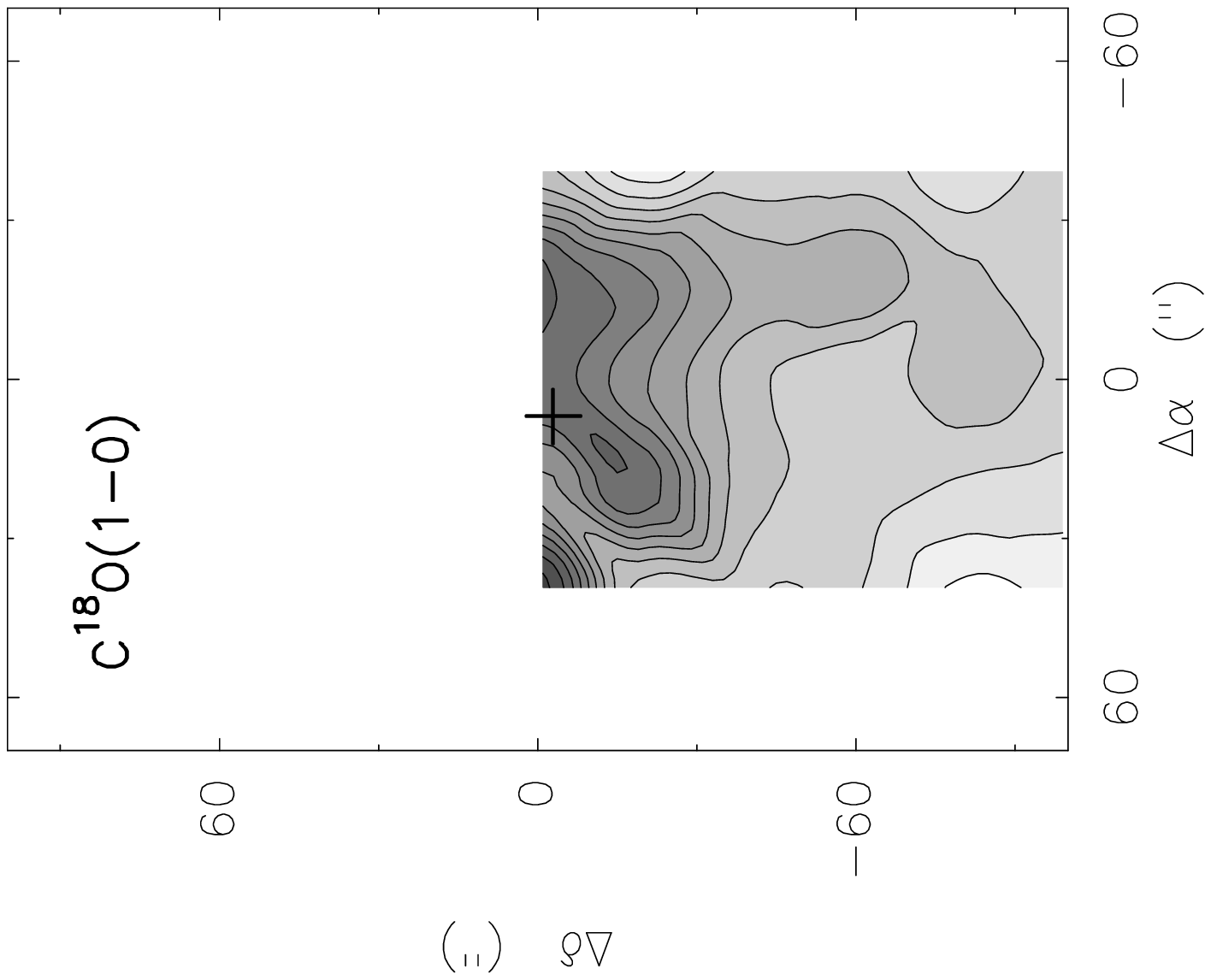}}}
\end{minipage}
\hfill
\begin{minipage}[b]{0.24\textwidth}
\centering
\resizebox{\hsize}{!}{\rotatebox{-90}{\includegraphics{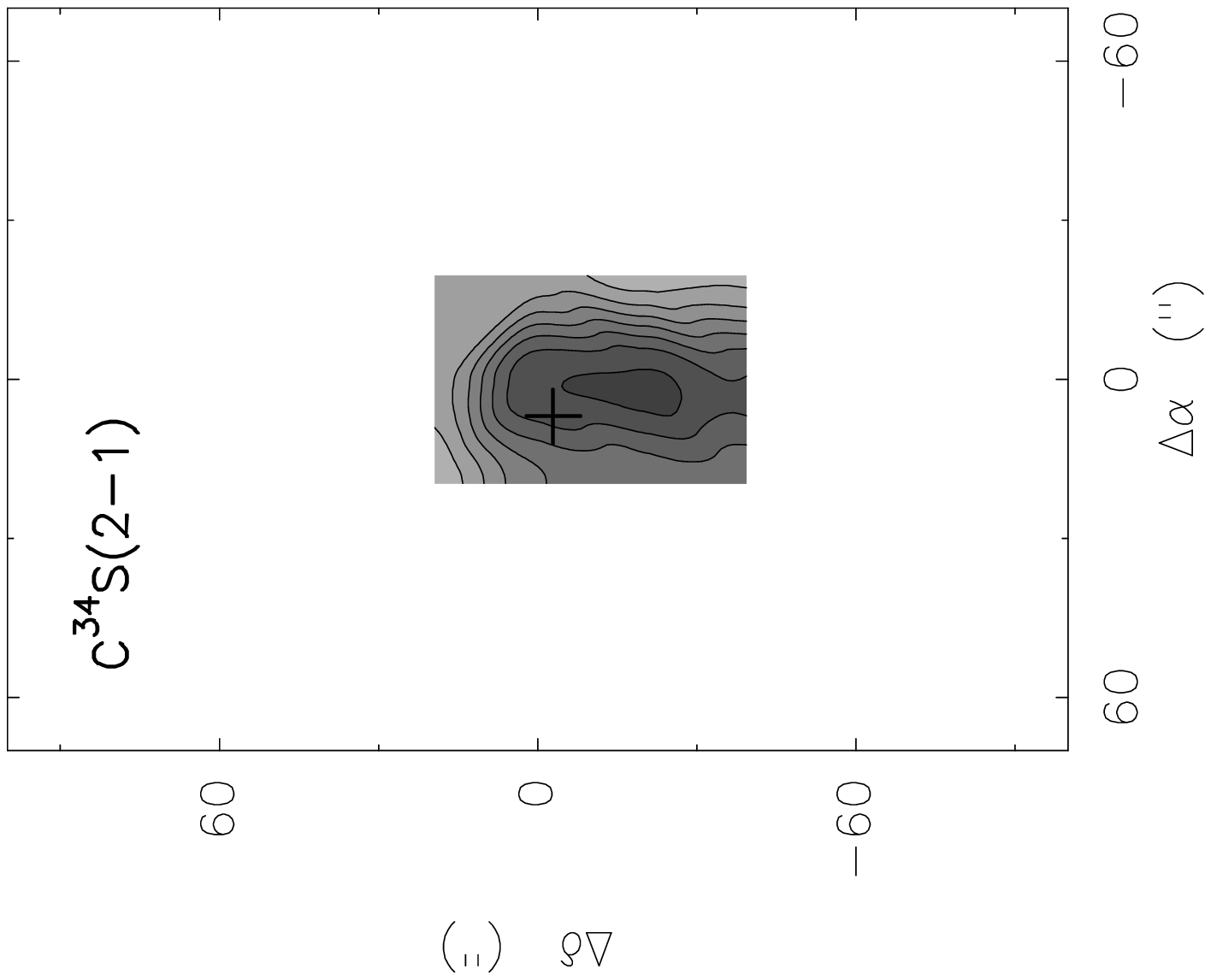}}}
\end{minipage}
\hfill
\begin{minipage}[b]{0.24\textwidth}
\centering
\resizebox{\hsize}{!}{\rotatebox{-90}{\includegraphics{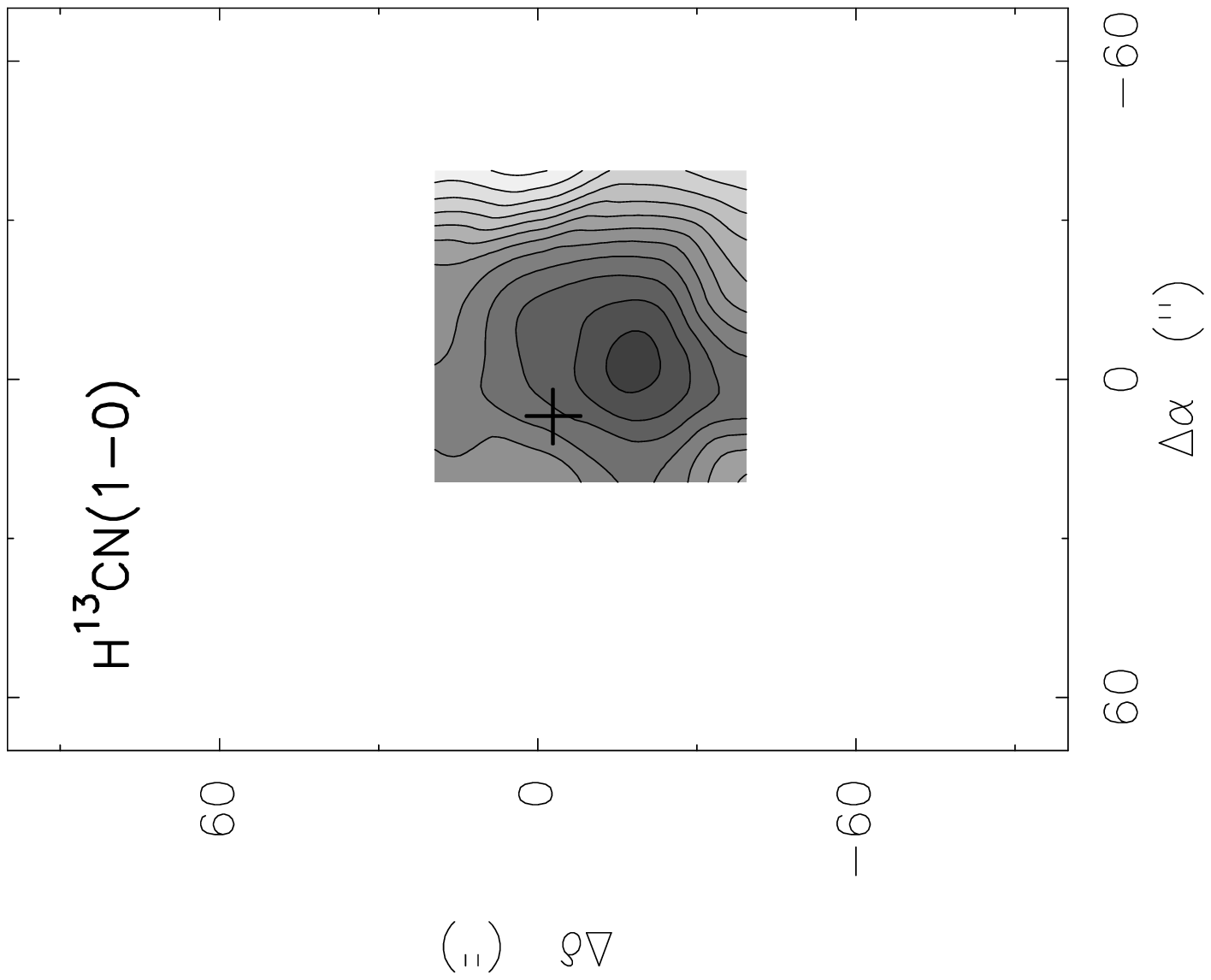}}}
\end{minipage}
\hfill
\begin{minipage}[b]{0.24\textwidth}
\centering
\resizebox{\hsize}{!}{\rotatebox{-90}{\includegraphics{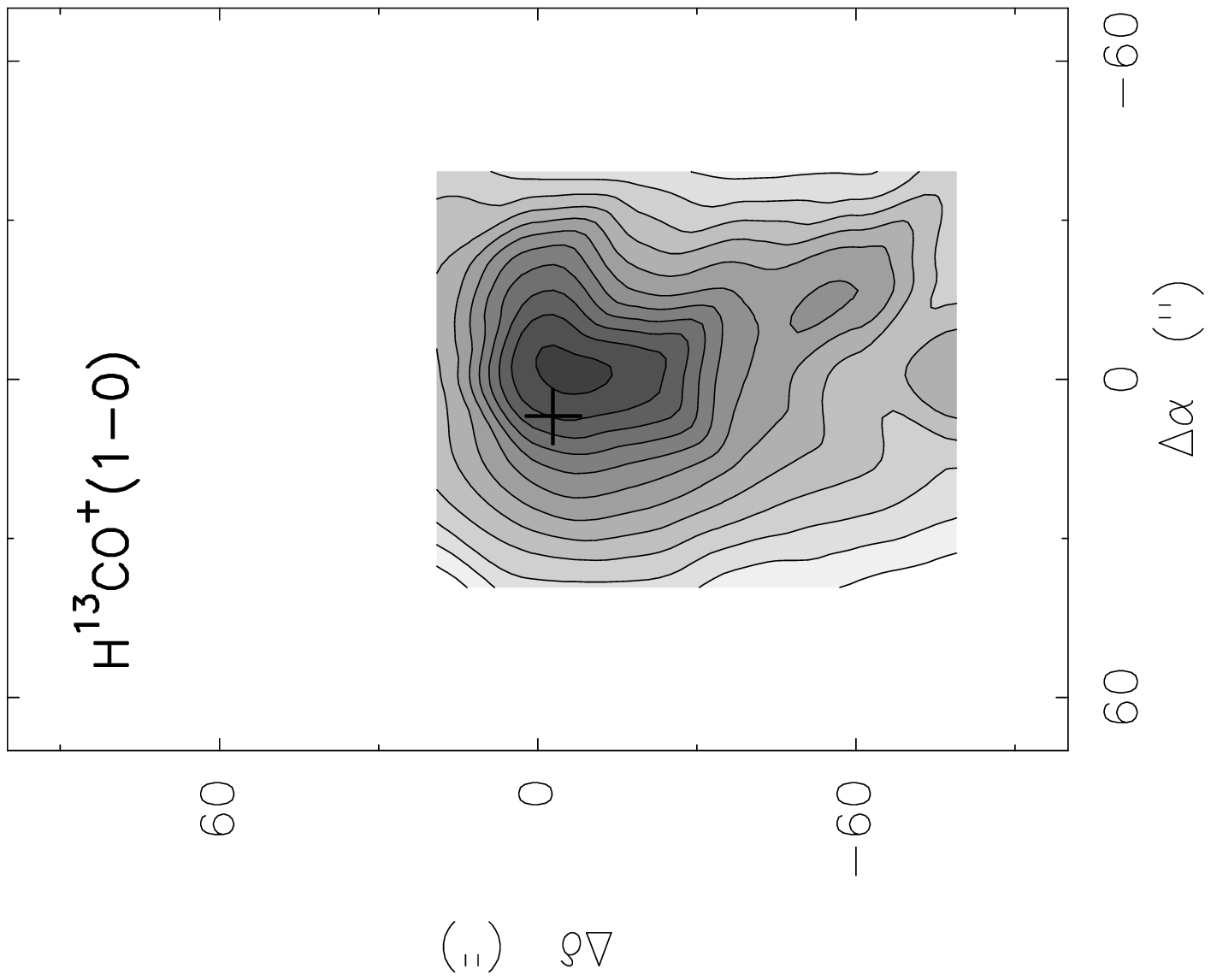}}}
\end{minipage}
\\[1ex]
\begin{minipage}[b]{0.24\textwidth}
\centering
\resizebox{\hsize}{!}{\rotatebox{-90}{\includegraphics{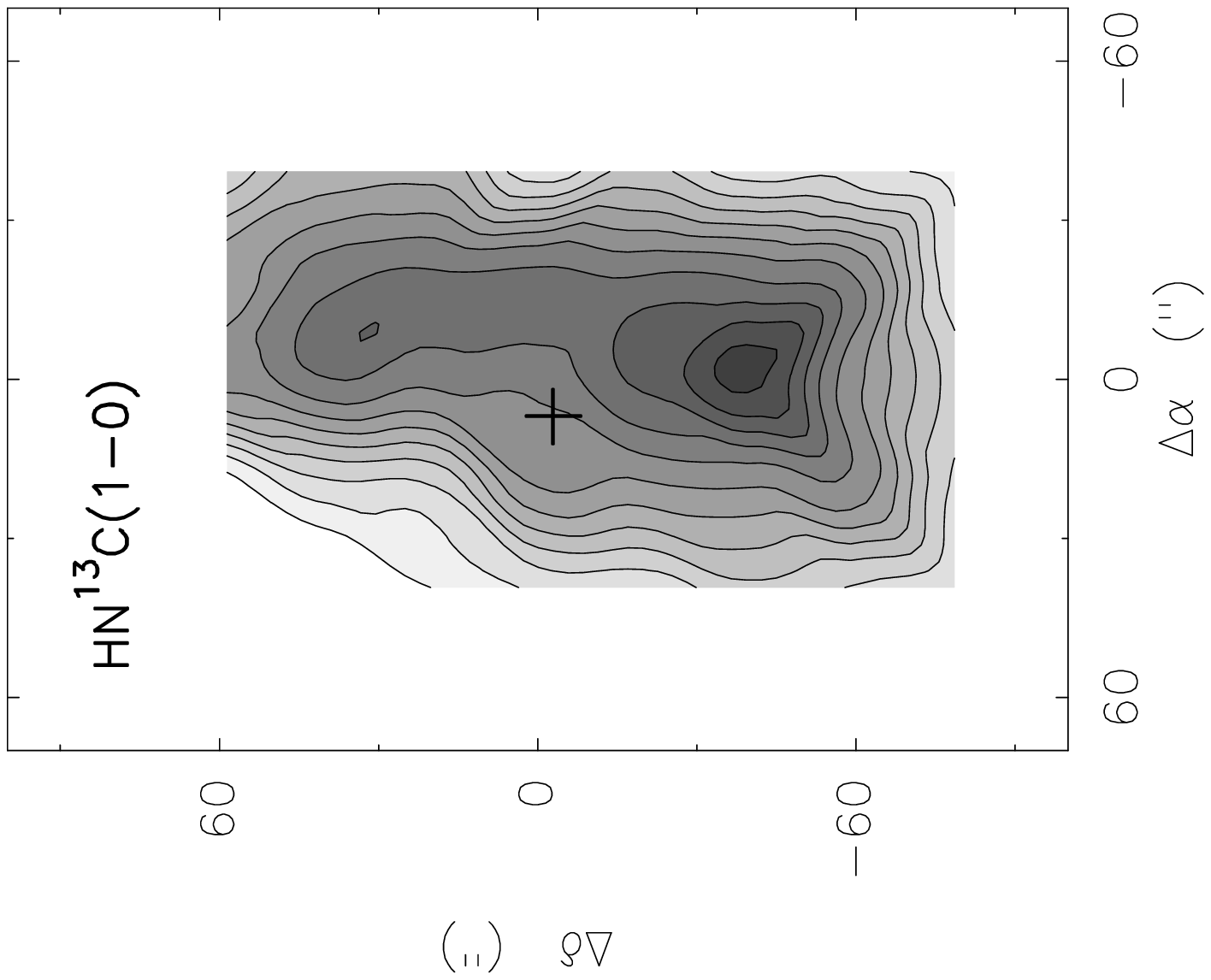}}}
\end{minipage}
\hfill
\begin{minipage}[b]{0.24\textwidth}
\centering
\resizebox{\hsize}{!}{\rotatebox{-90}{\includegraphics{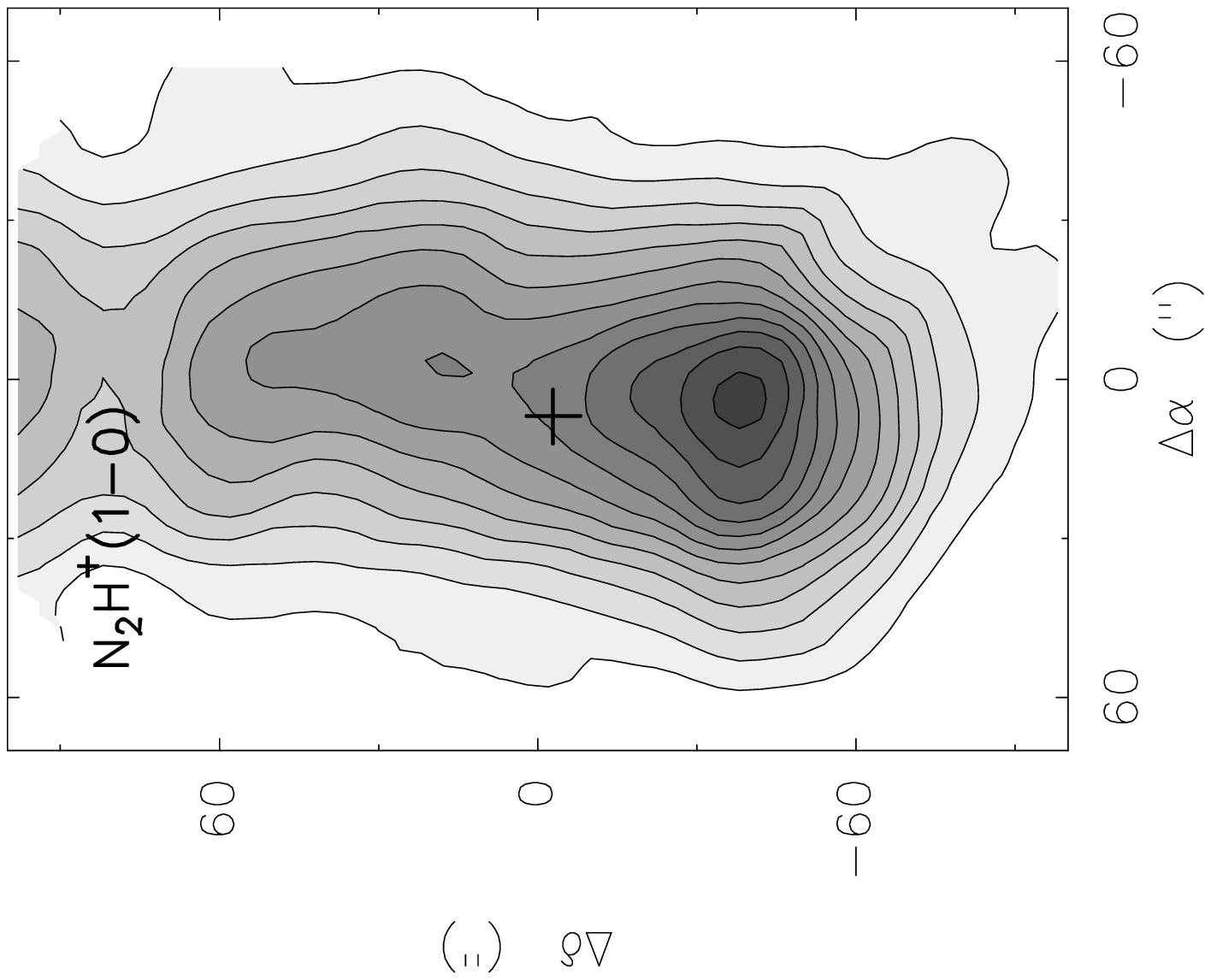}}}
\end{minipage}
\hfill
\begin{minipage}[b]{0.24\textwidth}
\centering
\hspace*{\textwidth}
\end{minipage}
\hfill
\begin{minipage}[b]{0.24\textwidth}
\centering
\hspace*{\textwidth}
\end{minipage}
\caption{Maps of DR-21 NH$_3$ in various molecular lines.
The contour levels
span from 20\% to 100\% of the peak intensity in steps of 10\%. The cross marks the position of DR21 (OH).}
\label{fig:maps-dr21}
\end{figure*}

This is another well known site of active star formation. The CS, HCN
and HCO$^+$ spectra here suffer from a very strong red-shifted self-absorption
(Fig.~\ref{fig:dr21-spectra}), which makes an analysis of corresponding
maps almost senseless. For this reason we present here and discuss only rarer isotopologue and N$_2$H$^+$(1--0) maps (Fig.~\ref{fig:maps-dr21}). We did not map the dust continuum emission in this area, given that it has been already observed by \citet*{Chandler93}.
It is easy to see that the N$_2$H$^+$ and HN$^{13}$C
peaks are shifted by about 0{\farcm}5 to the south from the emission peaks
of other species. The latter peaks practically coincide with the main dust
emission peak DR21(OH)M in the notation introduced by \citet*{Mangum91}. The N$_2$H$^+$ and HN$^{13}$C peaks lie near a weaker
dust peak DR21(OH)S. The gaussian line parameters are given in
Table~\ref{table:lines-dr21}.

\begin{table*}
\caption{Molecular line parameters at the CS and N$_2$H$^+$ emission
peaks in DR-21 NH$_3$.}
\label{table:lines-dr21}
\begin{tabular}{llllllll}
\hline%\hline
&\multicolumn{3}{l}{(0,$-$20$''$)}  &&\multicolumn{3}{l}{(0,$-$40$''$)}\\
\cline{2-4}
\cline{6-8}
Line &$T_{\rm mb}$ &$V_{\rm LSR}$ &$\Delta V$ & &$T_{\rm mb}$ &$V_{\rm LSR}$ &$\Delta V$ \\
&(K)          &(km/s)    &(km/s)     & &(K)          &(km/s)    &(km/s)     \\
\hline
C$^{18}$O(1--0) &4.99(14) &$-$2.99(05) &3.84(13) &&4.69(15) &$-$3.37(05) &3.28(12)\\
C$^{34}$S(2--1) &2.62(06) &$-$3.04(05) &4.68(12) &&2.89(06) &$-$3.01(04) &3.83(10)\\
H$^{13}$CN(1--0) &2.34(06) &$-$3.24(06) &4.07(10) &&2.34(06) &$-$3.47(05) &3.58(09)\\
H$^{13}$CO$^+$(1--0) &3.17(07) &$-$3.46(04) &3.67(09) &&2.73(07) &$-$3.57(04) &3.25(10)\\
HN$^{13}$C(1--0) &1.34(09) &$-$3.05(14) &4.38(32) &&1.70(09) &$-$3.25(10) &3.56(23)\\
N$_2$H$^+$(1--0) &4.99(04) &$-$3.11(02) &4.04(03) &&6.93(04) &$-$3.25(01) &3.51(02)\\
\hline
\end{tabular}
\end{table*}

\subsubsection{S140}

S140 is one of the best studied sites of active star formation. In particular it was observed at various instruments in the HCN, HCO$^+$ and CO isotopic lines \citep[e.g.][]{Park95}. Nevertheless, we present here our own data on these
lines too, to better compare with other molecular maps (all maps have been obtained at Onsala). The N$_2$H$^+$ results have been published earlier \citep{Pirogov03}.

\begin{figure*}
\begin{minipage}[b]{0.24\textwidth}
\centering
\resizebox{\hsize}{!}{\rotatebox{-90}{\includegraphics{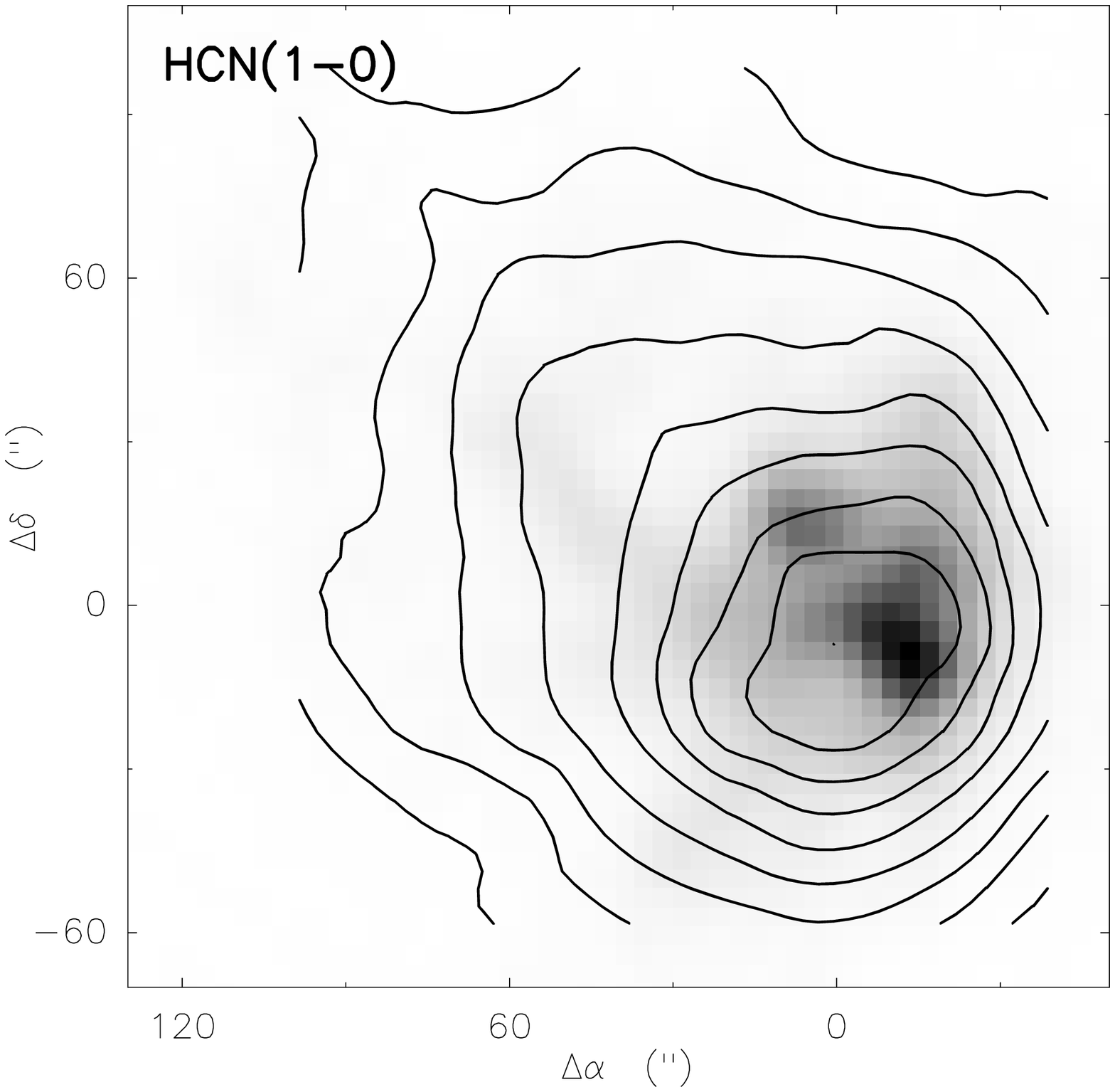}}}
\end{minipage}
\hfill
\begin{minipage}[b]{0.24\textwidth}
\centering
\resizebox{\hsize}{!}{\rotatebox{-90}{\includegraphics{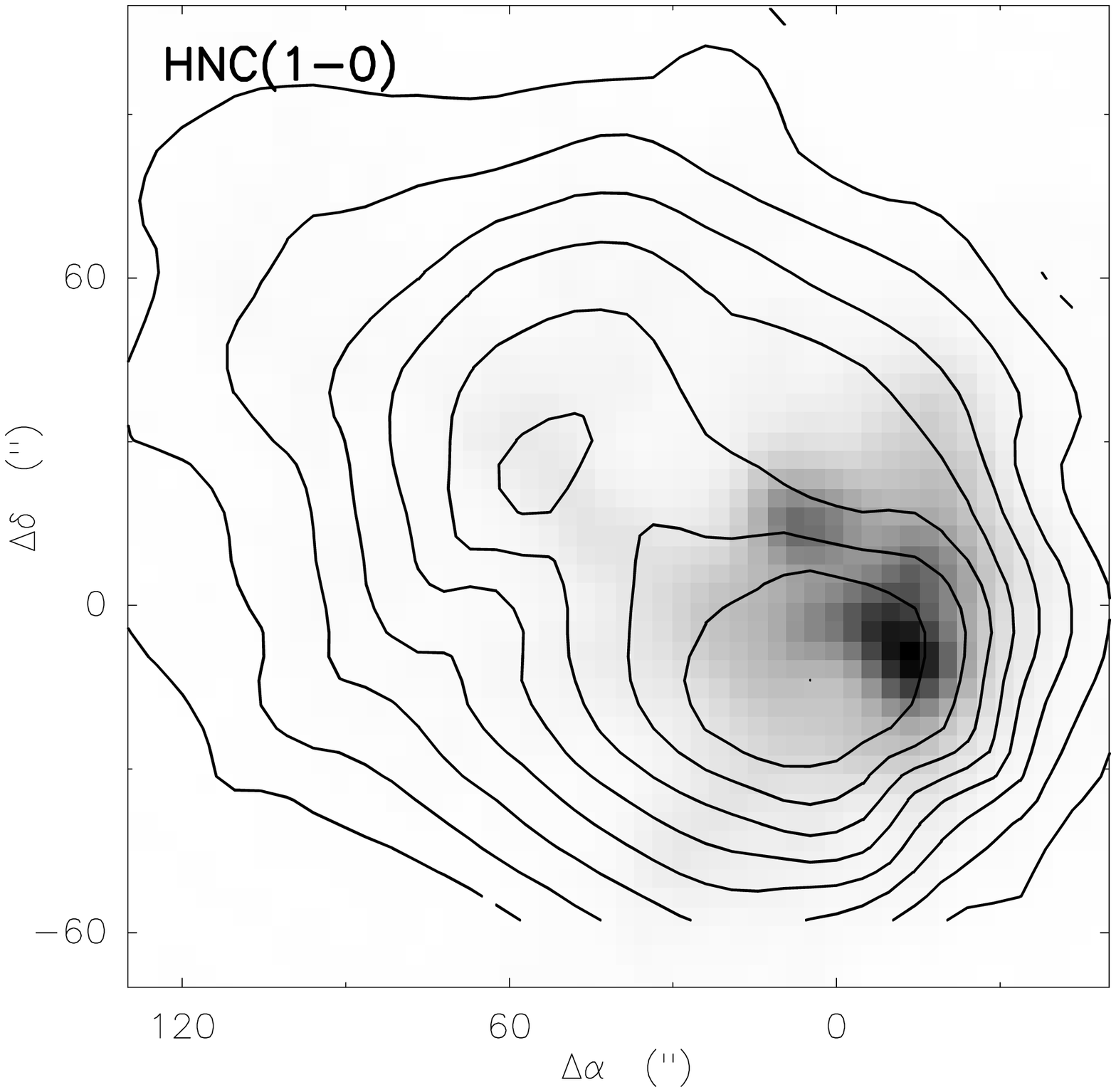}}}
\end{minipage}
\hfill
\begin{minipage}[b]{0.24\textwidth}
\centering
\resizebox{\hsize}{!}{\rotatebox{-90}{\includegraphics{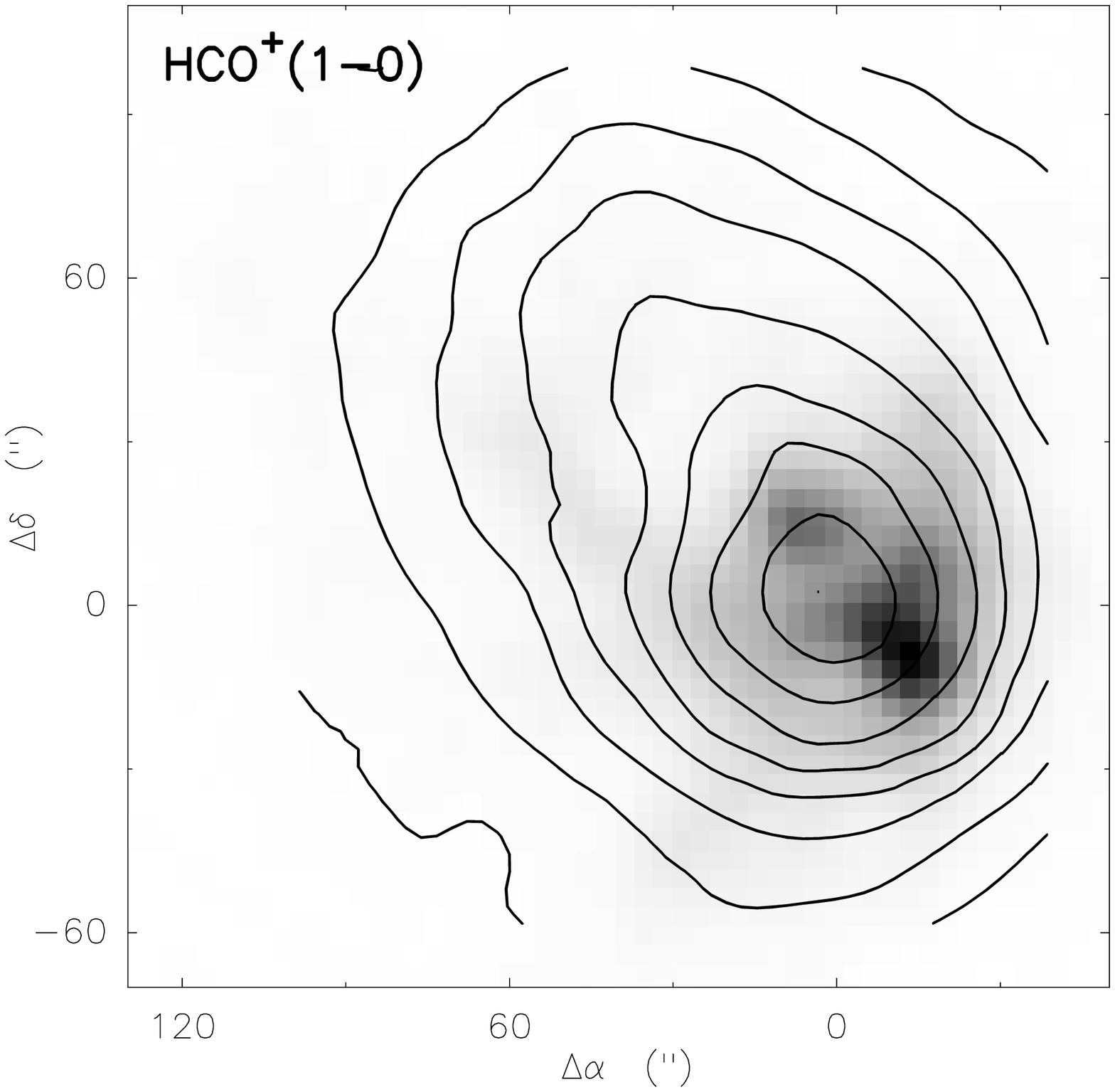}}}
\end{minipage}
\hfill
\begin{minipage}[b]{0.24\textwidth}
\centering
\resizebox{\hsize}{!}{\rotatebox{-90}{\includegraphics{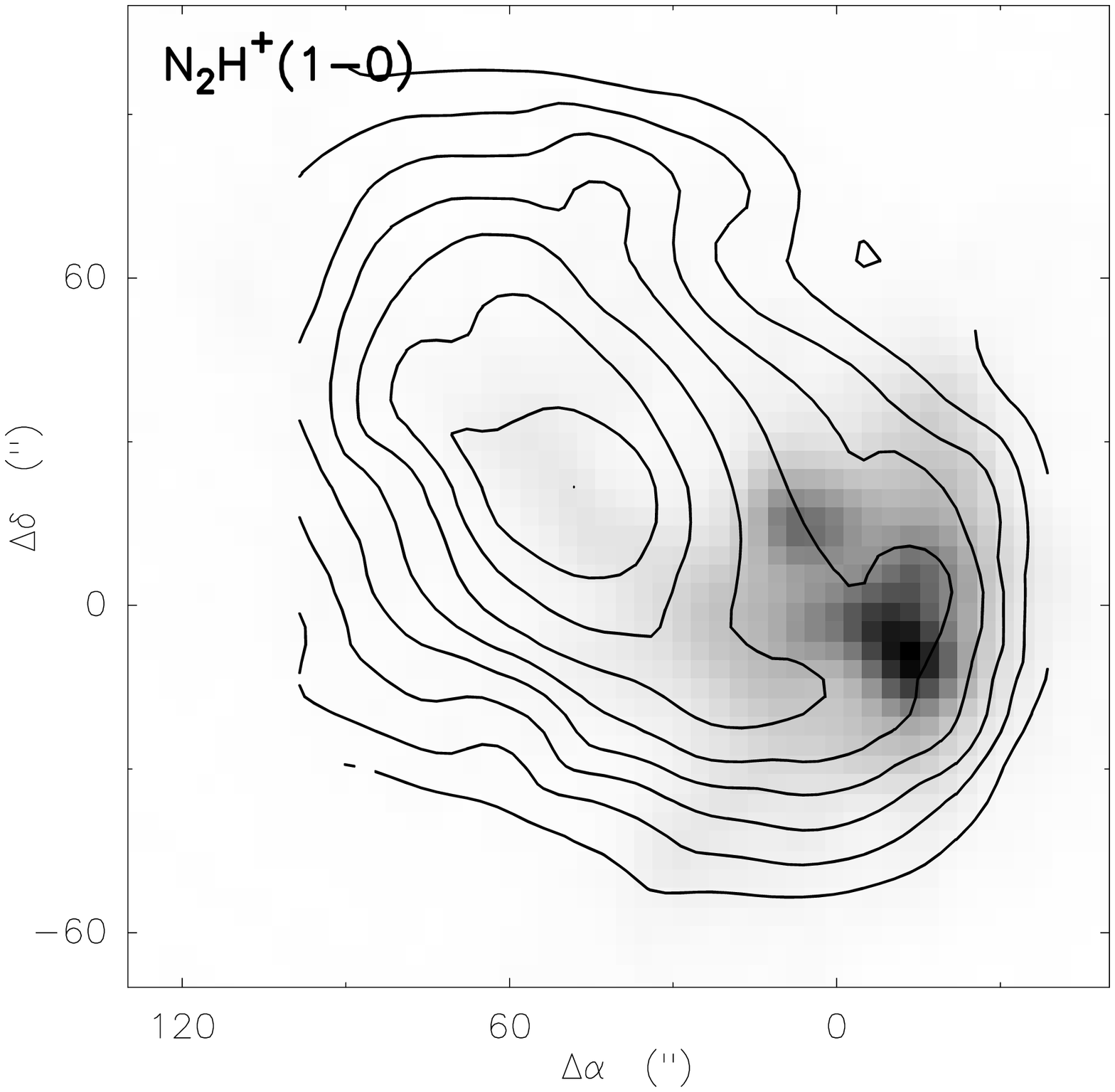}}}
\end{minipage}
\\[1ex]
\begin{minipage}[b]{0.24\textwidth}
\centering
\resizebox{\hsize}{!}{\rotatebox{-90}{\includegraphics{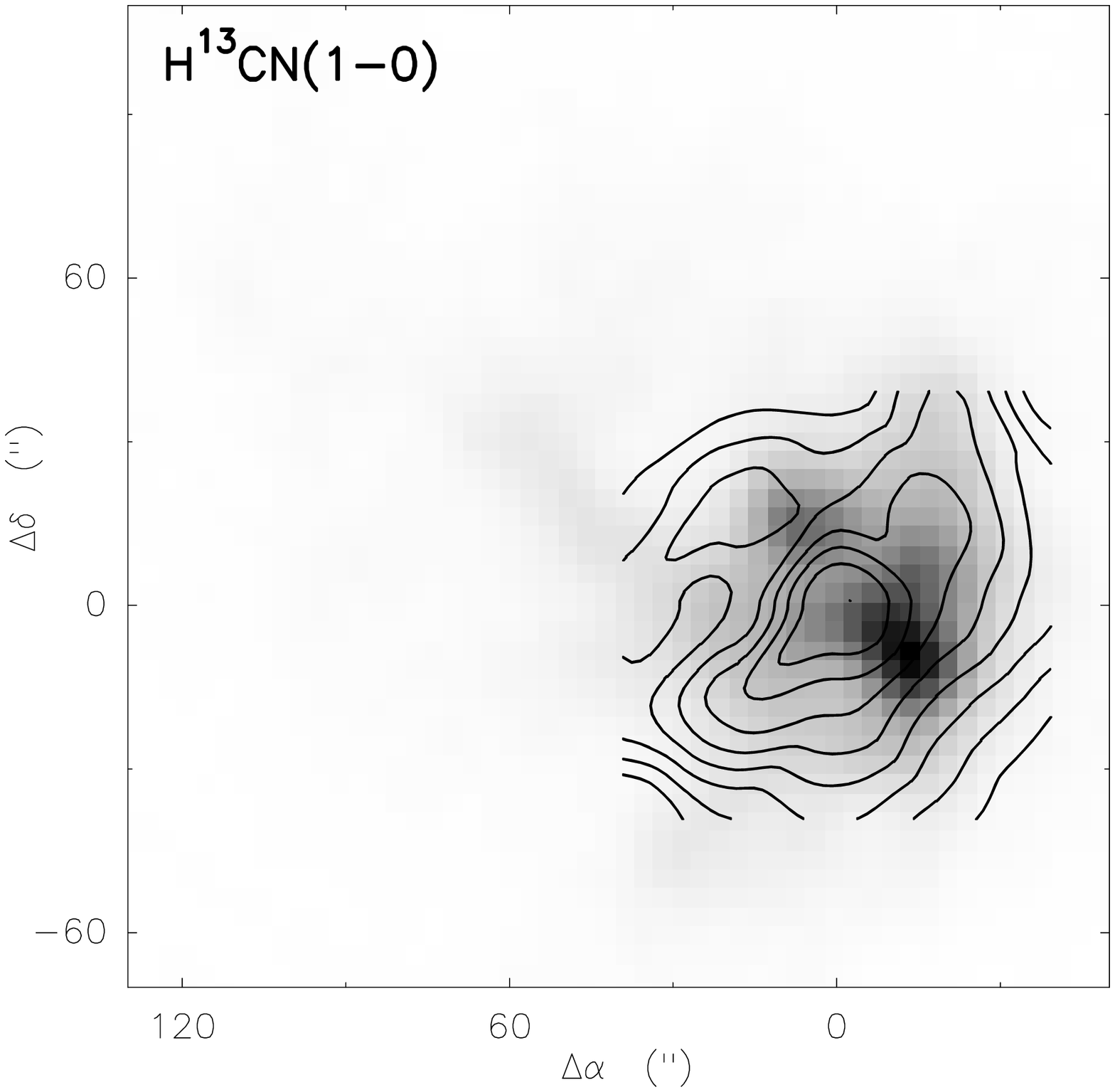}}}
\end{minipage}
\hfill
\begin{minipage}[b]{0.24\textwidth}
\centering
\resizebox{\hsize}{!}{\rotatebox{-90}{\includegraphics{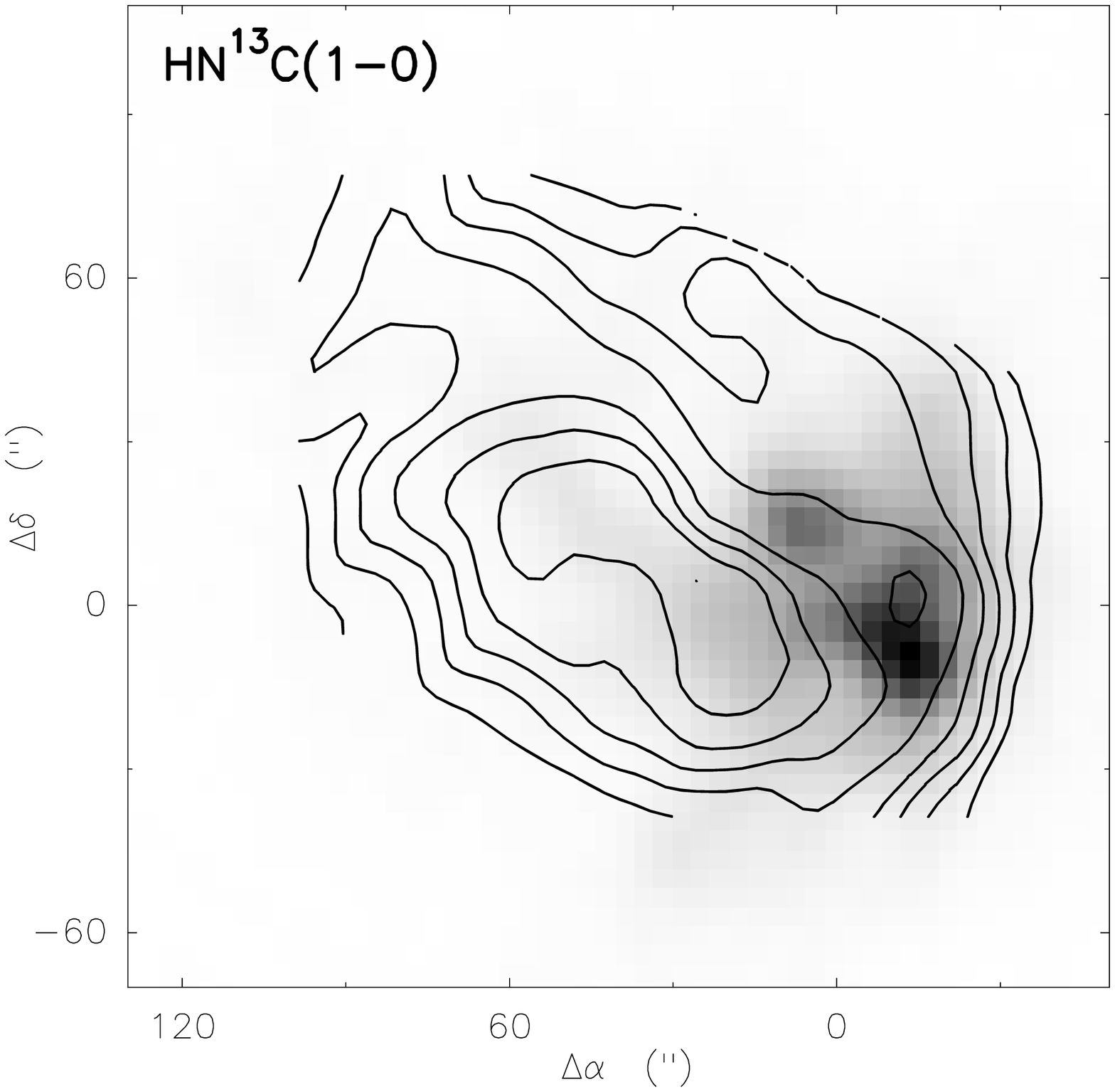}}}
\end{minipage}
\hfill
\begin{minipage}[b]{0.24\textwidth}
\centering
\resizebox{\hsize}{!}{\rotatebox{-90}{\includegraphics{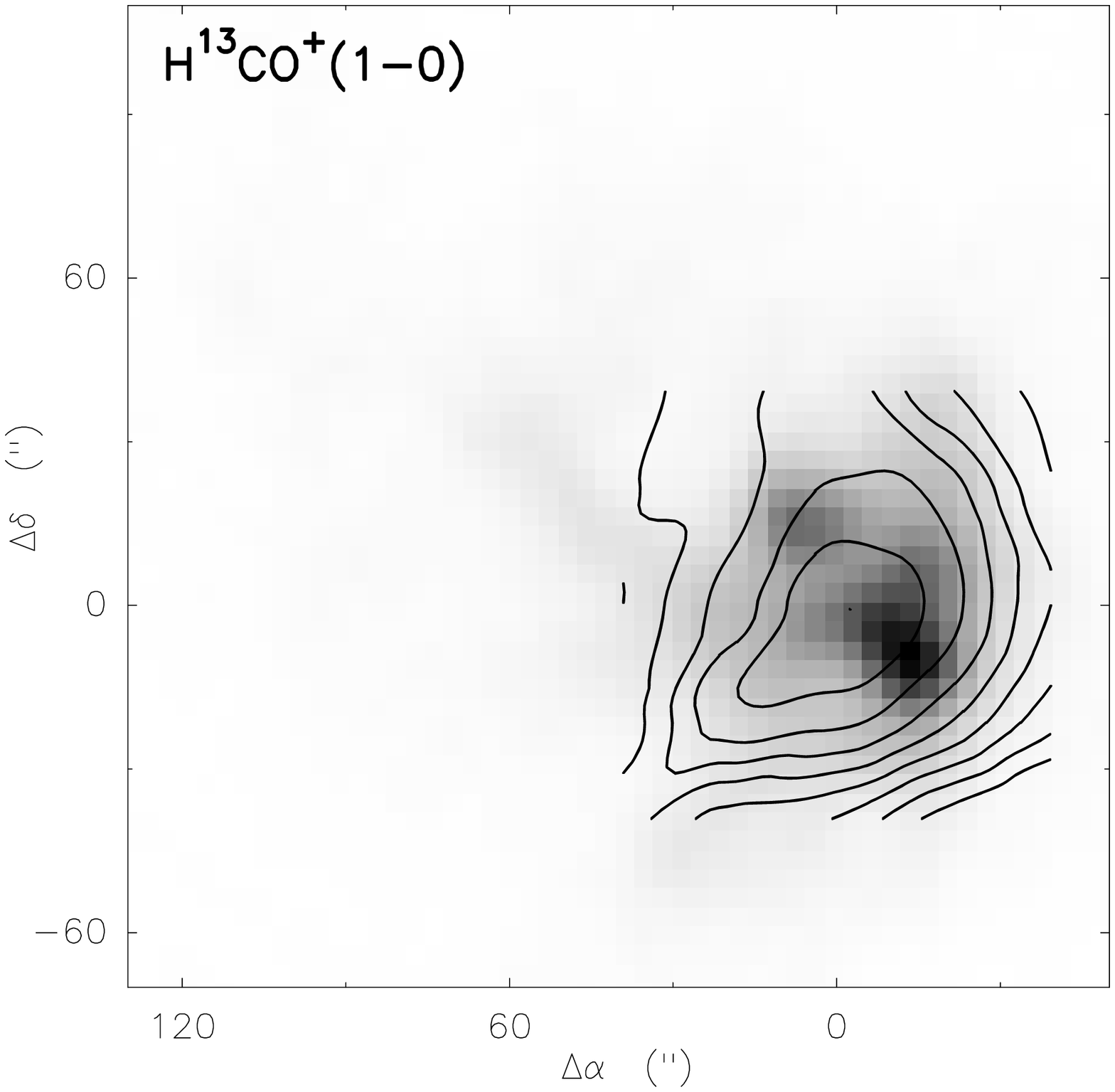}}}
\end{minipage}
\hfill
\begin{minipage}[b]{0.24\textwidth}
\centering
%\hspace*{\textwidth}
\resizebox{\hsize}{!}{\rotatebox{-90}{\includegraphics{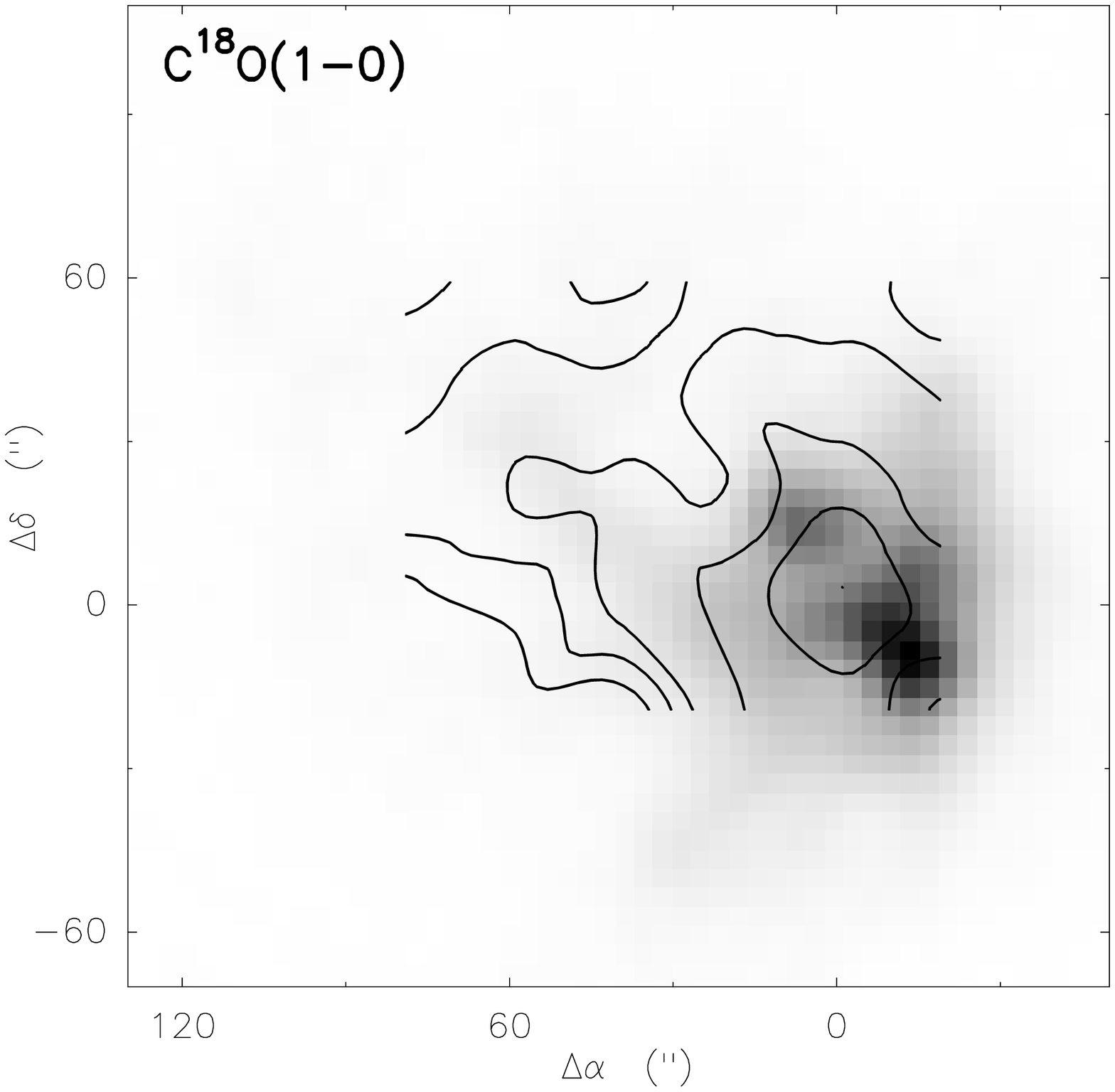}}}
\end{minipage}
\caption{Maps of S140 in various molecular lines (contours) overlaid on
the map of 1.2 mm dust continuum emission (grayscale). The contour levels
span from 20\% to 100\% of the peak intensity in steps of 10\%.}
\label{fig:maps-s140}
\end{figure*}

The maps (Fig.~\ref{fig:maps-s140}) clearly show that HCN and HCO$^+$
emissions peak near the (0,0) position while the N$_2$H$^+$ peak is shifted
to about (+50$''$,+20$''$). A multi-transitional CS study \citep{Zhou94} shows that CS emission peak is near the (0,0) position.
HNC is an intermediate case and there is a
secondary HNC peak near the N$_2$H$^+$ peak. These differences in
emission distributions are not caused by opacity effects because the maps
in the lines of rarer isotopic modifications of these molecules
show the same features
(the N$_2$H$^+$ optical depth is rather small, $\sim 0.4$,
as shown by \citealt{Pirogov03}). The gaussian line parameters are presented in
Table~\ref{table:lines-s140}.

\begin{table*}
\caption{Molecular line parameters at the CS and N$_2$H$^+$ emission
peaks in S140.}
\label{table:lines-s140}
\begin{tabular}{llllllll}
\hline%\hline
&\multicolumn{3}{l}{(0,0)}  &&\multicolumn{3}{l}{(+40$''$,+20$''$)}\\
\cline{2-4}
\cline{6-8}
Line &$T_{\rm mb}$ &$V_{\rm LSR}$ &$\Delta V$ & &$T_{\rm mb}$ &$V_{\rm LSR}$ &$\Delta V$ \\
&(K)          &(km/s)    &(km/s)     & &(K)          &(km/s)    &(km/s)     \\
\hline
C$^{18}$O(1--0) &3.19(16) &$-$6.83(09) &3.46(22)
&&2.53(18) &$-$7.51(10) &2.95(24)\\
HCN(1--0) &18.37(11) &$-$6.80(01)&3.05(02)& &11.43(11) &$-$6.92(01) &2.84(03) \\
H$^{13}$CN(1--0) &1.50(07) &$-$7.00(05) &2.22(10) &&0.79(08) &$-$6.96(08) &1.53(16)\\
HCO$^+$(1--0) &25.63(09) &$-$6.91(01) &3.05(01)& &15.13(09) &$-$6.88(01) &2.86(02)\\
H$^{13}$CO$^+$(1--0) &2.64(10) &$-$7.18(05) &2.58(11) &&1.94(10) &$-$7.21(06) &2.26(14)\\
HNC(1--0) &16.62(17) &$-$6.77(01)&2.83(03)&&14.95(18) &$-$7.03(02) &2.51(04)\\
HN$^{13}$C(1--0) &0.68(05) &$-$6.42(12)&3.27(27) &&1.24(07) &$-$6.94(05)&1.75(11)\\
N$_2$H$^+$(1--0) &3.41(07) &$-$6.90(02) &2.20(05) &&6.33(09) &$-$7.00(01) &1.87(03)\\
\hline
\end{tabular}
\end{table*}

%\subsection{Continuum brightness and fluxes}

%In Table~\ref{table:cont-res} we present continuum brightness (in Jy/beam) and
%fluxes in 1\arcmin\ diameter circle towards the same positions.

\section{Physical parameters of the sources and molecular abundances}

\subsection{Kinetic temperatures}

The kinetic temperatures of the sources have been estimated from the CH$_3$C$_2$H
observations. As shown e.g. by \citet{Bergin94} this
symmetric-top molecule is a good ``thermometer'' for gas densities
$n\ga 10^4$~cm$^{-3}$. Our main goal is to compare the temperatures
at the peaks of the CS and N$_2$H$^+$ emission. At first, we derived
the temperatures from the CH$_3$C$_2$H $J=6-5$ data obtained at Onsala.
These measurements were done only at the peak positions. Later the
CH$_3$C$_2$H $J=13-12$ emission was mapped at IRAM with the multibeam
receiver. The details of the temperature estimates are published
elsewhere \citep[][Zinchenko et al., in preparation]{Malafeev05}. Here we
present the main results. In S187 the CH$_3$C$_2$H emission was too weak.

The kinetic temperatures at the CS and N$_2$H$^+$ peaks, derived from the Onsala and IRAM CH$_3$C$_2$H data, are listed in Table~\ref{table:Tkin}.
It is easy to see that in most cases there is no significant temperature
difference between the CS and N$_2$H$^+$ emission peaks, although
on average the N$_2$H$^+$ peaks are somewhat colder than the CS ones.

\begin{table}
\caption{Kinetic temperatures at the CS and N$_2$H$^+$ peaks
derived from the Onsala and IRAM CH$_3$C$_2$H data
($T_{\rm kin}^{6-5}$ and $T_{\rm kin}^{13-12}$, respectively).}
\label{table:Tkin}
\begin{tabular}{lrrlll}
\hline%\hline
Source & $\Delta \alpha$ & $\Delta \delta$ & $T_{\rm kin}^{6-5}$   & $T_{\rm kin}^{13-12}$ &Emission   \\
 & $('')$ & $('')$ & (K) & (K) &peak  \\
\hline
W3 & 20 & -40 & $52.6\pm 3.1$ &$58.4\pm 1.5$ & CS \\
   & 160 & -160 & $30.7\pm 0.8$ &$37.3\pm 2.2$ & N$_2$H$^+$ \\
%\hline
%AFGL6366 & 0   & 20 & $39.4\pm 2.8$ & &  \\
%         & 100 & 60 & $35.7\pm 4.6$ & &   \\
%\hline
S255 & 0 & 60 & $34.9\pm 1.4$ &$39.6\pm 0.4$& N$_2$H$^+$ \\
 & 0 & 0 & $34.5\pm 1.0$ &$38.8\pm 0.9$ & CS \\
DR21 & 0   & -40 & $28.8\pm 0.1$ &$34.8\pm 1.1$ & N$_2$H$^+$ \\
     & 0   & 0   & $33.4\pm 2.7$ &$44.0\pm 1.4$ & CS \\
S140  & 40 & 20 & $27.8\pm 1.6$ &$37.7\pm 2.2$ & N$_2$H$^+$ \\
      & 0  & 0  & $30.6\pm 0.7$ &$42.4\pm 1.0$ & CS \\
\hline
\end{tabular}
\end{table}

CH$_3$C$_2$H maps of sufficiently high quality were obtained at IRAM for
DR21, S140 and the CS peak in W3.
The IRAM data indicate somewhat higher peak temperatures than obtained
from the Onsala data, as expected due to a higher angular resolution
and higher excitation requirements for the $J=13-12$ transition if temperature increases towards an embedded heating source. Our data do show such temperature gradients consistent with theoretical expectations \citep{Zin05} but we do not discuss them here.
At the same time the IRAM data confirm our main conclusion that there is no
significant temperature difference between the CS and N$_2$H$^+$ peaks
in most cases, although
the N$_2$H$^+$ peaks are somewhat colder than the CS ones, on average.

It is interesting to compare our estimates of kinetic temperature with
other available data for these sources. Most of them have been actively
studied. Such comparison was made by
\citet{Malafeev05} and here we repeat the main points.

\subsubsection{W3}
Based on NH$_3$(1,1) and (2,2) lines,  \citet{Tieftrunk98} derived kinetic temperatures of 44  K and 25 K toward our CS and N2H+ peaks, respectively. A comparison with Table~\ref{table:Tkin}
shows that these temperatures are somewhat lower
than those obtained from the CH$_3$C$_2$H data, although the relation is
the same.

%\subsubsection{AFGL 6366 S}
%The IR source AFGL 6366 S (IRAS 06056+2131)
%was observed in the (1,1) and (2,2) ammonia lines
%by Schreyer et al. (\citep{Schreyer96}) with the 100-m radio telescope in
%Effelsberg. However, it was not mapped. The derived kinetic temperature
%in the centre is $\sim 26$~K. The dust temperature from the IRAS fluxes
%at 60 and 100~$\mu$m is 32.3~K. Our estimate of the kinetic temperature
%is close to this value and again is somewhat higher than the estimate
%obtained from ammonia observations. The temperatures of the observed peaks
%are almost the same.

\subsubsection{S255}
From our ammonia observations \citep{Zin97} the kinetic
temperature at the (0,+80$''$) position (near the NH$_3$ and N$_2$H$^+$
peaks) is $23\pm 1$~K. At the central position the uncertainty in the
kinetic temperature is too high. \citet{Schreyer96}
obtained the kinetic temperature in the centre from their ammonia data of
30.8~K and the dust temperature from the IRAS data of 31.1~K.
\citet{Mezger88} derived almost the same dust
temperatures ($\sim 30$ K) for the two peaks discussed here from their
1.3 mm and 350 $\mu$m continuum observations.

\subsubsection{DR21 NH$_3$}
This source was observed (without mapping)
in the CH$_3$C$_2$H lines at the 11-m NRAO
telescope \citep{Kuiper84}. They observed the same $J=6-5$
transition and derived practically the same kinetic
temperature at the centre as in our present work, 33.0~K. There are
VLA ammonia observations by \citet*{Mangum92}. They derived
kinetic temperatures of 32~K at the (0,0) position and 25~K near the
(0,$-$40$''$) position. This is rather close to our estimates.

\subsubsection{S140}
S140 was also observed in CH$_3$C$_2$H at the 11-m NRAO
telescope by \citet{Kuiper84}. They obtained for the cloud
centre the kinetic temperature of $32.1\pm 6.7$~K which coincides with our
$J=6-5$ estimate within the uncertainties. From ammonia observations at Effelsberg, \citet*{Ungerechts86}  found a temperature of  about 20~K,  once again a value significantly lower  than found with CH$_3$C$_2$H. The dust temperature from IRAS data is 34 K \citep{Zhou94}, more similar to our findings. Thus, CH$_3$C$_2$H appears to better trace the dust.

\subsection{Masses and densities}
There are several ways to estimate masses and densities of interstellar
clouds. Here we derive the source masses and column densities
primarily from the dust
continuum observations. This is considered to be one of the most reliable methods, partly due to the fact that dust/gas mass ratio is rather constant.
At the same time the uncertainties in the dust opacities are still rather high.
Additional uncertainties are related to uncertainties in the dust
temperature. Here we use the temperature estimates obtained from our
methyl acetylene observations, found to be similar to the dust temperature. It is known that at the typical densities
of dense cores the dust and gas kinetic temperatures are close to each other.
The results of these estimates are presented in Table~\ref{table:mass}.
It is worth noting that the masses here are not really total masses of the
sources because they are derived from the fluxes integrated over
1\arcmin\ diameter circle around the indicated positions. The dust opacities
were adopted from \citet{Ossenkopf94}. The gas/dust
mass ratio was assumed to be equal to 100. In addition we present also estimates of the average volume density along the line of sight obtained as $\bar{n}=N_{\rm L}/L$ where the size $L$ is derived from the angular size (Table~\ref{table:cont-res}) and distance to the source (Table~\ref{table:sources}).

In the last column we give estimates of the virial masses of the clumps based on these sizes and widths of optically thin lines at these positions. These estimates are rather uncertain, at least by a factor of 2 due to large uncertainties in the parameters. Nevertheless, within this factor they agree well with the masses derived from continuum observations. This shows that the clumps are close to gravitational equilibrium.

\begin{table}
\caption[]{Estimates of the gas column densities, average volume densities and masses from
the continuum observations. In the last column estimates of the virial masses are given.}
\footnotesize
\begin{tabular}{lcccll}
\hline%\hline
Source &Peak &$N_{\rm L}\times 10^{-23}$ &$\bar{n}\times10^{-5}$ &$M$ &$M_{\mathrm{vir}}$\\
& &(cm$^{-2}$) &(cm$^{-3}$) &(M$_{\sun}$) &(M$_{\sun}$) \\
\noalign{\smallskip}\hline\noalign{\smallskip}
S~187         &CS &0.4 &0.7 &30  &50\\
              &N$_2$H$^+$ &0.5 &1.0 &14  &10 \\
W3            &CS &2.0 &1.6 &405  &310 \\
              &N$_2$H$^+$ &1.2 &1.5 &74 &90 \\
%AFGL~6366 S   &0  &+20  &1.62 &180  & \\
%              &+100&+60 &0.50 &90  & \\
S~255         &CS  &2.0 &1.9 &290 &210\\
              &N$_2$H$^+$ &1.9 &1.6 &280 &220 \\
S~140         &CS   &2.8 &3.9 &86  &90\\
              &N$_2$H$^+$ &0.6 &0.7 &32  &70 \\
%\noalign{\smallskip}\noalign{\hrule}\noalign{\smallskip}}
\noalign{\smallskip}\hline\noalign{\smallskip}
\end{tabular}
\label{table:mass}
\end{table}

The average gas volume densities are $\bar{n} \sim 10^5$~cm$^{-3}$ in all cases. However, it is well known that such mean
densities can be significantly lower than densities found
from molecular excitation analysis \citep[e.g.][]{Zin98}.
This discrepancy is most probably explained by small-scale gas clumpiness. Such clumpiness is indicated by many studies \citep*[e.g.][]{Bergin96}. At the same time our data do not provide sufficient material
for excitation analysis in most cases. Only for S255 we have 3 mm and 1.3 mm data in
C$^{18}$O, CS, C$^{34}$S and methanol transitions towards both CS and N$_2$H$^+$ peaks which can be used
for modeling. In particular, our LVG estimates of gas density from the C$^{34}$S data give values of about $n\sim 5\times 10^5$~cm$^{-3}$ for both components. Methanol data modeling gives $n\sim 3\times 10^5$~cm$^{-3}$ also for both peaks (S.~Salii, private communication). Estimates based on C$^{18}$O are highly uncertain but are consistent with these values and show similar densities for both components because the line intensity ratios are similar. These estimates are rather close  to the average densities presented in Table~\ref{table:mass}, although somewhat higher as expected. This shows that the volume filling factor in this density range is rather high. The continuum data also indicate practically equal mean gas densities for the components.

\subsection{Abundances}
We derive here abundances of several species which were observed in all our sources and are the most informative ones for our purposes: C$^{18}$O, H$^{13}$CN, H$^{13}$CO$^+$, HN$^{13}$C, N$_2$H$^+$ and C$^{34}$S. In most cases we use only rare isotope data which are presumably not affected by optical depth effects. The optical depth in the N$_2$H$^+$ lines is also small or moderate as shown by \citet{Pirogov03}. The column densities are estimated from integrated line intensities in the LTE approximation assuming the excitation temperatures equal to the kinetic temperatures as given in Table~\ref{table:Tkin} (approximate values close to $T_{\rm kin}^{6-5}$ and $T_{\rm kin}^{13-12}$ were used). For S187 we assume $T_{\rm kin}$ of 20~K for the CS peak and $T_{\rm kin}$ of 10~K for the N$_2$H$^+$ peak (taking into account narrow line widths and very weak CH$_3$C$_2$H emission). Then, we derive abundances from comparison of the molecular column densities and total gas column densities obtained from dust continuum observations (Table~\ref{table:mass}). In order to take into account different beam sizes we correct the abundance estimates assuming Gaussian beams and Gaussian brightness distributions. For DR21 where we do not have our own dust continuum data, the abundances were derived assuming $X(\mathrm{C^{18}O})= 2\times 10^{-7}$. The results are summarized in Table~\ref{table:abundances}. \citet{Pirogov03} derived N$_2$H$^+$ abundances in a different way: from the N$_2$H$^+$ column densities and virial masses of the clouds. Nevertheless, their estimates are very close to those presented in Table~\ref{table:abundances} when the positions coincide (within a factor of 2).

\begin{table*}
\caption{Estimates of molecular and electron abundances towards selected positions in the sample sources. For C$^{18}$O, abundances derived from the $J=1-0$ and $J=2-1$ transitions are shown separately.}
\label{table:abundances}
\begin{tabular}{lrrllllllll}
\hline%\hline
\noalign{\smallskip}
Source & $\Delta \alpha$ & $\Delta \delta$ &$X$(C$^{18}$O)$_{10}$ &$X$(C$^{18}$O)$_{21}$ &$X$(H$^{13}$CN) &$X$(H$^{13}$CO$^+$) &$X$(HN$^{13}$C) &$X$(C$^{34}$S) &$X$(N$_{2}$H$^+$)  &$X_\mathrm{e}$\\
 & $('')$ & $('')$ & \\
\hline\noalign{\smallskip}
S187 &160 &0 &4.2E$-$07	&1.5E$-$07	&1.0E$-$10&5.8E$-$11 &6.5E$-$11	 &6.0E$-$10 &6.5E-11 &1.7E-07
 \\
     &0  &80 &2.5E$-$07	&&&7.6E$-$11	&6.2E$-$11	&&1.2E-10 &1.3E-07
  \\
W3 & 0 & $-$40 &2.9E$-$07 &&4.8E$-$10&8.6E$-$11	&7.6E$-$11	&&6.5E-11 &1.2E-07
  \\
   & 160 & $-$160 &2.0E$-$07 &&&1.9E$-$10 &8.8E$-$11 &&2.7E-10 &5.4E-08
  \\
S255 & 0 & 0 &2.4E$-$07 &1.4E$-$07 &2.8E$-$10 &5.7E$-$11 &4.9E$-$11	 &9.2E$-$10 &1.7E-10 &1.8E-07
 \\
 & 0 & 60 &1.8E$-$07 &1.3E$-$07 &2.3E$-$10 &1.9E$-$10 &1.2E$-$10 &7.7E$-$10 &3.7E-10 &5.4E-08 \\
DR21 & 0   & $-$20 & &&2.2E$-$10 &1.6E$-$10 &1.5E$-$10	&6.3E$-$10 &3.2E-10 &6.3E-08 \\
     & 0   & $-$40   &	&&2.4E$-$10	&1.5E$-$10 &1.9E$-$10 &7.2E$-$10 &4.7E-10 &6.7E-08 \\
S140  & 0 & 0 &1.3E$-$07	&&1.0E$-$10	&1.2E$-$10	&7.1E$-$11 &&1.5E-10 &8.4E-08
 \\
      & 40  & 20  &3.4E$-$07 &&1.4E$-$10 &2.9E$-$10 &2.7E$-$10 &&9.1E-10 &3.4E-08
 \\
\hline
\end{tabular}
\end{table*}

Non-LTE modeling for some clumps in our sample using the LVG approximation or the RADEX code (based on the escape probability method; \citealt{Vandertak07}) gives systematically lower (by a factor of 1.5--2) column densities for all species considered here, except C$^{18}$O, assuming temperature and density as described above. This is a natural result because C$^{18}$O is practically thermalized at these conditions while the other species are still far from equilibrium and population of lower levels exceeds that expected in LTE. Nevertheless, we believe that non-LTE modeling is not justified here due to the absence of the necessary data for most of the sources. One can bear in mind that the abundances derived here are somewhat overestimated.

\subsection{Ionization fraction}
The ionization fraction can be estimated from the HCO$^+$ abundance using the results of relevant chemical models. It is known that in dense clouds HCO$^+$ is formed mainly from H$_3^+$ \citep[e.g.][]{Turner95}:
\begin{equation}\label{eq:H3+}
    \mathrm{H_3^+} + \mathrm{CO} \rightarrow \mathrm{HCO^+} + \mathrm{H_2}
\end{equation}
and is destroyed by dissociative recombination:
\begin{equation}\label{eq:HCO+}
    \mathrm{HCO^+} + e \rightarrow \mathrm{CO} + \mathrm{H}
\end{equation}

In addition, one has to take into account HCO$^+$ recombination onto negatively charged dust grains \citep[e.g.][]{Caselli08}. The rate of this reaction is given by the product $k_\mathrm{g} X_\mathrm{g}$ where the rate coefficient $k_\mathrm{g}$ and the fractional abundance of dust grains $X_\mathrm{g}$ are determined by grain size distribution and gas kinetic temperature \citep{Draine87}. For the typical temperatures in our objects ($\sim 30$~K) and MRN \citep{Mathis77} grain size distribution $k_\mathrm{g} X_\mathrm{g} \approx 3\times 10^{-15}$~s$^{-1}$.

In this model for HCO$^+$ abundance in steady state we can write:
\begin{equation}\label{eq:X_HCO+}
    X(\mathrm{HCO^+}) = \frac{k_1 X(\mathrm{H_3^+}) X(\mathrm{CO})}{\alpha(\mathrm{HCO^+}) X_e + k_\mathrm{g} X_\mathrm{g}}
\end{equation}
where $k_1$ is the rate of reaction (\ref{eq:H3+}) and $\alpha(\mathrm{HCO^+})$ is the HCO$^+$ dissociative recombination rate.

H$_3^+$ is formed by cosmic-ray ionization of molecular hydrogen and is destroyed in dense clouds primarily by reaction (\ref{eq:H3+}) \citep[e.g.][]{Black00}. For regions where CO is not frozen onto dust grains, recombination onto negatively charged grains is a negligible H$_3^+$ destruction process compared with this reaction. Therefore, for H$_3^+$ abundance in this regime we obtain:
\begin{equation}\label{eq:X_H3+}
    X(\mathrm{H_3^+}) = \frac{\zeta/n}{k_1 X(\mathrm{CO})}
\end{equation}
where $\zeta$ is the cosmic-ray ionization rate and $n$ is the total gas density. Now combining Eq.~(\ref{eq:X_HCO+}) and (\ref{eq:X_H3+}) we come to the following simple expression:
\begin{equation}\label{eq:X_HCO+_2}
    X(\mathrm{HCO^+}) = \frac{\zeta/n}{\alpha(\mathrm{HCO^+}) X_e + k_\mathrm{g} X_\mathrm{g}}
\end{equation}
Estimates of the electron abundances based on this formula are presented in the last column of Table~\ref{table:abundances}. The cosmic-ray ionization rate was assumed to be equal to $\zeta = 3\times 10^{-17}$~s$^{-1}$, $\alpha(\mathrm{HCO^+}) = 7.5\times 10^{-7}$~s$^{-1}$cm$^3$ \citep{Turner95}. The gas density was assumed to be $n= 10^5$~cm$^{-3}$ and $[\mathrm{HCO^+}]/[\mathrm{H^{13}CO^+}]=40$. The $k_\mathrm{g} X_\mathrm{g}$ term leads to corrections less than 10\% in the $X_\mathrm{e}$ values and we neglected it.
Strictly speaking, in this way we derive more reliably the parameter $nX_\mathrm{e}$, i.e. the electron density, not abundance. However, for a better comparison with other results we will discuss further mainly the $X_\mathrm{e}$ values.

\section{Discussion}

\subsection{Variations of molecular abundances}

An inspection of the observational results and estimates of the physical parameters presented above leads to several important conclusions:

(1) There are strong variations of the relative intensities in the lines of different molecular tracers across the investigated high mass star forming regions. In Paper~I we mentioned already striking differences between N$_2$H$^+$ and CS maps and the fact that the CS distribution follows that of the dust emission while N$_2$H$^+$ does not. Now we see that C$^{18}$O and HCN, like CS, are good tracers of the dust emission which presumably shows the total mass distribution. At the same time the behavior of HCO$^+$ and HNC resembles that of N$_2$H$^+$. Significant differences in distributions of various species in particular sources have been noticed in some other studies \citep[e.g.][]{Ungerechts97} but here we see systematic effects common for the sources in the sample.

(2) There is no sign of CO and/or CS depletion in these objects (in contrast to cold dark clouds). The abundances derived for C$^{18}$O from the $J=1-0$ and $J=2-1$ transitions more or less agree with each other. They are practically constant within each source and among the whole sample (with small deviations which can be caused e.g. by temperature uncertainties) and are close to the ``canonical'' values derived earlier (e.g. $X(\mathrm{C^{18}O}) \approx 1.7\times 10^{-7}$ obtained by \citealt*{Frerking82}).

\begin{figure}
  \includegraphics[width=\columnwidth]{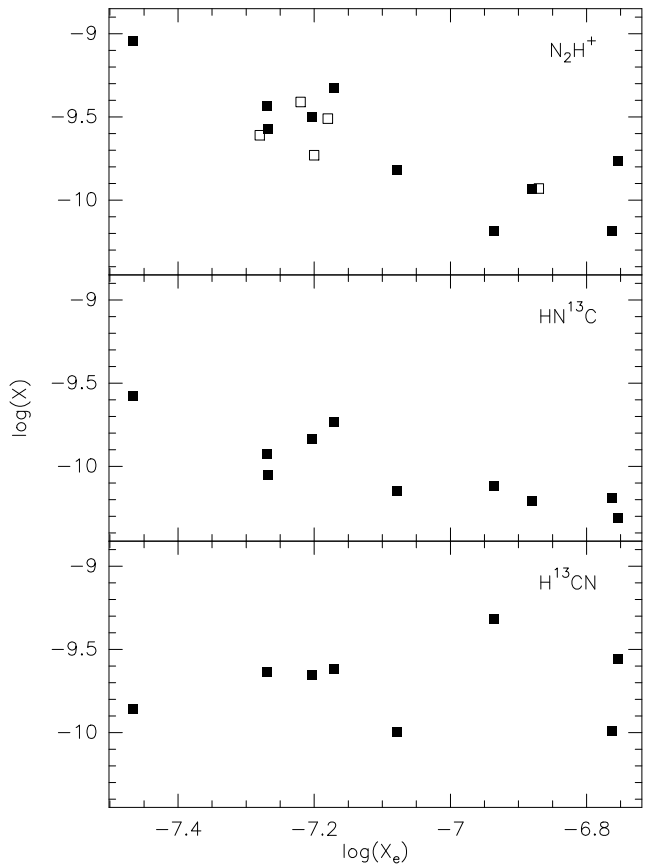}\\
  \caption{Relative abundances of N$_2$H$^+$, HN$^{13}$C and H$^{13}$CN (from top to bottom) in dependence on the electron abundance. Our estimates are plotted by the filled symbols. The open symbols correspond to the data presented by \citet{Bergin99}.}\label{fig:Xe-abund}
\end{figure}

(3) The N$_2$H$^+$ and HNC abundances significantly decrease with increasing ionization fraction (Fig.~\ref{fig:Xe-abund}). The best fit gives $X(\mathrm{N_2H^+})\propto X_\mathrm{e}^{-1.3\pm 0.3}$ and $X(\mathrm{HN^{13}C})\propto X_\mathrm{e}^{-0.8\pm 0.2}$ with the correlation coefficients of $\vert\rho \vert \approx 0.85$ for both dependences. At the same time the HCN abundance does not show significant variations. It is interesting that the data for N$_2$H$^+$ and $X_\mathrm{e}$ presented by \citet{Bergin99} are in excellent agreement with the $X(\mathrm{N_2H^+}) - X_\mathrm{e}$ relation obtained here, as one can see in Fig.~\ref{fig:Xe-abund}, where the open symbols are the $X(\mathrm{N_2H^+})$ values derived from the N$_2$H$^+$ and C$^{18}$O column densities presented by \citet{Bergin99}, and assuming $X(\mathrm{C^{18}O}) = 1.7\times 10^{-7}$ \citep{Frerking82}.

These correlations should not be affected much by the uncertainties in the abundances due to the LTE approximation, given that both variables scale in a similar way if the non-LTE approach is adopted.

(4) The derived ionization fraction in these objects in general increases towards the strong embedded IR sources which coincide with the main peaks of the dust and CS emission (by a factor of 2$-$3). The N$_2$H$^+$, HCO$^+$ and HNC abundances decrease correspondingly in these areas. This is consistent with our results in Paper~I where we found that the N$_2$H$^+$ abundance is systematically lower towards the CS/dust emission peaks which coincide with strong IR sources. However, it is important to emphasize that estimates of the electron abundance were performed under the assumption of a constant gas density. Our estimates of the gas density towards S255 IR and N give practically the same values. At the same time it is not unreasonable to expect density variations which can smooth the derived variations of the ionization fraction.

It is worth noting that these variations of the ionization fraction refer to the values averaged over the line of sight and an increase of the electron abundance in vicinities of a young star can be significantly larger. \citet{Ungerechts97} found similar variations of molecular abundances in the Orion molecular cloud and suggested that the abundances of molecular ions can be reduced by a higher electron abundance caused by UV radiation propagating in a clumpy photodissociation region. It is well known that in clumpy media UV photons can penetrate deep into dense clouds leading to significantly enhanced ionization \citep[e.g.][]{Bethell07}.
At the same time in their study of the ionization fraction in massive cores, \citet{Bergin99} did not find any noticeable increase of the electron abundance from the edge to the centre of massive cores. This can be caused perhaps by insufficient luminosities of their sample sources.

Thus, we believe that UV radiation from young massive stars in clumpy media can be the primary cause of the observed ionization enhancement. In addition, X-rays detected already from many massive stars \citep[e.g.][]{Zhekov07} can be also responsible for this effect.

%\begin{figure}
%  % Requires \usepackage{graphicx}
%  \includegraphics[width=\columnwidth]{Tk-abund.eps}\\
%  \caption{Relative abundances of N$_2$H$^+$, HN$^{13}$C and H$^{13}$CN (from top to bottom) in dependence on the kinetic temperature.}\label{fig:Tk-abund}
%\end{figure}

(5) We see no dependence of the relative abundances on the velocity dispersion, as derived from the observed line widths. In principle such dependence could be expected in particular for N$_2$H$^+$ if it really escapes perturbed regions as suggested e.g. by \citet*{Womack92}.

(6) There may be a trend for decreasing N$_2$H$^+$ and increasing H$^{13}$CN abundances with increasing temperature. However, this is based on only one point (W3 data) and cannot be considered reliable.

%\subsection{Individual sources}

%\subsubsection{S255} \label{sec:S255}

\subsection{Chemical implications}

Apparently the observed differences in molecular distributions cannot be
explained by molecular freeze-out
as in low-mass cores. The kinetic (and dust) temperatures at the CS and N$_2$H$^+$
peaks are similar in most cases and rather high, $\ga 30$~K, which
probably makes freeze-out ineffective. This is confirmed by the absence of any indication of the CO and CS depletion as mentioned above.

One possible explanation for the observed chemical differentiations was
proposed by \citet{Lintott05}. They
suggested that the enhancement of CS and reduction in
$\mathrm{N_2H^+}$ abundance found in regions of high-mass star
formation may be related to the high dynamical activity in these
regions which could enhance the rate of collapse of cores above the
free-fall rate. Consequently, high gas densities would be achieved
before freeze-out had removed the molecules responsible for
$\mathrm{N_2H^+}$ loss, while the high densities promote CS
formation.

However, this model has several drawbacks. In particular, its predictions
for some species (e.g. SO) are not supported by observations. Then,
it predicts a decrease in the abundance accompanied by CS
enhancement. Our data show no sign of the CS abundance variations.
Nothing to say that an accelerated collapse itself has no more or less
advanced physical basement.

Our results presented in the previous section indicate that ionization fraction is probably more important in establishing the steady-state N$_2$H$^+$ and HNC abundance in massive cores. The only important process which forms N$_2$H$^+$ is \citep[e.g.][]{Turner95}:
\begin{equation}\label{eq:N2H+}
    \mathrm{H_3^+} + \mathrm{N_2} \rightarrow \mathrm{N_2H^+} + \mathrm{H_2}
\end{equation}

There are two main processes which destroy N$_2$H$^+$: the reaction with the CO molecule
\begin{equation}\label{eq:N2H+-CO}
    \mathrm{N_2H^+} + \mathrm{CO} \rightarrow \mathrm{HCO^+} + \mathrm{N_2}
\end{equation}
and dissociative recombination
\begin{eqnarray}\label{eq:N2H+-e1}
    \mathrm{N_2H^+} + e &\rightarrow& \mathrm{N_2} + \mathrm{H}\\
\mathrm{N_2H^+} + e &\rightarrow& \mathrm{NH} + \mathrm{N} \label{eq:N2H+-e2}
\end{eqnarray}
which mainly lead to the formation of N$_2$ (90\% of the total reaction), as recently found by \citet{Molek07}.

Then, we have to take into account N$_2$H$^+$ recombination onto negatively charged dust grains as in the case of HCO$^+$.
The steady state N$_2$H$^+$ abundance in this model will be given by the formula:
\begin{equation}\label{eq:X_N2H+}
    X(\mathrm{N_2H^+}) = \frac{k_2 X(\mathrm{H_3^+}) X(\mathrm{N_2})}{k_3 X(\mathrm{CO})+\alpha(\mathrm{N_2H^+}) X_e + k_\mathrm{g} X_\mathrm{g}}
\end{equation}
where $k_2$ is the rate of the reaction (\ref{eq:N2H+}), $k_3$ is the rate of the reaction (\ref{eq:N2H+-CO}) which is $8.8\times 10^{-10}$~cm$^3$s$^{-1}$ according to the UMIST database and $\alpha(\mathrm{N_2H^+})$ is the summary rate of the reactions (\ref{eq:N2H+-e1}),  (\ref{eq:N2H+-e2}). Its value is significantly different in the UMIST and OSU databases. We accept the value of $8\times 10^{-7}$~cm$^3$s$^{-1}$ at 30~K (E.~Herbst, private communication). This means that the dissociative recombination will dominate at $X_\mathrm{e} \ga 10^{-3}X(\mathrm{CO})$, i.e. at $X_\mathrm{e} \ga 10^{-7}$ for the standard CO abundance $X(\mathrm{CO}) \sim 10^{-4}$. This is close to an average electron abundances derived in this work. Therefore, in principle it is possible that the dissociative recombination of N$_2$H$^+$ can dominate at least in part of our objects. Then, if the H$_3^+$ and N$_2$ abundances are more or less constant (as can be expected) we obtain for $X(\mathrm{N_2H^+})$ the dependence on $X_\mathrm{e}$ similar to that presented in Fig.~\ref{fig:Xe-abund}.

It is less clear how to explain the behavior of the HCN and HNC abundances. Their formation pathways are very different. HCN definitely forms from
the neutral-neutral processes \citep*{Turner97}
\begin{equation}
\mathrm{N} + (\mathrm{CH_2},\mathrm{CH_3}) \rightarrow \mathrm{HCN}
\label{eq:HCN}
\end{equation}
HNC forms via the distinctly independent sequence
\begin{equation}
\mathrm{C^+} + \mathrm{NH_3} \rightarrow \mathrm{H_2CN^+} + {e}
\rightarrow \mathrm{HNC}
\label{eq:HNC}
\end{equation}

The main destruction processes for HNC are probably reactions with C$^+$ and H$_3^+$ \citep{Turner97}. Calculations of steady state abundances of these species require additional chemical modeling.
We note that the HNC formation is closely linked to NH$_3$. Given that NH$_3$ and N$_2$H$^+$ both form from N$_2$, thus they are chemically related, one expects similar morphologies of HNC and N$_2$H$^+$, as in fact we observe.

\subsection{Chemical indicators of massive protostars}
One of the most intriguing problems in studies of high mass star formation is the identification of massive protostars at the earliest phases of evolution. Some clumps from our sample represent probably such protostellar objects. For example the SE clump in W3 area is associated with a water maser but has no embedded IR sources and/or UC H~II regions. The HCO$^+$ line profile shows a red-shifted self-absorption feature typical for a collapsing cloud. The mass of this clump from continuum data is about 70~M$_\odot$ (Table~\ref{table:mass}). Therefore, it can be a massive protostar on a rather early evolutionary stage. It is also very pronounced in N$_2$H$^+$ emission. Another example of this kind is the N$_2$H$^+$ emission peak in S187 (although we did not see signs of contraction on the line profiles). In the S255 area, the northern component which is dominant in N$_2$H$^+$, is apparently much younger than S255 IR. These examples show that a relatively strong N$_2$H$^+$ emission can be considered as an indicator of such objects. The species which correlate with N$_2$H$^+$ (HCO$^+$ and HNC) can be also useful in this respect.

\section{Conclusions}
We presented and discussed here observations of five regions of active high mass star formation in various molecular lines (at 3 mm and at 1.3 mm) and in continuum at 1.3 mm. On the basis of these observations we estimated physical parameters of the sources and molecular abundances.
The main results of this study can be summarized as follows:
\begin{enumerate}
\item The typical physical parameters for the sources in our sample are: kinetic temperature in the range $\sim 30-50$~K, masses from tens to hundreds solar masses, gas densities $\sim 10^5$~cm$^{-3}$, ionization fraction $\sim 10^{-7}$. In most cases the ionization fraction slightly (a few times) increases towards the embedded YSOs. The observed clumps are close to gravitational equilibrium. Our temperature estimates are systematically lower (by a factor of about 1.5--2) compared to those obtained with NH$_3$ observations. However, temperatures measured with CH$_3$C$_2$H are similar to dust temperatures, suggesting that the observed methyl acetylene transitions better trace the dust than ammonia (1,1) and (2,2) lines.
\item There are systematic differences in distributions of various molecules in regions of high mass star formation. The abundances of CO, CS and HCN are more or less constant and optically thin lines of rare isotopes of these species are good tracers of the dense gas distribution in these regions. There is no sign of CO and/or CS depletion as in cold low mass cores.
\item At the same time, the abundances of the high density tracers HCO$^+$, HNC and especially N$_2$H$^+$, strongly vary in these objects. They anti-correlate with the ionization fraction ($X(\mathrm{N_2H^+})\propto X_\mathrm{e}^{-1.3\pm 0.3}$ and $X(\mathrm{HN^{13}C})\propto X_\mathrm{e}^{-0.8\pm 0.2}$) and as a result decrease towards the embedded YSOs. For N$_2$H$^+$ this can be explained by dissociative recombination to be the dominant destroying process. This conclusion is more or less consistent with the data on chemical reaction rates. There is no correlation of these abundances with the line width.
\item The described variations of the HCO$^+$, HNC and N$_2$H$^+$ abundances make them potentially valuable indicators of massive protostars. In our sample there are some clumps which represent probably massive protostars at very early stages of evolution and they are very pronounced in the lines of these species, especially N$_2$H$^+$.
\end{enumerate}

\section*{Acknowledgements}
%\begin{acknowledgements}
The invaluable contributions to obtaining observational data and to initial discussions were made by Lars E.B. Johansson and Barry Turner who recently passed away. We acknowledge the support from the telescope staff at OSO, NRAO and IRAM. Estimates of the physical parameters from the methanol data were made by Svetlana Salii and kinetic temperature estimates by Sergey Malafeef who also participated actively in the observations at Onsala. We are grateful to Alexander Lapinov for providing his LVG code and to Eric Herbst who provided the rate coefficient for the dissociative recombination of N$_2$H$^+$. The constructive comments by the referee, Ted Bergin, helped to improve the manuscript.

The work was supported by
Russian Foundation for Basic Research grants 03-02-16307 and 06-02-16317, by the Russian Academy research program ``Extended objects in the Universe'' and by the INTAS grant 99-1667.
The research has made use of the SIMBAD database,
operated by CDS, Strasbourg, France.

%\end{acknowledgements}

\bsp

\label{lastpage}

\end{document}